\documentclass[twocolumn,aps,prx,floatfix,superscriptaddress]{revtex4-2}

\usepackage{amsmath,amssymb,amsfonts,graphicx,braket,hyperref}
\hypersetup{colorlinks = true}

\graphicspath{{img/},{img/nisq/}{./}}
\usepackage{wrapfig}
\usepackage{fontawesome5}
\usepackage{hyperref}

\usepackage[ruled,lined]{algorithm2e}

\usepackage{physics}

\usepackage{bbold}

\usepackage{xcolor}

\usepackage[acronym]{glossaries}
 
\newacronym{rl}{RL}{Reinforcement learning}
\newacronym{ml}{ML}{Machine learning}
\newacronym{ac}{AC}{actor critic}
\newacronym{nisq}{NISQ}{noisy intermediate-scale quantum}
\newacronym{qst}{QST}{quantum state tomography}

\begin{document}

\title{
Reinforcement Learning to Disentangle Multiqubit Quantum States \\ from Partial Observations
}

\author{Pavel Tashev}
\thanks{equal contribution}
\affiliation{Department of Mathematics and Informatics, St.~Kliment Ohridski University of Sofia, 5 James Bourchier Blvd, 1164 Sofia, Bulgaria}

\author{Stefan Petrov}
\thanks{equal contribution}
\affiliation{Department of Mathematics and Informatics, St.~Kliment Ohridski University of Sofia, 5 James Bourchier Blvd, 1164 Sofia, Bulgaria}

\author{Matthew~T.~Diaz}
\affiliation{Joint Quantum Institute and Department of Physics, University of Maryland, College Park, Maryland 20742, USA}
\affiliation{National Quantum Laboratory (QLab), University of Maryland, College Park, MD 20742 USA}

\author{Friederike~Metz}
\affiliation{Institute of Physics, École Polytechnique Fédérale de Lausanne (EPFL), CH-1015 Lausanne, Switzerland}
\affiliation{Center for Quantum Science and Engineering, École Polytechnique Fédérale de Lausanne (EPFL), CH-1015 Lausanne, Switzerland}
\affiliation{Quantum Systems Unit, OIST Graduate University, Onna, Okinawa 904-0495, Japan}

\author{Alaina~M.~Green}
\affiliation{Joint Quantum Institute and Department of Physics, University of Maryland, College Park, Maryland 20742, USA}
\affiliation{National Quantum Laboratory (QLab), University of Maryland, College Park, MD 20742 USA}

\author{Norbert~M.~Linke}
\affiliation{Joint Quantum Institute and Department of Physics, University of Maryland, College Park, Maryland 20742, USA}
\affiliation{National Quantum Laboratory (QLab), University of Maryland, College Park, MD 20742 USA}
\affiliation{Duke Quantum Center and Department of Physics, Duke University, Durham, North Carolina 27701, USA}

\author{Marin Bukov}
\email{mgbukov@pks.mpg.de}
\affiliation{Max Planck Institute for the Physics of Complex Systems, N\"othnitzer Str.~38, 01187 Dresden, Germany}

\date{\today}

\begin{abstract}

Using partial knowledge of a quantum state to control multiqubit entanglement is a largely unexplored paradigm in the emerging field of quantum interactive dynamics with the potential to address outstanding challenges in quantum state preparation and compression, quantum control, and quantum complexity.
We present a deep reinforcement learning (RL) approach using an actor-critic algorithm for constructing short disentangling circuits for states with up to 16 qubits.
With access to only two-qubit reduced density matrices, our agent decides which pairs of qubits to apply two-qubit gates on; requiring only local information makes it directly applicable on modern NISQ devices, as we demonstrated experimentally on a trapped-ion quantum computer.
Utilizing a permutation-equivariant transformer architecture, the agent can autonomously identify qubit permutations within the state, and adjusts the disentangling protocol accordingly. 
Once trained, it provides circuits from different initial states without further optimization. 
We demonstrate the agent's ability to identify and exploit the entanglement structure of multi-qubit states.
We analyze the disentangling circuits constructed by the agent for 4- and 5-qubit Haar-random states, and observe strong correlations between consecutive gates and among the qubits involved.
Through extensive benchmarking, we show the efficacy of the RL approach to find disentangling protocols with minimal gate resources. 
We explore the resilience of our trained agents to noise, highlighting their potential for real-world quantum computing applications.
Analyzing optimal disentangling protocols, we report a general circuit to prepare an arbitrary 4-qubit state using at most 5 two-qubit (10 CNOT) gates.
\end{abstract}

\let\oldaddcontentsline\addcontentsline
\renewcommand{\addcontentsline}[3]{}
\maketitle

\begin{figure*}[t!]
\centering
\includegraphics[width=0.95\textwidth]{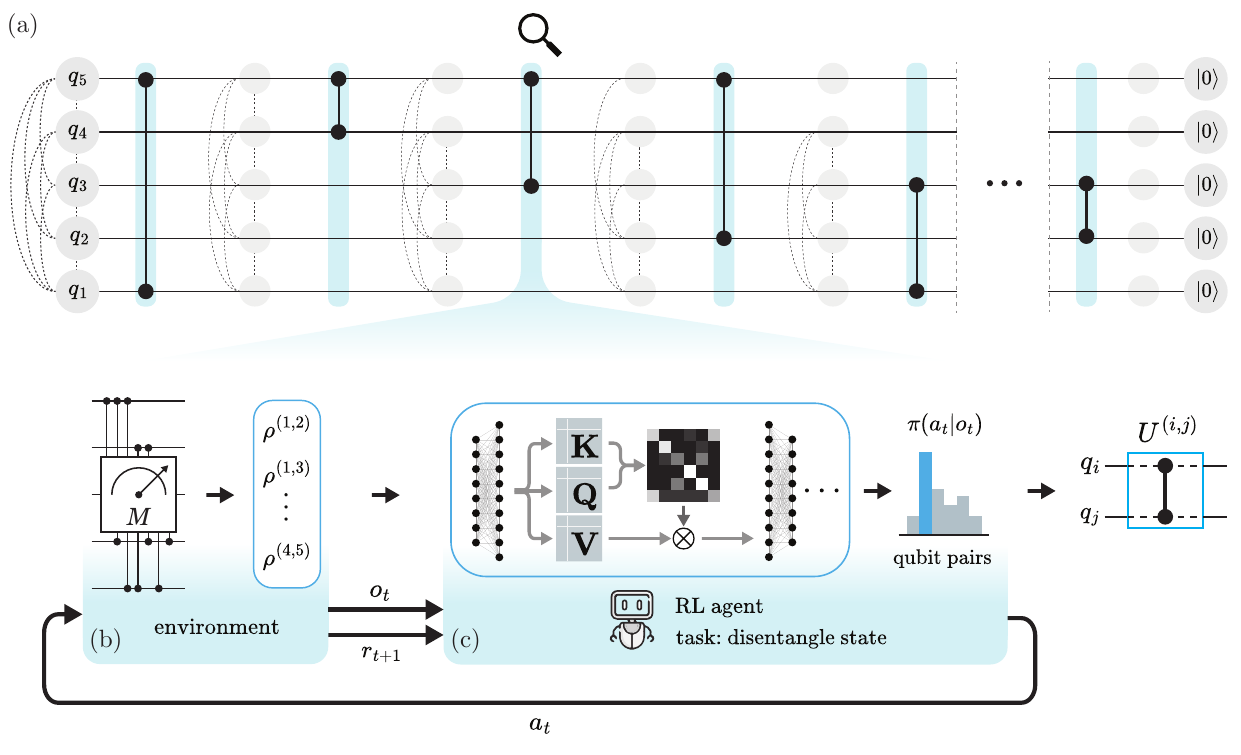}
\caption{
    Schematic representation of the pipeline used in this work. 
    (a) Given an arbitrary entangled state, we design a noise-resilient RL agent to construct a protocol that disentangles it in a minimum number of steps. Blue shaded rectangles indicate a single step of the RL feedback loop which produces a two-qubit unitary gate to apply onto the state (black dots indicate the two qubits acted on). Light grey circles and dashed arcs sketch pictorially the process of reducing entanglement in the multi-qubit state. 
    (b) Reinforcement learning environment: at each step $t$, the agent is given 
    (i) a partial observation $o_t$, consisting of all two-qubit reduced density matrices of the state, and
    (ii) a reward signal $r_{t+1}$, used to train the agent to find optimal disentangling gates. 
    (c) The RL agent consists of the so-called policy $\pi(a_t|o_t)$ -- a model for a probability distribution over action space (here the qubit pairs $(i, j)$). At each time step the agent selects that qubit pair to apply a quantum gate $U^{(i,j)}$ on, which maximizes the policy; the gate itself is determined analytically. 
    To do this, the observation $o_t$ is fed into a permutation equivariant transformer neural network, which is trained using the rewards $r_t$ to approximate the optimal disentangling policy. 
    This procedure is repeated in a feedback loop until the state is disentangled.
}
\end{figure*}

\section{Introduction} \label{sec:intro}

Quantum entanglement is central to modern quantum technologies. It is widely perceived as a proxy for the quantum nature of physical processes and phenomena involving more than one particle. 
Entanglement is currently contemplated as an instrumental resource for quantum computing~\cite{nielsen2010quantum}, and plays a fundamental role in quantum optics~\cite{raimond2001manipulating} and strongly-correlated condensed matter systems~\cite{laflorencie2016quantum}. 
Therefore, finding protocols to control the dynamics of entanglement is of primary importance in modern physics.   
 
On the fundamental side, maximally entangled two-qubit states can be generated from initial product states using perfect entanglers~\cite{kraus2001optimal,watts2015optimizing,goerz2015optimizing}. 
Scaling up such protocols to produce ordered many-body entangled states is a cornerstone of quantum simulation~\cite{elredge2017fast,omran2019generation,madjarov2020high,holland2023demand,iqbal2023topological,bluvstein2024logical}.
The difficulty in controlling entanglement among many quantum particles with only experimentally accessible local gates, lies in the exponential size of their joint Hilbert space dimension: states exist that require an exponential (in the system size) number of two-qubit gates to disentangle.
Moreover, it was demonstrated that a unique (state-independent) quantum gate cannot disentangle arbitrary mixed two-qubit states~\cite{terno1999nonlinear}, implying that an ideal universal disentangling machine does not exist~\cite{mor1999disentangling, zhou2000disentangling,bandyopadhyay1999disentanglement, ghosh2000optimal}.

More practically, manipulating many-body entanglement~\cite{dehaghani2024enhancing} is currently being investigated within the framework of matrix product states~\cite{hyatt2017extracting,hauschild2018finding, ljubotina2022optimal}, flow equations~\cite{kehrein2017flow}, and using optimal control~\cite{lu2023persistent}. 
Since unitary gates are invertible, disentangling circuits present a way of constructing algorithms to initialize arbitrary states on quantum computers~\cite{shende2006}; hence, the complexity of many-body state preparation can be quantified by the amount of entanglement present in the target state~\cite{chamon2014emergent,shaffer2014irreversibility,yang2017entanglement,araujo2021approximated}. 
Beyond state preparation, disentangling routines serve as a building block to implement arbitrary unitaries using local two-qubit gates~\cite{shende2006,mottonen2004transformation}.

Recent progress notwithstanding, very little is known about how to exploit the distribution of entanglement among qubits to optimally disentangle quantum states, while at the same time keeping the number of two-qubit gates minimal. In the era of noisy intermediate-scale quantum (NISQ) computing, especially pressing is the necessity to design algorithms that can identify the structure of the entanglement distribution of a state, and use it to improve disentangling protocols in the presence of noise and decoherence. 
In this work, we investigate the question as to whether or not, and under which conditions, it is advantageous to exploit partial knowledge of the quantum state to gain an advantage in addressing these issues.

As we will show, a particularly suitable framework for this purpose is
Reinforcement Learning (RL)~\cite{sutton_barto_rl} -- a branch of Machine Learning (ML)~\cite{mehta2019highbias,carleo2019machine,carrasquilla2020machine,krenn2023artificial} that uses interactive feedback dynamics between an agent and its environment to control a physical system. 
RL has been employed to design quantum circuits for preparing specific classes of 4-qubit entangled states \cite{Giordano22}, for determining the entanglement structure of spin systems by disentangling a single qubit from the rest of the system \cite{an2022learning}, and to engineer entanglement in optomechanical systems~\cite{ye2024entanglement}. Low-entanglement protocols that transfer the population between ordered quantum many-body ground states have been discovered using tensor-network-based reinforcement learning~\cite{metz2022self}. More generally, RL has been successfully leveraged for a wide range of quantum state preparation \cite{foesel2018reinforcement,bukov2018,bukov2018b,dalgaard2020global,Haug2020,mackeprang2020,yao2020,yao2020b,guo2021,he2021,guo2021,cao2021,Porotti2022,Porotti2022b,Sivak2022,Reuer2022,Yao22,Zhang2019,Qiu2022,Wauters20,kolle2024reinforcement} and circuit compilation \cite{zhang2020,bolens2021,moro2021,he2021b,fosel2021quantum,Li-2023,Quetschlich23,Pozzi22,HerreraMarti2022policygradient,chen2022efficient,nguyen2024reinforcement,ivanova2024discovery, kremer2024practical} tasks, fundamentally related to the inverse process of disentangling quantum states~\cite{bukov2026reinforcement}.

In this work, we investigate the problem of disentangling arbitrary multi-qubit quantum states [Sec.~\ref{sec:setup}] by having access to only partial information of these states. We make use of a deep RL algorithm known as actor critic (AC) in order to train an agent to find the shortest protocol that transforms an initial entangled state into a product state. Importantly, our agent needs access only to the two-qubit reduced density matrices of a state to generate a two-qubit disentangling gate; it can, thus, be applied to realistic NISQ devices [App.~\ref{app:nisq}]. Moreover, the agent is equipped with a permutation equivariant transformer architecture that allows it to learn policies insensitive to arbitrary qubit permutations, cf.~Sec~\ref{sec:RL_framework}.

We leverage the agent's interpolation and extrapolation capabilities to construct approximately optimal circuits that disentangle 4-, 5-, and 6-qubit Haar-random states which lack any obvious spatial entanglement structure in the computational basis: 
for arbitrary 4-qubit states, we report a perfect disentangling sequence consisting of at most five 2-qubit (yet only ten CNOT, cf.~App.~\ref{subsec:small_sys_exact}) gates; inverting it gives a protocol to prepare an arbitrary 4-qubit state with unit fidelity.
For 5- and 6-qubit Haar-random entangled initial states, the agent requires on average 20 and 56 two-qubit gates, respectively. 
We test our agent's performance on a trapped-ion quantum computer, and find that, even in the presence of noise, it manages to completely reduce the entanglement between the qubits despite the state being mixed with the environment.

We further explore the agent's capabilities to construct disentangling circuits for large, weakly entangled subsystems of Haar-random states of up to 16 qubits.
We analyze the protocols the RL agent learns and show that they feature both spatial (i.e., among the qubits) and temporal (i.e., between consecutive gates) correlations. Moreover, whenever present, the trained agent successfully recognizes local entanglement structures and uses them to construct shorter protocols, see Sec.~\ref{sec:RL-agent}. Our results demonstrate that, compared to deterministic disentangling algorithms, state-aware disentangling RL agents learn to utilize measurable partial information about the quantum state to reduce the number of required CNOT operations.

Finally, we quantify and analyze the capability of our RL agent to adjust to noisy environments both in simulation and on realistic devices in Sec.~\ref{sec:applns_noisy}. We show that the RL protocols are robust to moderate levels of sampling and NISQ-hardware noise.

\section{Multiqubit Disentangling Problem} \label{sec:setup}

Consider an arbitrary pure $L$-qubit state $|\psi\rangle$. The multiqubit disentangling problem seeks to find a unitary operation $U$ that brings $|\psi\rangle$ to the product state $|\psi_\ast\rangle=|0\rangle^{\otimes L}$ in the computational $z$-basis, i.e., $U|\psi\rangle=|\psi_\ast\rangle$.
Thus, $U$ necessarily disentangles $|\psi\rangle$ 
\footnote{Note that the most general product (i.e., zero-entanglement) pure state can be mapped to $|\psi_\ast\rangle$ by applying at most $L$ single-qubit rotations. Since these extra operations do not alter the entanglement structure of the state, we will consider them part of $U$.}.  

We formulate the search for $U$ as an entanglement minimization problem.
Note the combinatorially large number of ways we can partition $L$ qubits into two subsystems.
Hence, a simple measure for the total amount of entanglement in a given state is the maximum single-qubit entanglement:
\begin{equation} \label{eq:max_sqe}
    S_\text{tot} = \max_{1 \leq j \leq L} S_\text{ent}[\rho^{(j)}],
\end{equation}
where $\rho^{(j)} = \mathrm{tr}_{\{1,\dots,j-1,j+1,\dots,L\}} |\psi\rangle\langle\psi|$ is the reduced density matrix of qubit $j$. Since $S_\text{ent}[\rho^{(j)}]\geq 0$, $S_\text{ent}=0$ if and only if $|\psi\rangle$ is a product state.

The system entanglement serves as a natural criterion for assessing whether a given state has been successfully disentangled. However, this quantity is not continuous and therefore not directly amenable to gradient-based optimization. To circumvent this limitation, we instead define a smooth cost function based on the average single-qubit entanglement:

\begin{equation} \label{eq:avg_entanglement}
    S_\text{avg} = \frac{1}{L}\sum_{j=1}^L S_\text{ent}[\rho^{(j)}]\, .
\end{equation}
In what follows, we use the von Neumann entropy for $S_\text{ent}$, although other measures, e.g., the Renyi entropies or the Fisher information, can also be considered.

To comply with the constraints of modern NISQ architectures, we furthermore require that $U$ be represented as a sequence of two-qubit gates $U^{(i,j)}$. We let  $U^{(i,j)}$ act on any pair of qubits $(i,j)$ (not just neighboring ones), which is native to trapped-ion or Rydberg atom platforms. In addition, we also seek to find an algorithm that exhibits resilience to noise.

The disentangling problem then amounts to: 
(1) identifying the sequence of pairs of qubits to apply two-qubit gates on, and 
(2) finding the corresponding optimal two-qubit unitary gates, 
which together give the shortest quantum circuit to reduce the entanglement of the initial $L$-qubit state. 
Note that (1) is a discrete combinatorial problem, while (2) is a continuous optimization problem. The different nature of these two problems suggests using different techniques to tackle them. In this work, we use RL to solve (1), see Sec.~\ref{sec:RL_framework}, and propose an analytical solution to (2) in Sec.~\ref{sec:two_qubit_U}: for a fixed pair of qubits $(i,j)$, we determine a locally optimal disentangling unitary $U^{(i,j)}$ by diagonalizing the corresponding reduced density matrix; this gate depends on the current quantum state only through the relevant two-qubit reduced density matrix, and can be computed using local measurements on NISQ devices [Sec.~\ref{app:qst}]. 
However, note that the modularity of the disentangling framework we develop makes it straightforward to consider other prescriptions for computing locally optimal two-qubit gates. 

We refer the reader to App.~\ref{app:beam_search} for an alternative approach to solve (1) using a tree search algorithm suitable for noise-free environments.

\begin{figure}[t!]
    \centering
    \includegraphics[width=0.49\textwidth]{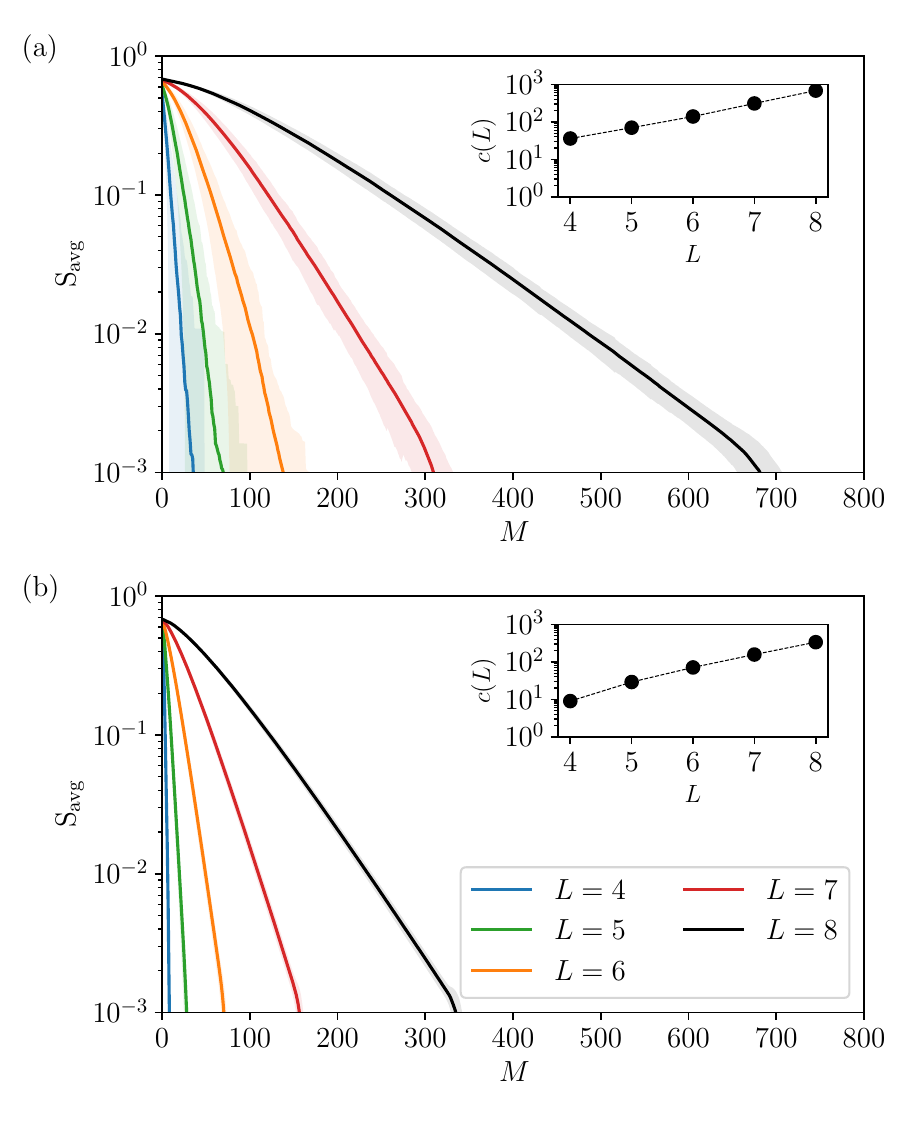}
    \caption[Random agent performance]{
    Exponential difficulty of the multiqubit disentangling problem starting from an $L$-qubit Haar-random state.
    (a) We place locally-optimal two-qubit gates on randomly chosen pairs of qubits (so-called random agent in Sec.~\ref{sec:RL-agent}), and monitor the average entanglement $S_\text{avg}$ after every gate for quantum states of different sizes. The number of applied gates increases exponentially with the system size. While $S_\text{avg}$ decays exponentially with the number of gates $M$, the corresponding decay timescale $c(L)$ to reach a threshold value of $10^{-3}$ diverges exponentially in the system size $L$ [inset]. 
    (b) Same as in (a) but for a locally greedy protocol (so-called greedy agent in Sec.~\ref{sec:RL-agent}): at each step, we compute the entanglement after acting on all pairs of qubits and postselect the pair which leads to the smallest value of $S_\text{avg}$. 
    For each system size, the curves are averaged over $2048$ different states; the shaded area shows the corresponding standard deviation.
    }
    \label{fig:rnd_agent}
\end{figure}

\section{\label{subsec:large_sys_rnd}Randomly and greedily placed locally optimal gates} 

For 3- and 4-qubit systems, it is possible to find exact optimal disentangling sequences analytically, cf.~App.~\ref{subsec:small_sys_exact}. Starting from 5-qubit states onwards, however, it is no longer clear how to construct disentangling sequences by having only partial information about the state, and even less so -- how to do this using a minimal number of gates. The difficulty arises from the exponential scaling~\cite{shende2006} of the size of the sequence space with the total number of two-qubit gates, which renders exhaustive search algorithms inapplicable. To demonstrate this, we show in Fig.~\ref{fig:rnd_agent} quantitative evidence for the difficulty of the problem, using Haar-random initial states.

Applying locally optimal disentangling gates [Sec.~\ref{sec:two_qubit_U}] to randomly selected pairs of qubits leads to an approximate exponential decrease of the average entanglement with the number $M$ of gates applied, $S_\text{avg}\sim \exp(-M/c)$ for large $M$, cf. Fig.~\ref{fig:rnd_agent}(a). However, the onset of this decay regime is itself exponentially delayed with increasing system size $L$, $c(L)\propto \exp(L)$ [Fig.~\ref{fig:rnd_agent}(a), inset]. Intuitively, this behavior arises since two-qubit reduced density matrices of Haar-random states are exponentially better approximated by a maximally mixed state with increasing the number of qubits; for maximally mixed 2-qubit reduced density matrices ${\propto}\mathbb{1}$, any local 2-qubit gate becomes ineffective. This behavior reflects the known exponential scaling of the disentangling process.

Opposite in spirit to this random protocol, at each step, one can try out all possible $L(L-1)$ pairs of qubits, and postselect the one which reduces $S_\text{avg}$ by the largest amount. The behavior of this greedy algorithm exhibits similar properties, cf.~Fig.~\ref{fig:rnd_agent}(b). This locally greedy protocol manages to roughly halve the number of unitaries compared to the random protocol, and it shares the same scaling behavior: exponential decay as a function of the number of applied gates $M$, together with an exponential increase of the decay timescale with the number of qubits $L$ in the Haar-random state. 

The exponential scaling of resources with system size is inevitable and occurs as a direct consequence of the nature of quantum mechanics. Hence, when designing disentangling strategies for Haar random states, one can merely try to reduce the constant prefactor of the scaling. Nevertheless, this is an important task since, using additional information, the total number of required gates can be reduced substantially, as illustrated in the example above, which is crucial for practical applications. Moreover, the optimal prefactor that gives rise to the smallest number of two-qubit gates is still unknown. This motivates us to optimize and analyze the disentangling process in the multiqubit regime where the effects of the exponential scaling are still manageable.

In the following, we discuss the design of a Reinforcement Learning (RL) agent that 
(i) learns to exploit partial information of the state of the system to make an informed choice about selecting the optimal order of qubit pairs in the disentangling sequence, while 
(ii) keeping the number of gates minimal.
In addition, we demonstrate and discuss (iii) the desired noise-resilient properties of our agent.

\section{Reinforcement Learning to Disentangle Quantum States} \label{sec:RL_framework}

    Due to the arbitrariness of the initial state $|\psi\rangle$, there exists no single universal transformation $U$ which disentangles any input state $|\psi\rangle$~\cite{terno1999nonlinear}. Intuitively, this is related to the lack of spatial structure in the entanglement distribution within a generic $|\psi\rangle$.

    An algorithm for disentangling quantum states was already presented in \cite{shende2006}; however, the quantum circuits produced are not optimal in the number of operations required and, to the best of our knowledge, the optimal scaling prefactor remains unknown.
    Moreover, the algorithm requires knowledge of the complete initial quantum state in order to produce the disentangling circuit.
    To resolve these issues we train an RL agent, which constructs the disentangling protocol in real time. At each step, given partial (yet experimentally measurable) information about the current state, the agent produces a two-qubit quantum gate. Applying this gate reduces the entanglement in the state and brings it closer to a fully disentangled state. In a sense, employing a learning algorithm allows us to construct an informed (state-dependent) disentangling machine~\cite{mor1999disentangling, zhou2000disentangling,bandyopadhyay1999disentanglement, ghosh2000optimal} that is functional in the face of uncertainty and noise.

    Let us begin by casting the optimal disentangling problem from Sec.~\ref{sec:setup} in the RL framework.

    \subsection{Reinforcement learning framework}

\begin{figure}[t!]
        \centering
        \includegraphics[width=0.45\textwidth]{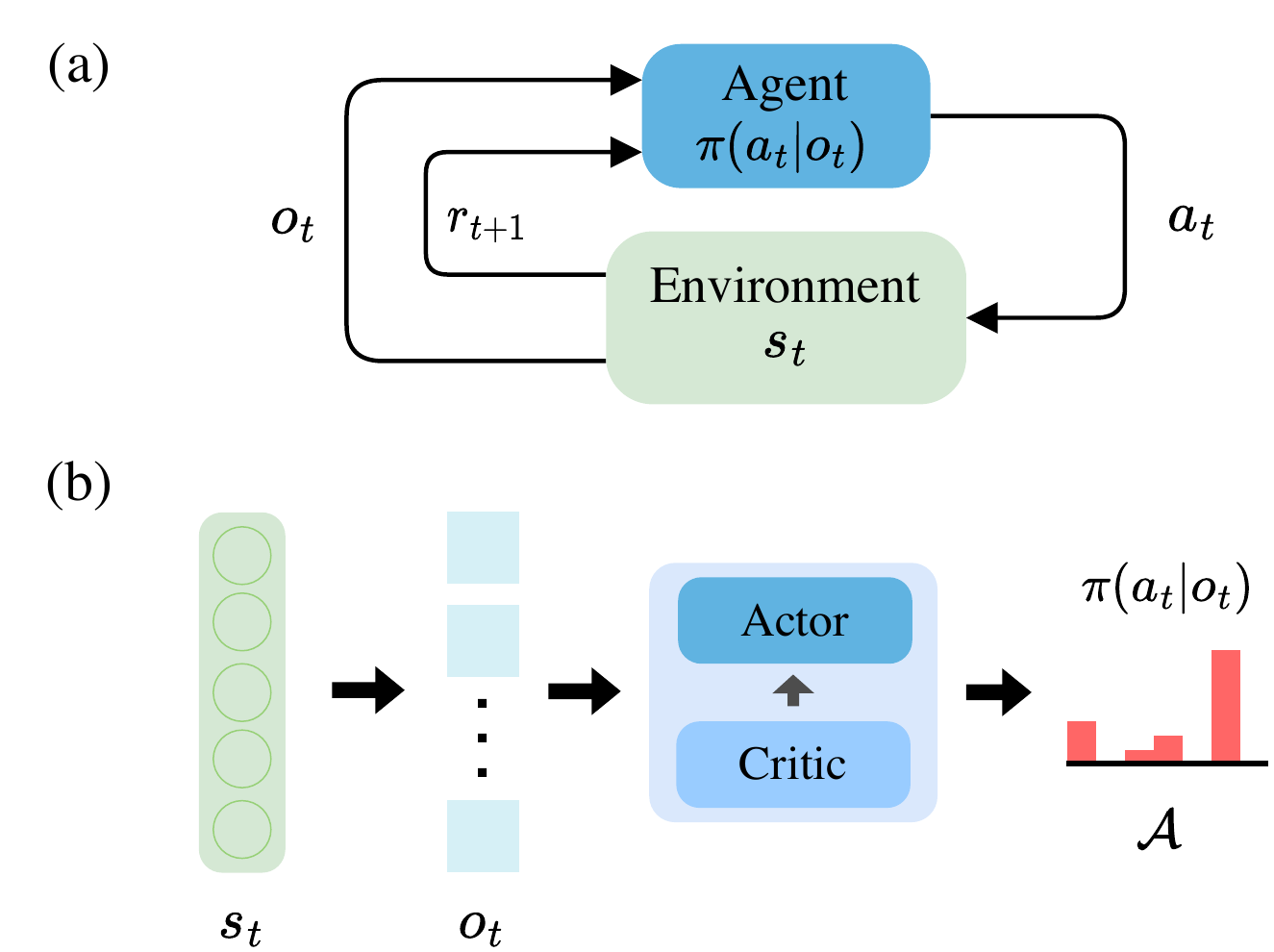}
        \caption[Interaction loop between agent and environment]{
        (a) Interaction loop between the RL agent and the environment. At each time step the agent makes an observation $o_t$ containing partial information about the current state of the environment $s_t$; using that observation it selects an action $a_t$. After the action is applied the environment emits a reward signal $r_{t+1}$ and transitions into a new state $s_{t+1}$, which is subsequently used by the agent to select the next action $a_{t+1}$. 
        (b) Action selection: an observation of the environment state $o_t$ comprises a measurement of all symmetrized two-qubit reduced density matrices (see text). The observation is then fed to both the actor (policy network) and the critic (value network). The actor learns a probability distribution $\pi(a_t|o_t)$ over the action space, called policy, and the action is sampled from that distribution. The critic learns a scalar number that evaluates the actions of the actor and is used for improving its performance during training.
        }
        \label{fig:agent_env}
    \end{figure}

    In Reinforcement Learning (RL), an agent learns to solve a task by interacting with its environment in a trial-and-error approach~\cite{sutton_barto_rl}. 
    Key components of RL are the agent and the environment, and their repeated interaction which is modeled by a feedback loop (Fig.~\ref{fig:agent_env}). At every step of the loop, the agent observes the current state of the environment and, based on that observation $o_t$, decides on an action to take. The agent uses a policy $\pi(\cdot | o_t)$ to choose an action given the current observation. When the environment is acted upon it transitions to a new state and also emits a reward signal.

    A step of the agent-environment interaction loop is referred to as a time step, and one run of the loop from beginning to end of the task is called an episode. Running the loop for $T$ time steps following the policy $\pi$ yields a sequence of (observation, action, reward) triples, called a trajectory $\tau$:
    \begin{eqnarray*}
        \tau =&& [(o_0, a_0, r_1),\; (o_1, a_1, r_2),\; (o_2, a_2, r_3),\cdots \\
        && \cdots, (o_{T-1}, a_{T-1}, r_T),\; o_T],    
    \end{eqnarray*}
    where $o_0$ is the observation of the environment state at the start of the episode. Usually, it is assumed that the environment obeys the Markov property; thus, every new state $s_{t+1}$ depends only on the action taken by the agent and the last observed state $s_t$, but not on the preceding ones.

    The goal of the RL agent is to maximize the expected cumulative reward $R$, also called the expected return:
    \begin{equation} \label{eq:exp_return}
        \underset{a \sim \pi}{\mathbb{E}} \big[ R \big] = \underset{a \sim \pi}{\mathbb{E}} \bigg[ \sum_{t=0}^{T} \gamma ^{t} r_{t+1} \bigg],
    \end{equation}
    where $a {\sim} \pi$ denotes that actions are drawn according to the policy $\pi$, $\gamma$ is a discount factor for future rewards, and $T$ is the number of steps until a terminal state is reached ($T$ could be $\infty$ in non-episodic tasks).

    \textit{Environment.} The environment for our problem comprises a physical simulation of a quantum system. The state space of a quantum system of $L$ qubits is described by a $2^L$-dimensional complex Hilbert space. Hence, the number of components in the wavefunction amplitude grows exponentially, and storing the entire state exactly becomes quickly infeasible. Moreover, quantum states are theoretical constructs that cannot be naturally measured in the lab, while full-state tomography is exponentially expensive (in the number of qubits) for multi-qubit systems. Therefore, we restrict to the use of partial information about the state which can be obtained from local quantum measurements.

    \textit{Observation space.} The idea is to replace the RL state space with an observation space $\mathcal{O}$, whose size is only polynomial in the number of qubits. Given an $L${-}qubit quantum state $|\psi\rangle$, we denote by $\rho^{(i,j)}$ the reduced density matrix for subsystem $\{i,j\}$, containing only the fixed qubits $i$ and $j$. Discarding the information about the rest of the qubits reduces the exponential scaling of tomography to only quadratic. We define an observation to comprise all symmetrized two-qubit reduced density matrices of the current quantum state $\rho$ (cf.~App.~\ref{app:obs_space} for details):
    \begin{equation*}
        o(\rho) = \left\{ \tfrac 1 2 (\rho^{(i,j)} + \rho^{(j,i)}) \quad | \quad 1 \leq i < j \leq L \right\}.
    \end{equation*}
    Hence, when the agent ``observes" the environment, it makes a full tomographical measurement of the two-qubit reduced density matrices only.
    Whereas using only two-qubit reduced density matrices enables the applications in experiments, cf.~Sec.~\ref{sec:exp_impl}, we note that, for certain numbers of qubits, there exist so-called absolutely maximally entangled states~\cite{rather2022thirtysix}, for which all two-qubit reduced density matrices are maximally mixed, and hence they cannot be disentangled within the present framework.

    \textit{Action space.} For a fixed observation of the current state of the environment, the agent selects the pair of qubits to which a locally disentangling unitary is applied. The corresponding locally optimal quantum gate $U^{(i,j)}$ can then be calculated by diagonalizing the relevant two-qubit reduced density matrix (see Sec.~\ref{sec:two_qubit_U}). For simplicity, we consider unordered pairs only~\footnote{We experimented also with ordered pairs but found no improvement in the learning behavior of the agent; however, the computational cost increased considerably.} (cf.~App.~\ref{app:act_space}). Thus, the action space is the set of all combinations of two-qubit pairs:
    \begin{equation*}
        \mathcal{A} = \left\{ (i,j) \quad | \quad 1 \leq i < j \leq L \right\},
    \end{equation*}
    and there are a total of $|\mathcal{A}|=L(L-1)/2\propto L^2$ actions. 

    \textit{Policy function.} To choose an action, the agent uses a policy function, which maps each observation to a probability distribution over action space:
    \begin{equation*}
        \pi(\cdot|o) : \mathcal{O} \rightarrow [0, 1]^{|\mathcal{A}|},
        \qquad 
        \displaystyle \sum_{a\in\mathcal{A}} \pi(a|o) {=}1.
    \end{equation*}
    We parametrize the policy $\pi_\theta$ using a deep neural network with parameters $\theta$. A feature of deep RL is that, once properly trained, the agent is able to produce close to optimal disentangling sequences given any initial state. Therefore, we are interested in network architectures that exhibit permutation equivariance (or covariance): re-arranging the qubits in the quantum state should lead to a corresponding re-arranging of the output probabilities for each action, respectively. An example of an architecture, known to have a built-in permutation equivariance, is the transformer~\cite{vaswani2023attention}. We thus model the policy of the agent using a stack of transformer encoders (App.~\ref{app:net_arch}). We note in passing that alternative architectures have recently also been proposed for entanglement structure detection~\cite{li2024entanglement}.

    \textit{Reward function.} Once the selected action has been taken, the agent receives a reward signal from the environment. The task of the agent is to reduce the average entanglement entropy $S_\text{avg}(|\psi\rangle)$, cf.~Eq.~\eqref{eq:avg_entanglement}, of the initial quantum state, using as few gates as possible. Therefore, we design a reward function specifically to achieve this goal:
    \begin{equation} \label{eq:reward_fn}
        \displaystyle \mathcal{R}(s_t, a_t, s_{t+1}) =
        \sum_{j=1}^L \frac
            {S_\text{ent}[\rho_t^{(j)}] - S_\text{ent}[\rho_{t+1}^{(j)}]}
            {\max (S_\text{ent}[\rho_t^{(j)}], S_\text{ent}[\rho_{t+1}^{(j)}])}
        - n(s_{t+1}),
    \end{equation}
    where $\rho_t^{(j)}$ is the reduced density matrix of qubit $j$ at episode step $t$, and $n(s_{t+1})$ keeps track of the number of entangled qubits in the state $s_{t+1}$. We can see that the first term rewards the agent whenever the entanglement of one of the qubits with respect to the rest of the system is reduced, while the second term penalizes the agent for every taken action. For an in-depth discussion on the choice of the reward function please refer to App.~\ref{app:reward_fn}.

    \subsection{Actor critic algorithm} \label{subsec:RL_algo}

    To train the agent and obtain a policy network that produces optimal disentangling circuits for any input quantum state, we use a learning algorithm from the RL toolbox known as actor critic (AC). The algorithm works by utilizing two separate neural networks: a policy network -- used for selecting actions; and a value network -- used for evaluating states.
    
    The choice of algorithm is based on the following premises:
    (i) due to complexity constraints our agent makes partial observations without having access to the full quantum state; thus, the agent operates in a model-free setting;
    (ii) our goal is to optimize the policy network $\pi_\theta$, and this is the most direct and stable algorithm for this purpose \cite{sutton_barto_rl};
    (iii) we want to minimize costly environment interactions while retaining the stability of on-policy policy gradient methods.
    The specific AC algorithm that we make use of is a version of Proximal Policy Optimization~\cite{ppo}, augmented with a regularization term given by the entropy of the policy. For more details on the algorithm we refer the reader to App.~\ref{app:ppo_details}.

    We first train RL agents to disentangle arbitrary states for three different system sizes: 4-, 5-, and 6-qubit systems in a noise-free environment.
    We then go on and train on larger 12- and 16-qubit states, formed from subsystems (of up to 4 qubits) of Haar-random states, with a controlled amount of entanglement between the subsystems (see App.~\ref{app:large_systems}). Due to the increased complexity of the multi-body problem, training the agents to solve these systems proceeds in two stages, as discussed in Sec.~\ref{app:staged_training}.
    Finally, we test the performance of our agents in both noise-free and noisy environments (see Secs.~\ref{sec:RL-agent}, \ref{sec:applns_noisy}).
    To show that training works properly, we monitor:
    (1) the agent's accuracy in disentangling the states -- i.e., the percentage of states the agent can disentangle successfully;
    (2) the average episode length during training -- i.e., the average number of actions (or gates) the agent takes to disentangle a state; the average here is taken over the training batch (see App.~\ref{app:hyperparams}).

    \begin{figure}[t!]
        \centering
        \includegraphics[width=0.48\textwidth]{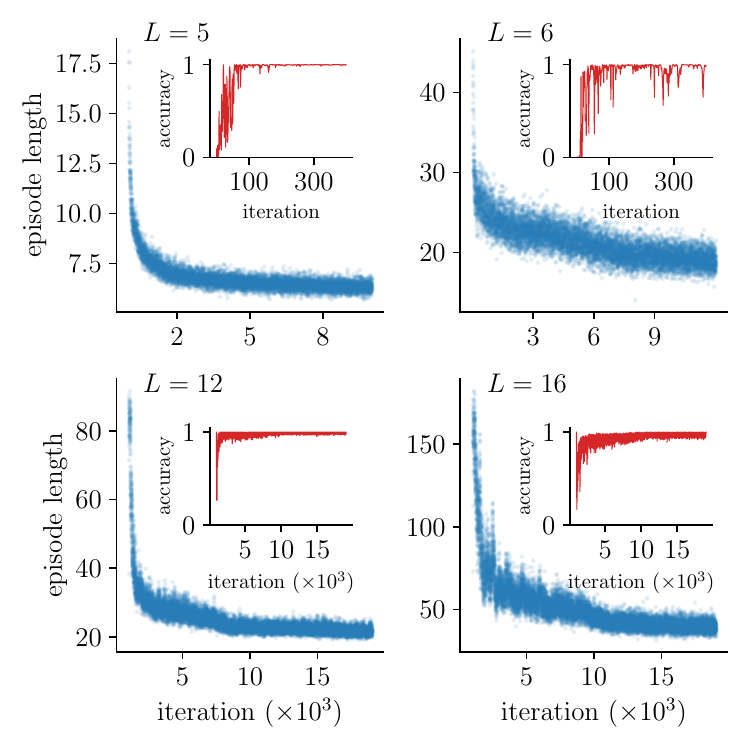}
        \caption[Average episode length during training]{
        Average episode length (main figure in blue), and average accuracy (insets in red) defined as the percentage of disentangled states, achieved by the RL agent during training on 5-, 6-, 12-, and 16-qubit systems (different panels). Training proceeds in two separate regimes: 
        (i) Accuracy improvement dominates learning at early iterations [inset] until the agent's accuracy reaches nearly 100\%.
        This is followed by, (ii), an efficiency improvement regime during which the agent reduces the average number of gates in the circuit needed to disentangle a state. 
        Training converges. Agents trained on larger systems require more iterations. 
        Trained agents can construct more compressed disentangling circuits than previously known algorithms (see Sec.~\ref{sec:cnot_count} for a detailed comparison). All hyperparameters used for training are listed in Table~\ref{table:hyperparams}.
        The $L=12$ and $L=16$ plots show only the training stage on weakly entangled states (the pretraining on product states is not plotted, see App.~\ref{app:staged_training}).
        }
        \label{fig:training_rl}
    \end{figure}

    \subsubsection{Training the RL Agent on 4-, 5- and 6-qubit Haar-random states}
    
    We observe that the training procedure exhibits two well-separated regimes. In the first regime, the agent's accuracy quickly increases and reaches close to 100\% for all four system sizes (cf.~Fig.~\ref{fig:training_rl}, inset). This implies that the agent is able to reliably disentangle any input state. Even though at each step we make only partial observations of the state, the policy network is still able to find patterns in the reduced density matrices and produce a disentangling circuit. 

    The training process proceeds with the second regime, where the average episode length is minimized. Shortening the episode length effectively implies that the agent learns to produce disentangling circuits requiring fewer and fewer gates. The absolute minimal number of gates that can be reached is determined by the disentangling speed limit, which is initial-state dependent. Key ingredients for the emergence of this second regime are the design of the reward function and augmenting the algorithm with entropy regularization. The combination of the two incentivizes the agent to explore different circuits during training and to search for shorter solutions without reducing accuracy. The results from training the agent on the system sizes $L=5,6$ can be seen in Fig.~\ref{fig:training_rl}.
    Note that the second regime of training ends with a prolonged converging period showing that although the procedure for optimizing the quantum circuit keeps converging, further training does not provide a sizeable gain.

    For a video of the training process for Haar-random 4-qubit states, see Sec.~\ref{sec:video}.

    \subsubsection{Scaling of the RL framework with number of qubits} \label{subsec:scaling_laws}

    \begin{figure}[t!]
        \centering
        \includegraphics[width=0.45\textwidth]{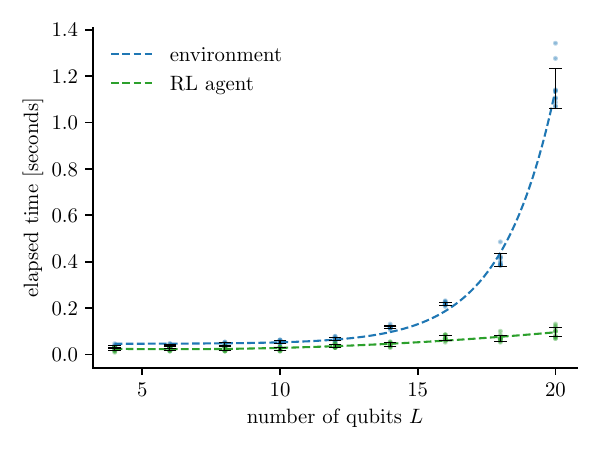}
        \caption{
        Scaling of computational resources against the number of qubits $L$. 
        We distinguish between running one step of the classical simulator (blue), and one forward/backward pass through the neural network that constitutes the RL agent (green).
        Running the environment scales exponentially with $L$, as reflected by the exponential fit (dashed blue line). 
        By contrast, the RL agent scales quadratically with $L$, as shown by the quadratic fit (green dashed line). 
        For each $L$, we show 10 independent runs using individual data points.
        Error bars show the standard deviation, centered at the average values used to do the fits. 
        }
        \label{fig:rl_scaling_laws}
    \end{figure}

    Increasing the number of qubits $L$ in a Haar-random quantum state significantly increases the number of actions that need to be applied to disentangle the state since the protocol length grows accordingly (cf.~Fig.~\ref{fig:rnd_agent}). As discussed in Sec.~\ref{subsec:large_sys_rnd}, any algorithm capable of finding optimal disentangling circuits that consist of only two-qubit gates will inevitably produce exponentially long protocols (in the number of gates applied). 
    To see why, notice first that a two-qubit gate $U$ can only modify the entanglement between the two qubits it acts on; hence, the relevant information of how it acts on the pure state of the full system is contained in the corresponding two-qubit reduced density matrix. However, for Haar-random states, the latter is exponentially close to the maximally mixed state, $\rho^{(2)}_{i,j}\propto 1 + \epsilon R$ with $\epsilon{\propto}\exp(-L)$ and some hermitian $4\times 4$ matrix $R$~\cite{page1993average}. Therefore, applying the two-qubit unitary, $\rho^{(2)}_{i,j}\to U \rho^{(2)}_{i,j} U^\dagger$, leaves the reduced density matrix intact, up to exponentially small corrections. For this reason, it takes exponentially many two-qubit gates to disentangle a Haar-random state.

    As a result, any disentangling algorithm capable of finding optimal disentangling circuits that consist of only two-qubit gates will inevitably produce exponentially long protocols.
    We emphasize that this exponential scaling is a mathematical statement about the structure of the optimal disentangling circuit and not about the algorithm used to find it. 
    Concerning our RL agent, this results in exponentially large episode lengths and, thus, prolongs the training process, making training prohibitively expensive.

    In addition, increasing the system size significantly increases the computational resources needed to run the classical simulations. Indeed, in Fig.~\ref{fig:rl_scaling_laws} we compare the scaling of resources for (i) the forward and backward pass of the RL agent (green), and (ii) the environment, which applies a single two-qubit gate on the simulator (blue). We observe that while our RL agent scales quadratically with the size of the quantum system $L$, running the simulator of the environment scales exponentially. This, of course, is expected since representing a quantum system on a classical computer requires exponential memory.

    These scaling considerations suggest that training on substantially larger quantum systems remains feasible provided the number of environment simulation steps is kept small. As stated in Sec.~\ref{subsec:RL_algo}, this motivates the use of a reinforcement learning algorithm that minimizes costly interactions with the environment while retaining the stability of policy-gradient methods. We therefore employ Proximal Policy Optimization (PPO), which alternates between a rollout phase for collecting simulation data and multiple optimization updates on the same batch (cf.~Sec.~\ref{app:train_algo_ppo}), thereby improving sample efficiency without sacrificing training stability.

     \begin{figure}[t!]
        \centering
        \includegraphics[width=0.45\textwidth]{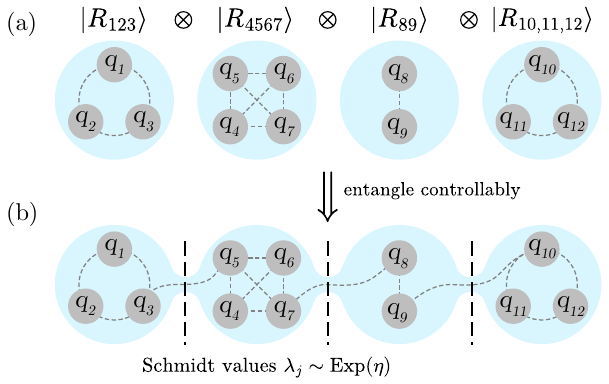}
        \caption{
        Initial state distribution used to train the RL agent to disentangle $12$- and $16$-qubit states.
        (a) We sample uniformly subsystem sizes of up to 4 qubits, and Haar-random states with the corresponding support. There is no entanglement between the subsystems. 
        (b) We then manually design weak entanglement between the subsystems: we enumerate the qubits, and sample the corresponding Schmidt values of the reduced density matrix w.r.t.~the cuts between subsystems from an exponential distribution with parameter $\eta$ (implemented in practice using matrix product states). For $\eta\gg 1$, the state is effectively the same as in (a), whereas for $\eta=0$ we have a highly entangled state on the entire system.
        }
        \label{fig:multiqubit-state}
    \end{figure}

    \subsubsection{Training the RL Agent on 12- and 16-qubit weakly-entangled clusters of Haar-random states}
    \label{subsubsec:weakly_ent_Haar}

    When it comes to scaling up the number of qubits $L$, we note first that Haar-random states do not represent a physically meaningful initial state distribution. The reason is that they do not contain useful entanglement (compared, e.g., to the GHZ state, which can be utilized for quantum sensing). Haar-random states of many qubits correspond to featureless infinite-temperature states with no immediate practical applications.       

    Therefore, to explore the capabilities of our RL agent to disentangle multi-qubit states, we consider a distribution of initial states which consists of weakly entangled subsystems of Haar-random states but supported on smaller subsystems, see Fig.~\ref{fig:multiqubit-state}. This is a meaningful~\footnote{though not particularly useful for experimental purposes} initial state distribution for the disentangling problem because entanglement is nontrivially distributed among all qubits, which makes the problem challenging. Moreover, it requires a disentangling algorithm capable of assessing which pair of qubits to apply gates on and in which order, by only having the reduced two-qubit density matrices as partial information (which is mathematically insufficient but physically relevant). 
    For comparison, the problem of identifying disentangled subsets of entangled qubits in an $L$-qubit state (so-called $m$-partite separability problem) scales exponentially with $L$ and is NP-hard~\cite{gharibian2008strong}. In the initial state distribution we use, these sets are, in addition, weakly entangled.
    
    The training procedure for such weakly entangled subsystems of maximally entangled Haar-random states runs in two stages, as detailed in App.~\ref{app:staged_training}.
    
    In the first stage, the agent is shown only product states of subsystems of maximally entangled Haar-random states. The subsystem size is chosen to be between 2 and 4 qubits, and is selected uniformly at random. We find that this task is relatively easy to solve and the agent quickly learns to disentangle such systems. As training progresses, the policy becomes more specialized to solve such states, and the agent focuses on reducing the number of gates in the circuit.

    In the second training stage, the agent is presented with only weakly entangled states of subsystems of maximally entangled Haar-random states. The subsystem size is again chosen between 2 and 4 qubits, and is selected uniformly at random. The entanglement between these subsystems is controlled using an external dimensionless parameter $\eta$: for $\eta\gg 1$ there is little-to-no entanglement, while for $\eta\to 0$ we have maximal entanglement; we use the same value of $\eta=4.1$ for each of the bonds between the subsystems to preserve the equivalence of the subsystems and to avoid bias arising from non-symmetric coupling. We refer the reader to App.~\ref{app:large_systems} for the technical details. 

    At the transition between the two training stages (iteration $1000$ for both 12- and 16-qubit systems), we observe a sharp drop in performance that showcases the difference in difficulty between the two tasks (product vs.~weakly entangled Haar-random subsystems). However, starting with the pre-trained initial policy from the first training stage, the agent is able to quickly adjust the policy for the new problem at hand. 
    We mention in passing that our attempts at training in a single stage directly on weakly entangled states, starting from a uniform policy, were unsuccessful. The two-stage setup, resembling \textit{transfer learning}, proves handy and allows us to train the agent to disentangle larger systems with a limited amount of compute and data.

\section{\label{sec:RL-agent}Analyzing the behavior of disentangling RL agents}

\begin{figure}[t!]
    \includegraphics[width=0.32\textwidth]{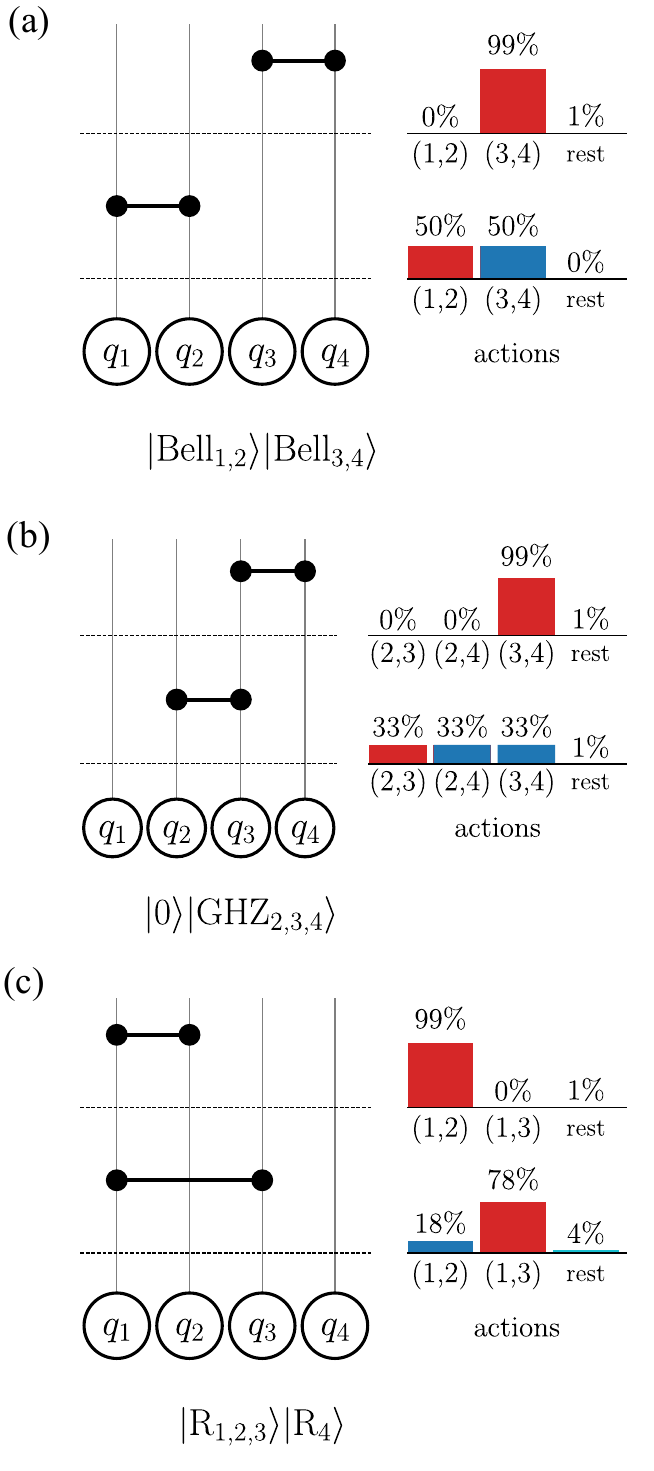}
    \caption{
    Benchmarking the trained RL agent on a set of four-qubit states starting from
    (a) a pair of Bell states $|\psi_{1,2,3,4}\rangle {=} |\text{Bell}_{1,2}\rangle |\text{Bell}_{3,4}\rangle$ featuring bipartite entanglement,
    (b) a tripartite entangled GHZ state $|\psi_{1,2,3,4}\rangle {=} |0_1\rangle|\text{GHZ}_{2,3,4}\rangle$, and
    (c) a product of a Haar-random tripartite entangled state and a Haar-random single-qubit state, $|\psi_{1,2,3,4}\rangle {=} |\text{R}_1\rangle|\text{R}_{2,3,4}\rangle$.
    The left-hand side shows the circuit diagram using locally optimal two-qubit gates (indicated by black lines connecting filled black circles). The right-hand side shows in the form of a histogram the RL agent's policy before applying each gate (probabilities rounded to percent). We display separately the most probable actions (i.e., qubit pairs). The `rest' column denotes the aggregated probability for all remaining actions; the selected action at each step is shown in blue.
    The RL agent correctly identifies the spatial structure of entanglement even for random states where the latter is not obvious, and disentangles the initial state using as few gates as possible.
    }
    \label{fig:validate_agent_4q}
\end{figure}

\begin{figure}[th!]
    \includegraphics[width=0.4\textwidth]{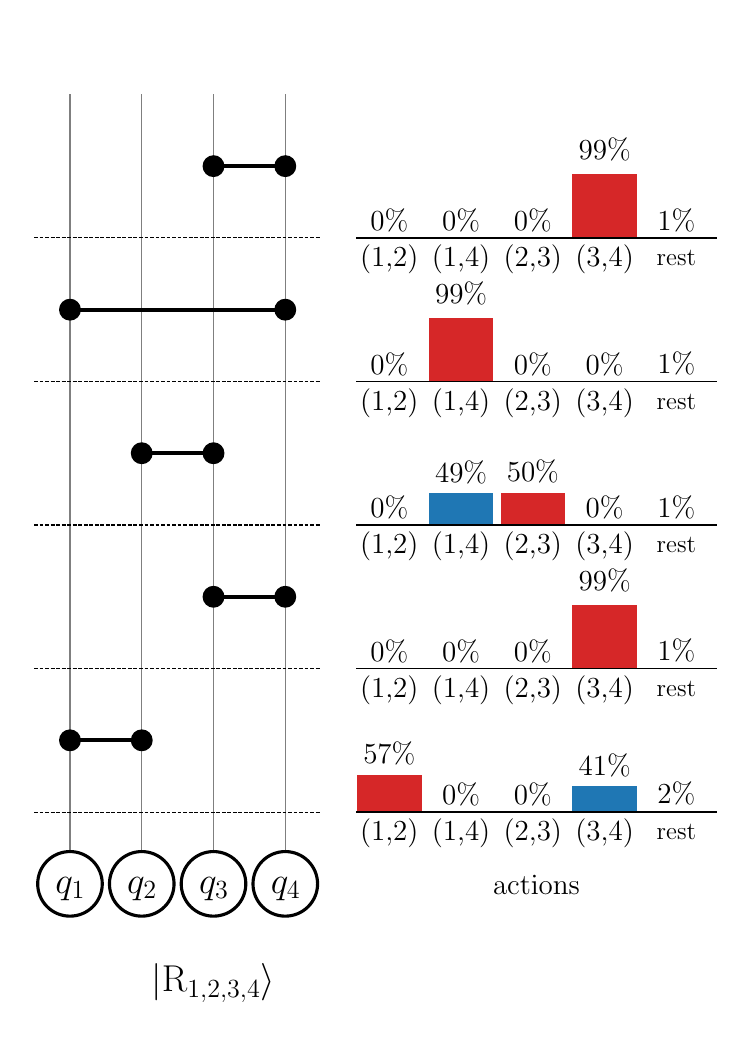}
    \caption{
    RL agent finds the shortest disentangling sequence for a Haar-random 4-qubit state without an obvious entanglement structure. See Fig.~\ref{fig:validate_agent_4q} for details. 
    }
    \label{fig:validate_agent_4q_RRRR}
\end{figure}

\subsection{Benchmarking the RL agent on entangled 4-qubit states}

Having successfully trained the RL agent, we now benchmark its performance. To this end, in Figs.~\ref{fig:validate_agent_4q}~and~\ref{fig:validate_agent_4q_RRRR} we first consider a few specific 4-qubit states and analyze the actions chosen by the agent to disentangle them.  In Fig.~\ref{fig:validate_agent_4q}(a) we consider a pair of Bell states of arbitrarily selected qubits. By observing the two-qubit reduced density matrices, the agent correctly identifies the entangled qubit pairs and assigns to each corresponding action $50\%$ probability. Upon applying the first action, the new state of the system contains only the remaining Bell pair, and the policy of the agent adjusts accordingly. Similar tests can be done using a GHZ state, cf.~Fig.~\ref{fig:validate_agent_4q}(b), or a W-state that is tripartite entangled [see accompanying Jupyter notebook].

Since the 4-qubit state space contains the 3-qubit state space as a subspace, we can also initialize the system in a product state between a fixed qubit and the remaining qubits, e.g., $|\psi\rangle {=} |\text{R}_1\rangle|\text{R}_{2,3,4}\rangle$ [cf.~Fig.\ref{fig:validate_agent_4q}(c)], where $|\text{R}\rangle$ stands for a Haar-random state on the corresponding qubit subspace indicated by the subscript. Clearly, the agent first correctly identifies the entangled subsystem, and then disentangles the state in two steps. Moreover, we see that it has learned to use the optimal three-qubit sequence determined analytically in App.~\ref{subsec:small_sys_exact} (cf.~Fig.~\ref{fig:3_qubits}). We also verified the permutation equivariance of the policy: shuffling the qubit labels results in a rearrangement of the corresponding qubits, as desired [not shown]. 
Finally, in Fig.~\ref{fig:validate_agent_4q_RRRR} we let the agent disentangle a 4-qubit Haar-random state. Once again, the RL agent learns to apply a permutation of the locally optimal sequence discussed in App.~\ref{subsec:small_sys_exact}  (cf.~Fig.~\ref{fig:4_qubits}(a)).
Unlike the Bell and GHZ states which possess a structured entanglement distribution among the qubits, for Haar-random states there is no obvious way to determine which pair of qubits to start the disentangling procedure from; this example clearly demonstrates the motivation of applying RL to the state disentangling problem.

The examples we discussed showcase the ability of the trained RL agent to identify and recognize local entanglement structure in multiqubit quantum states, and use it to find the shortest disentangling circuit. 

\begin{figure*}[t!]
    \centering
    \includegraphics[width=1.05\textwidth]{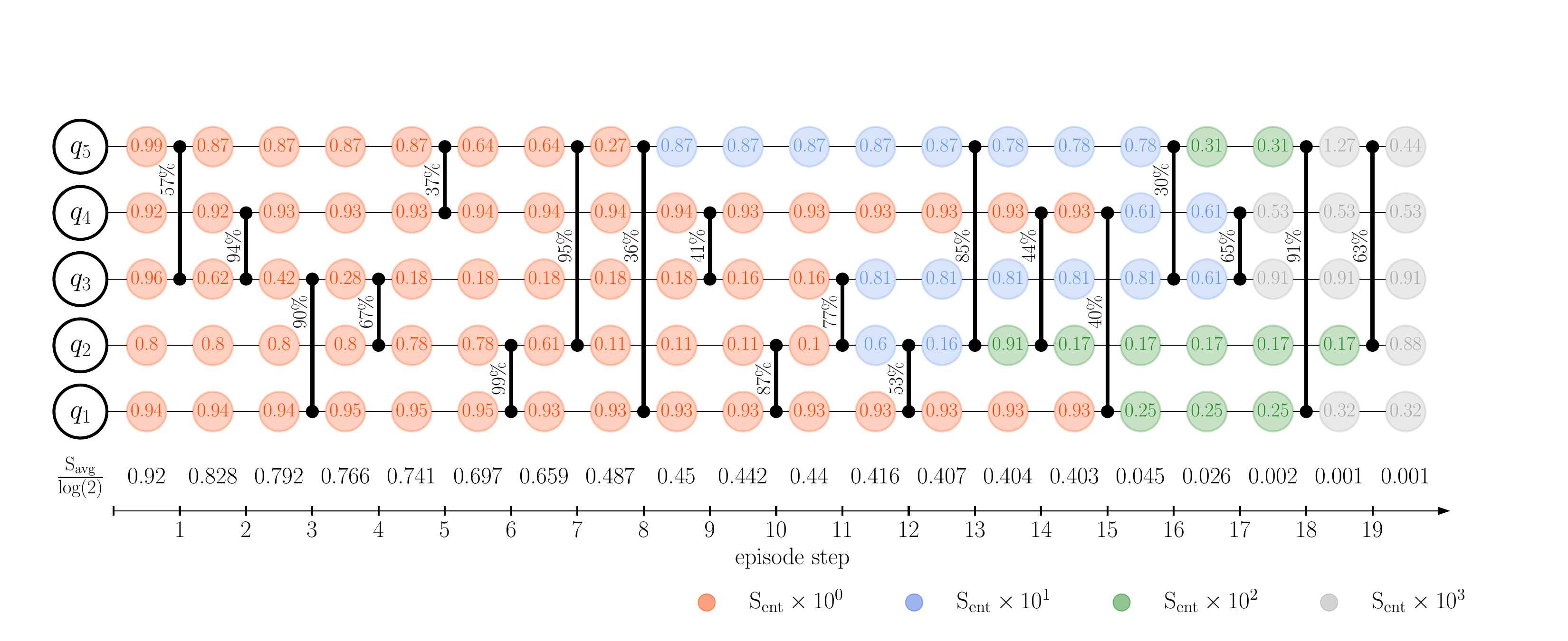}
    \caption{
    An example of a protocol sequence used by the RL agent to disentangle a Haar-random 5-qubit state $|\psi\rangle{=}|\text{R}_{1,2,3,4,5}\rangle$ using 19 locally-optimal two-qubit gates. The agent identifies first a suitable qubit (here $q_3$) and keeps applying different two-qubit gates on it to reduce the entanglement between $q_3$ and the rest. This process continues until it becomes advantageous to let go of the preferred qubit, and focus on another one. The entire sequence exhibits correlations between the support of consecutive gates which tend to share a common qubit, indicating that the RL agent has learned to implement and exploit interactions among all qubits in the system. 
    At step 17 the agent manages to factor out both qubits $q_3$ and $q_4$ by bringing their entanglement below the threshold with a single action, leaving the system in the product state $|\psi_3\rangle|\psi_4\rangle|\psi_{1,2,5}\rangle$; finally, subsystem $|\psi_{1,2,5}\rangle$ is disentangled with 2 gates. Hence, the agent learns sequences that exploit interactions between all qubits, rather than focusing on a specific qubit to greedily factor it out. 
    The resulting disentangling sequence can be viewed as a compression of the initial quantum state into a sequence of two-qubit gates. 
    Entanglement per qubit is normalized within the interval $[0,1]$, and shown in the colored circles; the circle color encodes the order of magnitude (see legend) with grey indicating values below the optimization threshold $\epsilon{=}10^{-3}$, cf.~Table~\ref{table:hyperparams}. The average entanglement over all qubits is displayed below the circuit for each episode step. The percentage value next to each selected gate shows the probability of this action in the policy. Other types of protocols are shown in Fig.~\ref{fig:5q-seqs}.
    }
    \label{fig:validate_agent_5q}
\end{figure*}

\subsection{Disentangling Haar-random 5-qubit states} \label{sec:disent_5q_haar_random}

To see the clear advantage of using the RL agent, we now test it on 5-qubit Haar-random states. Recall that finding the optimal disentangling sequence for $L{>}4$ qubits is a difficult combinatorial problem due to the large size $|\mathcal{A}|^{N_T}$ of the sequence space to be explored; $|\mathcal{A}|^{N_T}$ scales exponentially with the number of steps $N_T$ where $|\mathcal{A}|=L(L-1)/2$ is the size of the action space, see Sec.~\ref{subsec:large_sys_rnd}. Therefore, we set a threshold $\epsilon{=}10^{-3}$ (as per Table~\ref{table:hyperparams}) on the minimum single-qubit entanglement $\min\{{S_i}\}_{i=1,2,..,L}$ in the state, and stop the agent once this threshold has been reached.
Figure~\ref{fig:validate_agent_5q} shows that, starting from a Haar-random 5-qubit initial state $|\text{R}_{1,2,3,4,5}\rangle$, the agent takes $19$ steps to disentangle it (for the average number of steps see Fig.~\ref{fig:compare_agents}).

By analyzing the behavior of the RL agent over different random states, we make the following observations:
(i) it identifies a suitable qubit (here $q_{3}$) and applies a sequence of gates involving this qubit. This leads to a reduction of the entanglement between this qubit and the rest (shown by the value within the colored circles). The agent keeps applying gates involving this qubit until,
(ii), it becomes more advantageous to switch to a different qubit (step 5). The probability of taking the optimal action is denoted by the percentages on the gates: eventually, an action pair involving another qubit becomes preferable.

Curiously, subsequences that contain consecutive gates acting on the same qubit can be considered topologically distinct: e.g, episode steps $(1-4)$, $(5)$, $(6-8)$, $(9)$, $(10-15)$, $(16-17)$, $(18-19)$ in Fig.~\ref{fig:validate_agent_5q}. The topological feature can be seen by placing the qubits on the vertices of a graph and regarding the gates as connections drawn between the vertices in the order prescribed by the protocol: moving between subsequences, e.g., $(1-4)\to(5)$, requires a discontinuous jump from one vertex to another, whereas moving within a subsequence does not. The specific pairs of qubits selected depend on the structure of the initial state; since it was chosen randomly, it is meaningless to assign a concrete interpretation to them. Nevertheless, a clear pattern emerges in that most consecutive gates share a common qubit, which suggests that the protocol found by the agent has strong step-to-step (temporal) correlations. The physical meaning of this topological feature is tied to the compressibility of the circuit: gates within a topological class are incompressible, but (some) gates from consecutive topological subsequences can be executed in parallel, thus reducing circuit depth. 

By studying various initial states, we observe two generic types of behavior of our RL agent (see Fig.~\ref{fig:5q-seqs} for more examples).
Using the first approach (shown in Fig.~\ref{fig:validate_agent_5q}, also Fig.~\ref{fig:5q-seqs}a, b), the agent follows a protocol, which reduces the entanglement across the entire system. Once the entanglement of some qubit drops below the threshold (step 17 in Fig.~\ref{fig:validate_agent_5q}), to disentangle the remaining subsystem, the agent uses fewer gates than the generic 5-gate circuit required by the optimal 4-qubit sequence (App.~\ref{subsec:small_sys_exact}). This implies that the entire protocol constitutes a more efficient (yet more complex) strategy that exploits interactions among all qubits in the system, rather than fixating on a single qubit until it is disentangled from the rest.
Instead, we find this latter behavior in the second type of sequences (Fig.~\ref{fig:5q-seqs}c) where the agent manages to initially bring the entanglement of one of the qubits below the threshold; from this point on, it follows an optimal 4-qubit strategy. This divide-and-conquer strategy is intuitive to understand (although still difficult to find during optimization, especially for Haar-random states).
These behaviors demonstrate that the agent can learn to generate protocols that induce effective interactions among all five qubits, and exploits them to achieve its goal.

\subsection{Statistical properties of trained RL agents for 4-, 5-, and 6-qubit Haar-random states} \label{subsec:results_4q5q6q}

\begin{figure*}[t!]
    \centering
    \includegraphics[width=1.0\textwidth]{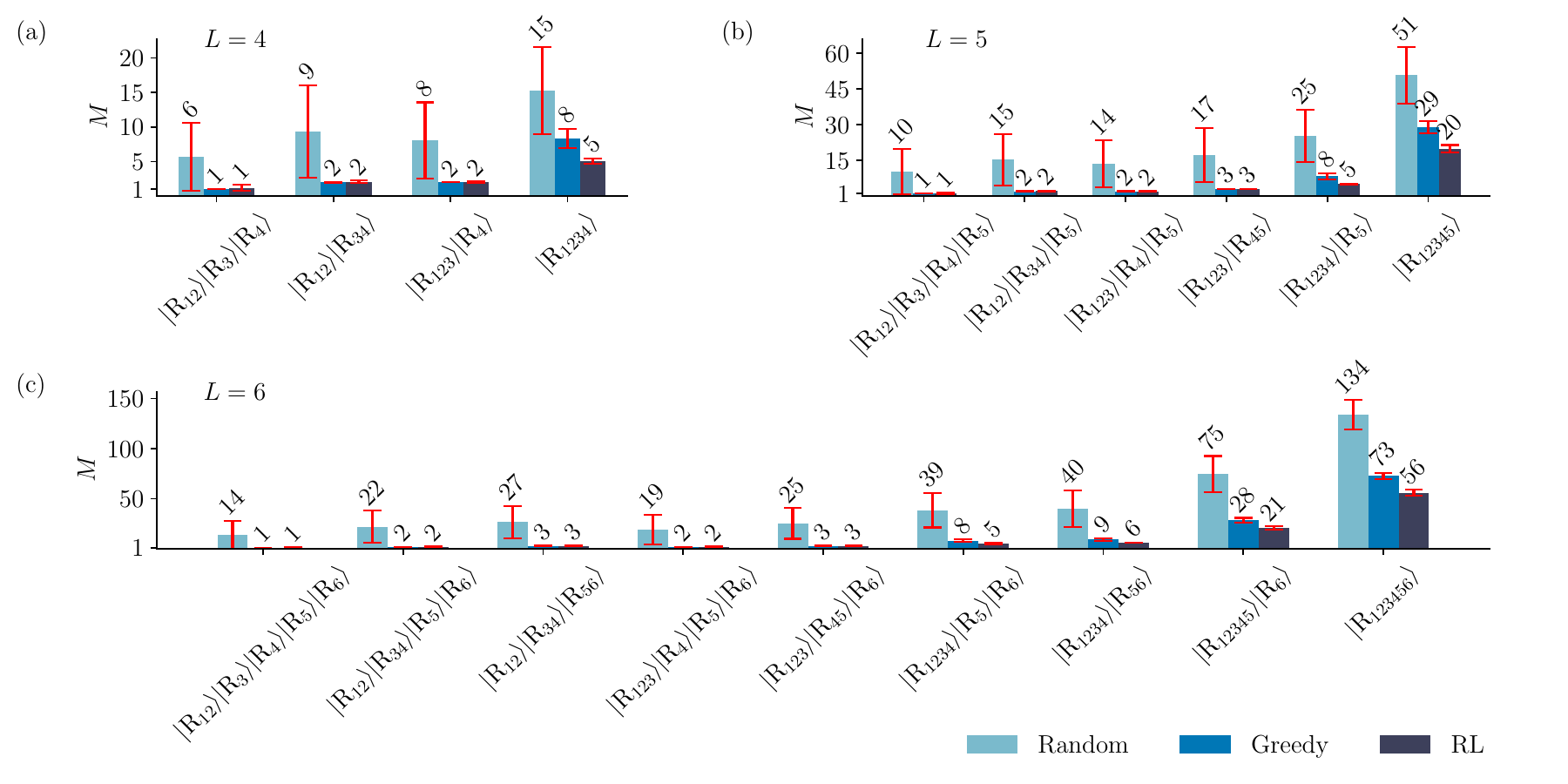}
    \caption{
    Average performance of the RL (black bars), greedy (dark blue bars), and random (cyan bars) agent, for $L=4$ (a), $L=5$ (b), and $L=6$ (c) multi-qubit states. 
    The statistics are computed using 1000 initial states, of the form shown on the $x$-axis, where $|\text{R}_\alpha\rangle$ is a Haar-random state over the Hilbert space of the qubits labeled by $\alpha$. The $y$-axis displays the number of gates $M$. The error bars show the standard deviation around the mean value (displayed on top of each bar and rounded to the nearest integer). 
    The ability of the RL agent to recognize the structure of the entanglement distribution over the qubits allows it to decisively outperform the random and greedy agents in terms of the average number of gates applied.
    }
    \label{fig:compare_agents}
\end{figure*}

Let us now turn to the statistical performance of trained RL agents for systems of $L=4,5,6$ qubits. Figure~\ref{fig:compare_agents} shows the average number of steps required to bring the average entanglement below the $\epsilon$ threshold (see Table~\ref{table:hyperparams} for the values of $\epsilon$). We perform the analysis using Haar-random initial states on all possible subsystems (up to permutations of the qubits that are built into the policy network architecture) while keeping the subsystems in a product state; this allows us to compare the results with agents trained directly on the smaller subsystems. 
For instance, whereas the 6-qubit agent takes about $20.96 \pm 1.40$ steps on average (Fig.~\ref{fig:compare_agents}c, $|\text{R}_{1,2,3,4,5}\rangle|\text{R}_6\rangle$ black bar), the 5-qubit agent takes $20.00 \pm 1.41$ steps (Fig.~\ref{fig:compare_agents}b, $|\text{R}_{1,2,3,4,5}\rangle$ black bar); this suggests that a further slight (though insignificant) improvement can be expected, most likely by increasing the number of training epochs and/or the neural network size. 

To quantify the advantages offered by a learning algorithm, we also compare our RL agents
(i) to a random agent (Fig.~\ref{fig:compare_agents}, cyan bars) that uses a uniformly distributed random policy, and
(ii) to a greedy agent (Fig.~\ref{fig:compare_agents}, dark blue bars) which applies the gate to all possible pairs at each step and postselects the action that minimizes the average entanglement. We studied the behavior of these agents for an increasing number of qubits in Sec.~\ref{subsec:large_sys_rnd}.
Most notably, compared to the random agent, for $L=6$ qubits the RL agent obtains, on average, an almost 3-fold reduction of the number of gates used in the disentangling sequence. Similarly, the RL agent requires $56$ gates, compared to the greedy agent which takes $72$ gates on average. This implies that the RL agent learns to occasionally take locally (in time) sub-optimal actions which however allow it to obtain a higher reward in the longer run. Although the number of steps taken by the RL agent increases rapidly (and very likely still exponentially) with the number of qubits, such a reduction can prove very useful when dealing with noisy NISQ devices (see Sec.~\ref{sec:applns_noisy}), as it brings down the number of required CNOT gates.

\subsection{Disentangling 12- and 16-qubit states} \label{subsec:results_12q16q}

\begin{figure}[t!]
    \includegraphics[width=0.52\textwidth]{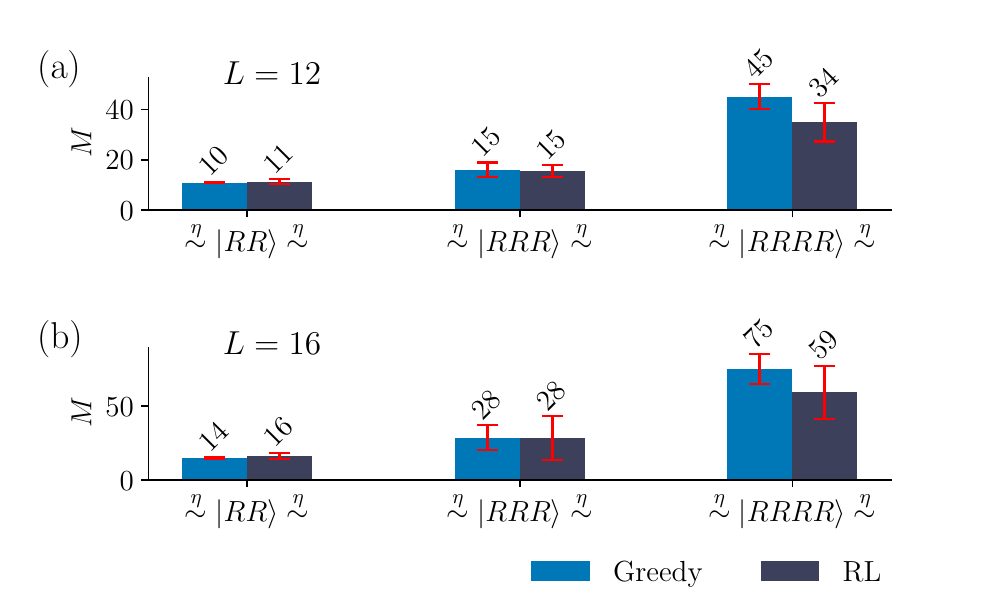}
    \caption{
    Average performance of the RL (dark blue bars) compared to the greedy agent (blue bars) for $L{=}12$ (a), and $L=16$ (b) multi-qubit weakly entangled ($\eta{=}4.1$) subsystems of Haar-random states supported on smaller subsystems.
    The $y$-axis displays the number of gates $M$. 
    The statistics are computed using 200 initial states, of the form shown on the $x$-axis, where $|R\dots R\rangle$ denotes a fully entangled Haar subsystem and $\overset{\eta}{\sim}$ denotes a weakly entangled bond, see Fig.~\ref{fig:multiqubit-state}. 
    Error bars show the standard deviation around the mean value (displayed on top of each bar and rounded to the nearest integer). 
    The ability of the RL agent to recognize the structure of the entanglement distribution over the qubits allows it to outperform the greedy agent in terms of the average number of gates applied.
    }
    \label{fig:compare_agents_12q_16q}
\end{figure}

    The statistical performance of the trained RL agents for systems of $L=12,16$ qubits is shown in Fig.~\ref{fig:compare_agents_12q_16q}. The figure shows the average number of steps required to bring the average entanglement of the system below the $\epsilon{=}10^{-3}$ threshold (see Table~\ref{table:hyperparams}). The analysis is performed using weakly entangled subsystems of Haar-random states [cf.~Sec.~\ref{subsubsec:weakly_ent_Haar}].

    We again compare our RL agents to a random agent and a greedy agent as described in Sec.~\ref{subsec:results_4q5q6q}; however, in Fig.~\ref{fig:compare_agents_12q_16q} we only show the results obtained from the greedy agent and omit the random agent since its performance was off-scale worse. We observe that for systems where the subsystem size is small (e.g., $2$ or $3$ qubits), both the RL agent and the greedy agent perform on par. This is because smaller systems require shorter (in time-steps) solutions and are, thus, easily solved by the greedy agent. Comparing the performance on larger subsystems, however, we see that the RL agent is capable of producing shorter circuits, again, pointing to the agent's ability to select sub-optimal actions in order to obtain better results in the long run.

    Referring back to our discussion on scaling up the system size in Sec.~\ref{subsec:scaling_laws}, we see that although the underlying quantum simulation scales exponentially with system size, the overall approach remains practical because the targeted systems require only a minimal number of simulation steps, cf.~Fig.~\ref{fig:rl_scaling_laws}. As a result, the quadratic scaling of the agent dominates the computational cost, enabling strong performance even for systems as large as $L=16$ qubits.
    
    We note in passing that the RL agent exhibits a degree of generalization beyond the specific class of systems on which it was trained. We observe that, although optimization was performed exclusively on states constructed from weakly entangled Haar-random subsystems with support up to 4 qubits at $\eta=4.1$, the trained policy retained nontrivial disentangling capability on more challenging instances; this includes systems with stronger inter-subsystem entanglement ($\eta<4.1$), as well as states composed of larger Haar-random components, such as two entangled 6-qubit subsystems [not shown]. While the success probability decreases away from the training distribution, the persistence of meaningful performance in these regimes is a notable and rather unexpected result.

\subsection{RL-informed circuit transpilation}\label{sec:cnot_count}

\begin{figure*}[t!]
    \centering
    \includegraphics[width=1.0\textwidth]{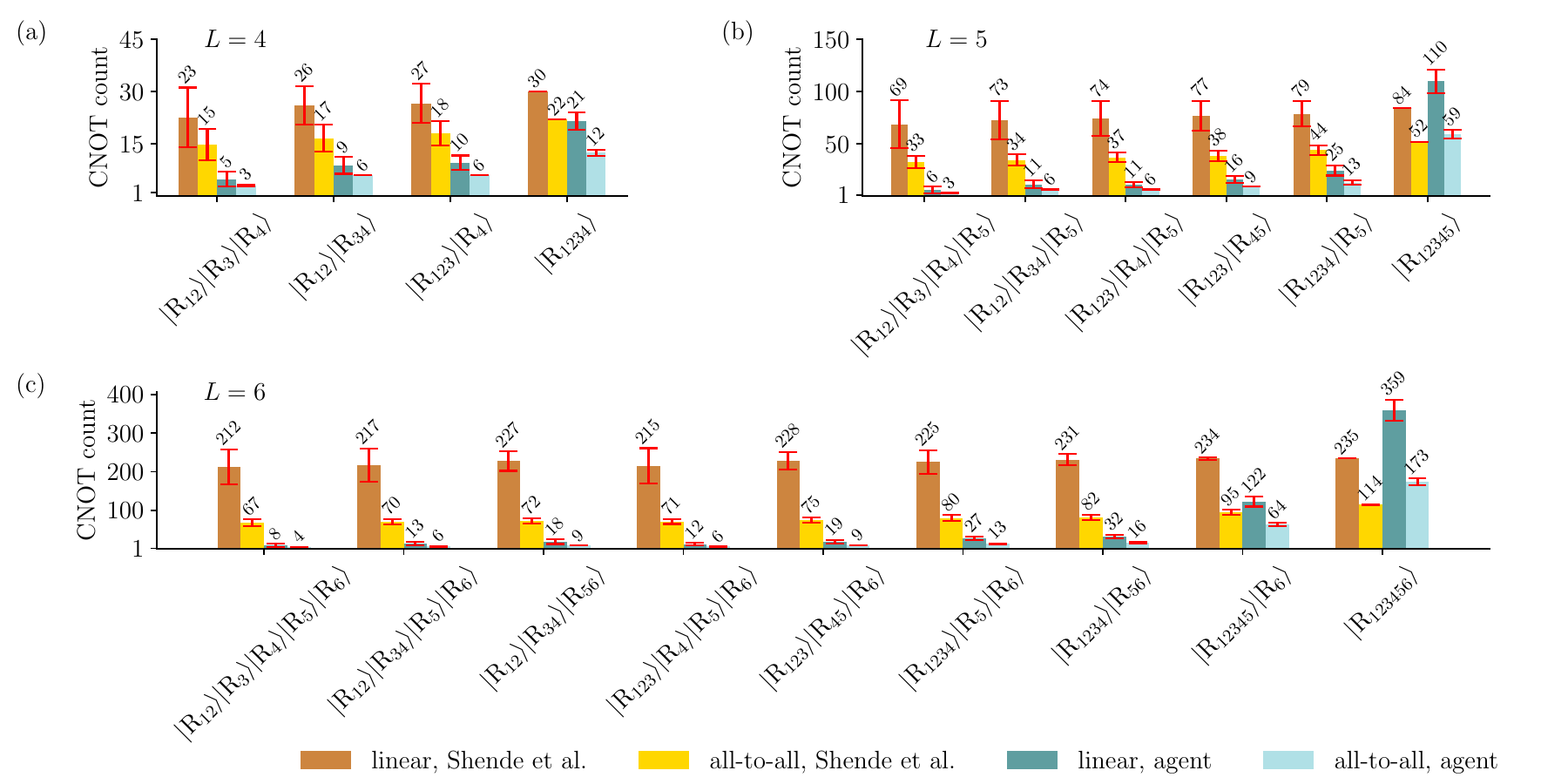}
    \caption{Number of CNOT gates in the transpiled circuit using the Qiskit transpiler for different $L=4,5,6$-qubit initial states. See Fig.~\ref{fig:compare_agents} for details. Each bar is an average over 100 random initial states with random qubit permutations. The error bars show the standard deviation. The orange/yellow colors indicate the number of CNOT gates when using the deterministic algorithm of Shende et al.~\cite{shende2006} for the decomposition. The green/blue colors display the resultant CNOT count when first using the agent to decompose the global circuit into a sequence of two-qubit gates and subsequent transpilation. In most cases, the agent can drastically reduce the number of CNOT gates in the final, transpiled circuit. The yellow/blue colors correspond to a device with all-to-all connectivity between qubits while the orange/green colors show the CNOT count when assuming a linear connectivity and thus, qubits have to be swapped. These additional SWAP gates give rise to a larger number of CNOT gates in the transpiled circuit.}
    \label{fig:nisq_0}
\end{figure*}

To showcase the practical usefulness of an algorithm that has the ability to recognize and exploit the local entanglement structure in a quantum state, we also compare our RL agent to the deterministic algorithm of Shende et al.~\cite{shende2006}, implemented in Qiskit v0.20.0-v1.0.0. To do so, we count the number of CNOT gates required to prepare (disentangle) the different ensembles of random states considered in Fig.~\ref{fig:compare_agents} and show the results in Fig.~\ref{fig:nisq_0}. 

First, we use the deterministic algorithm directly on the initial states and compute the average number of CNOT gates in the decomposed circuit assuming either an all-to-all qubit connectivity (yellow bars) or a linear connectivity (orange bars). The latter requires additional SWAP operations to be inserted into the circuit when two-qubit gates are applied to non-adjacent sites. Hence, we find an overhead in the required number of CNOT gates compared to all-to-all connected devices. 
Next, we employ the RL agent to decompose the same initial states into a sequence of two-qubit gates. In principle, every two-qubit gate can be implemented using at most 3 CNOT gates; however, additional optimization can reduce the overall number of CNOT gates even further. Thus, we employ the Qiskit transpiler on the RL decomposed circuit and infer the resulting number of CNOT gates assuming again either a linear qubit connectivity (green bars) or an all-to-all connectivity (light blue bars). 

When comparing the CNOT counts obtained using the two different approaches, we find that for a system of $L=4$ qubits, the RL agent can drastically reduce the number CNOT gates across all of the considered initial states. For example, the RL agent reduces the average CNOT count for disentangling a fully Haar random state $\ket{R_{1234}}$ from 22 to 12 gates and in the case of the $\ket{R_{123}}\ket{R_4}$ state the average gate count is reduced from 18 to 6. For $L=5,6$ qubits, the agent also outperforms the deterministic algorithm for all but the full-support Haar random initial states. Therefore, we deduce that as soon as the initial state possesses a non-trivial entanglement structure, using the RL agent as a pre-transpilation routine results in an advantage for the three considered system sizes. In App.~\ref{app:transpile} we show that this is also the case for fully-entangled but not Haar-random initial states. 

Finally, let us point out that such a comparison is not perfectly well-posed, since the algorithm of Ref.~\cite{shende2006} leads to perfect disentangling, while the latter is most probably not feasible using only the locally optimal two-qubit gates from Sec.~\ref{sec:two_qubit_U}; this is most likely the origin of the discrepancy for the largest-support Haar-random states. Another important difference is that the RL agent only has access to locally measurable two-qubit reduced density matrices. Nonetheless, this suffices to point out a major advantage of having an adaptive algorithm for reducing the circuit size. To our knowledge, the question of how to best use partial information of the quantum state for optimal disentangling is yet to be answered.

\section{Application on noisy NISQ devices} \label{sec:applns_noisy}

\begin{figure*}[t!]
    \centering
    \includegraphics[width=1.0\textwidth]{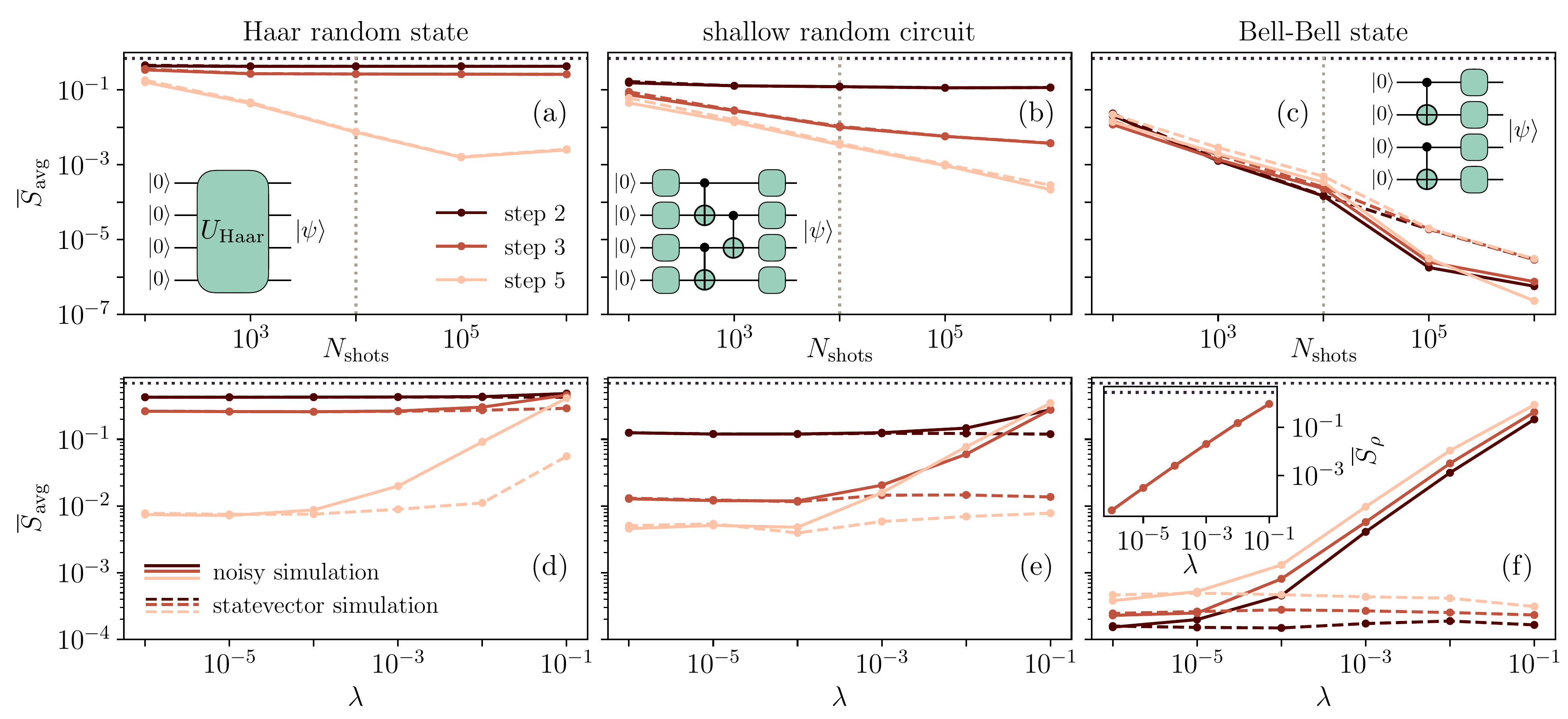}
    \caption{
    (a)-(c) Averaged single-qubit entanglement entropies as a function of the number of shots used to perform state tomography for three different initial state ensembles (three columns). $\overline{S}_\text{avg}$ denotes the mean of the averaged entropies for 100 random realizations of the initial state, including random qubit permutations. Insets show the circuits. Colors indicate the protocol step, i.e., the number of applied two-qubit unitaries. Haar random states can be disentangled in 5 steps, the shallow random circuit in 3 steps, and Bell-Bell states in 2 steps. The applied gates are chosen according to a noise-free, but shot-based simulation for quantum state tomography as they are determined from noisy measurements of the reduced density matrix. 
    Entanglement entropies (solid lines) are calculated from a shot-based reconstruction of the state; in contrast, dashed lines indicate the exact averaged entropies; black dotted lines mark the entropy of a maximally mixed state. 
    For Haar random states, the agent is not able to disentangle all 100 initial states in the 5 steps even in exact statevector simulations. 
    (d)-(f) Same as above, but as a function of the noise strength $\lambda$ when applying a depolarizing channel after every two-qubit gate. The number of shots is $10^4$. Gates are chosen according to the noisy reduced density matrices [App.~\ref{subsec:small_sys_exact}]. Entropies are calculated from noisy states. State vector simulations (dashed lines) show the entropies in exact noise-free simulations using gates produced by the agent based on noisy observations.
    Remarkably, protocols constructed in noisy simulations perform well when applied in the noise-free setting.
    Inset in (f) displays the average entanglement entropy of the full 4-qubit state: the residual entanglement entropy in the final noisy states is mostly due to decoherence rather than left-over entanglement between the qubits.
    }
    \label{fig:nisq_1}
\end{figure*}

In the following, we show that our RL-based disentangling algorithm can be applied to quantum states that are stored on a quantum computer, and that the agent as well as the resulting unitary sequences display a certain robustness to typical errors present in NISQ devices.

\subsection{Sampling noise}

To obtain the next unitary gate in a disentangling protocol, we need to provide the agent with an observation consisting of all two-qubit reduced density matrices (see. App.~\ref{app:obs_space}). Even though $\rho^{(j,i)}$ can be mathematically derived from $\rho^{(i,j)}$, this still necessitates performing quantum state tomography (QST) on $L(L-1)/2$ pairs of qubits to obtain all distinct reduced density matrices (see App.~\ref{app:qst} for a detailed explanation of the procedure behind QST).
Note, however, that once all reduced density matrices have been initially computed, all subsequent steps of the RL protocol require re-computing only those $2L-3$ density matrices that have been modified by the application of the two-qubit gate. Therefore, the number of circuit evaluations scales quadratically in the system size $L$ for the first step and linearly for all succeeding steps.

QST reconstructs the quantum state using stochastic sampling, i.e., the quantum state is repeatedly measured in the $3^2=9$ different directions of the Pauli basis (e.g. $XX,XY,ZY,$ etc.). Each of these measurements (shots) results in a bitstring and the accumulated statistics are classically post-processed to obtain an estimate of the true quantum state. Hence, a finite number of shots introduces a statistical error on the estimated density matrices. It is therefore interesting to examine whether the agent gets confused by the statistical noise and chooses wrong actions that prevent the state from being disentangled. 

In Fig.~\ref{fig:nisq_1}(a)-(c) we study the effect of shot-noise on the RL disentangling framework for three different initial state ensembles with $L=4$ qubits. In each case we consider 100 random realizations of the initial states and show the average single-qubit entanglement entropies at three different steps of the protocol as a function of the number of shots used when performing QST. The solid lines indicate the entanglement entropies that are calculated from the noisy reduced density matrices whereas dashed lines correspond to the exact values obtained from statevector simulations. In (a) we consider Haar random states that can be successfully disentangled within 5 steps; in (b) we consider an ensemble of shallow circuits (see inset) with random single-qubit rotations and randomly permuted qubits that can be disentangled after 3 steps; and finally in (c) we examine random permutations of Bell-Bell pairs with a final layer of random single-qubit rotations that only requires 2 steps of the disentangling protocol. 

Overall, we find that for a finite number of measurement shots, the entanglement entropies cannot be fully reduced to zero. However, the achieved final entanglement entropies decrease with increasing number of shots. Specifically, we observe a scaling of the final averaged values as $\overline{S}_{\mathrm{avg}}\propto N_{\text{shots}}^{-\kappa}$ where $N_{\text{shots}}$ is the number of shots and $0.5\lesssim \kappa\lesssim 1.0$ for the three considered states. Such a scaling can be expected as the applied unitaries are chosen according to the noisy reduced density matrices and thus, are noisy themselves. Hence, even a perfect RL agent cannot bring the entanglement arbitrarily close to zero. Nevertheless, it is robust in the sense that the presence of shot noise does not confuse the action selection of the agent. This robustness can be attributed to the policy network returning the probability of selecting a specific action. These probabilities are perturbed due to the noise; however, as long as the maximum probability in the noise-free case remains dominant in the presence of perturbations, the resultant action is unchanged. 

\begin{figure}[t!]
	\centering
    \hspace*{-0.15cm}
	\includegraphics[width=0.49\textwidth]{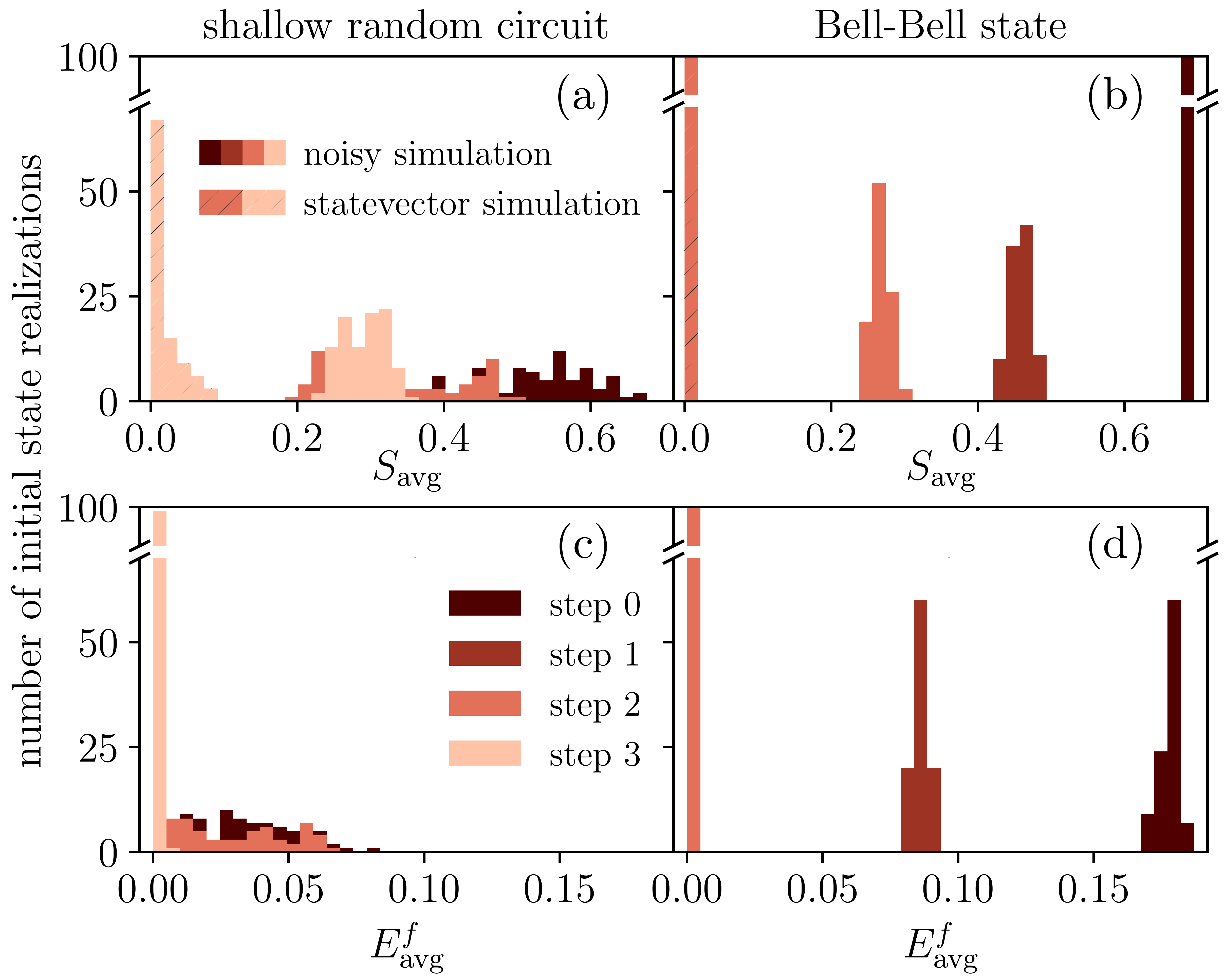}
	\caption{
    (a),(b) Distribution of averaged single-qubit entanglement entropies obtained from a noisy simulation mimicking the \texttt{ibm\_perth} quantum device, for 100 random realizations of the initial state. Colors indicate different steps of the protocol [legend in (c)]. 
    The shallow random circuit (a) and the Bell-Bell state (b) can be disentangled after 3 and 2 steps, respectively. The entanglement entropies at the end of the protocol are substantially larger than zero due to decoherence. The final entropies computed in a noise-free statevector simulation (dashed bars) show that in most cases the qubits have been successfully disentangled. 
    (c),(d) Same as above, but the distribution is plotted over the average entanglement of formation $E^f_\text{avg}$ which measures genuine 2-qubit entanglement. For both initial state ensembles, $E^f_\text{avg}$ is reduced to zero, and hence the entanglement between the qubits has been successfully removed; any residual entanglement entropy is due to entanglement between the system and its environment.}
    \label{fig:nisq_2}
\end{figure}

\subsection{Depolarizing noise channel}

The statistical noise caused by sampling is inherent to quantum computing and cannot be alleviated apart from increasing the number of shots. However, NISQ computations are also subject to a variety of other errors, most notably decoherence. To study its effect we consider an artificial depolarizing noise channel that is applied after each two-qubit gate of the disentangling protocol
\begin{equation}
    \mathcal{E}(\rho)=(1-\lambda) \rho+\lambda  \frac{I}{2^L} \ ,
\end{equation}
where $\rho$ is the density matrix of the full $L$-qubit quantum system, $\lambda\in[0,1]$ is the noise strength, and $I$ is the identity matrix that corresponds to a maximally mixed state. 
In Fig.~\ref{fig:nisq_1}(d)-(f) we study the effect of such a depolarizing noise channel on our disentangling scheme for varying noise parameters $\lambda$ and the same initial state ensembles as in (a)-(c). For small noise strengths $\lambda\lesssim 10^{-4}$, the entanglement entropies approach their values set by the statistical shot noise. For reference, we fixed the number of shots to $10^4$ which is indicated by a vertical dotted line in panels (a)-(c). 

For increasing noise strengths the entanglement entropies computed from the noisy density matrices increase as expected. This deviation can arise from either of two factors: (i) Due to decoherence, the qubits become entangled with their environment. This residual entanglement cannot be removed by the disentangling protocol as it only acts on the system qubits. (ii) The agent does not select the correct pairs of qubits and, thus, fails in disentangling the state within the optimal number of steps. 
To determine which of these scenarios dominates, we also plot the exact entanglement entropies calculated via state vector simulations (dashed lines in Fig.~\ref{fig:nisq_1}). For the shallow random circuit and the Bell-Bell state examples, the exact entanglement entropies are roughly independent of the noise strength. This shows that the agent still chooses the correct actions even in the presence of depolarizing noise. In contrast, for the Haar random states at very large noise strengths the entanglement entropy increases indicating that in this case the agent does fail to choose the right actions for some states. This is however not surprising since at this point the density matrix of the full quantum state is close to a maximally mixed state and retrieving any information about the remaining local entanglement structure becomes challenging. For comparison, we display the dependence of the entanglement entropy of the full density matrix on the noise parameter in the inset of Fig.~\ref{fig:nisq_1} (f). 

To summarize, we find that the agent is indeed robust to depolarizing noise as long as the quantum state is sufficiently distinct from a fully mixed state. Interestingly, this robustness implies that we can infer the optimal disentangling circuit from a noisy quantum computation which realizes the optimal unitary circuit (up to shot-noise errors) also in the noise-free setting.

\subsection{Hardware noise model}\label{sec:fake_noise}

Finally, we study a more realistic noise model provided by Qiskit which mimics the noise encountered in one of IBM's real quantum devices. We start from 100 random realizations of the shallow random circuit and the Bell-Bell state and plot the distribution of the measured entanglement entropies at different steps of the protocol in Fig.~\ref{fig:nisq_2}(a),(b). Note that we did not perform this experiment for the Haar random states as the preparation of the initial states already resulted in a fully mixed state due to the large depth of the state preparation circuit and the introduced noise. However, even for the shallow random circuits and the Bell-Bell pairs we obtain large values for the entanglement entropies at the end of the protocol. Hence, we also show the exact entanglement entropies computed via statevector simulations (striped bars) at the final time step. For the Bell-Bell example, all initial states have been successfully disentangled. The residual entanglement entropies encountered in the noisy simulations can therefore entirely be attributed to decoherence. For the shallow random circuits, the success probability is slightly diminished and thus, the noise can affect the action selection of the agent in this case.

On noisy quantum computers, the single-qubit entanglement entropy is not necessarily a good figure of merit for quantifying the performance of the agent as it also contains contributions from any entanglement of the system with its environment. To examine the entanglement strictly between the system qubits we therefore consider the entanglement of formation $E^f$ [cf.~Eq.~\eqref{eq:ent_formation}] measuring genuine two-particle entanglement. We compute an average over the entanglement of formation between any two pairs of qubits via their noisy two-qubit density matrices and plot the corresponding distribution in Fig.~\ref{fig:nisq_2}(c),(d). We indeed find that the entanglement of formation is minimized close to zero at the end of the protocol and thus, in most cases the qubits are perfectly disentangled from each other.

Note that the agent used in the simulations above has not been trained in a noisy environment. Nevertheless, we find that the agent can extrapolate its protocols to the case of noisy density matrices and thus, mixed states. We also trained an agent directly in noisy environments, however, we did not observe an improvement in the agent's performance.

\section{Implementation on a trapped-ion quantum computer}
\label{sec:exp_impl}

\begin{figure*}[t!]
    \centering
    \includegraphics[width=0.95\textwidth]{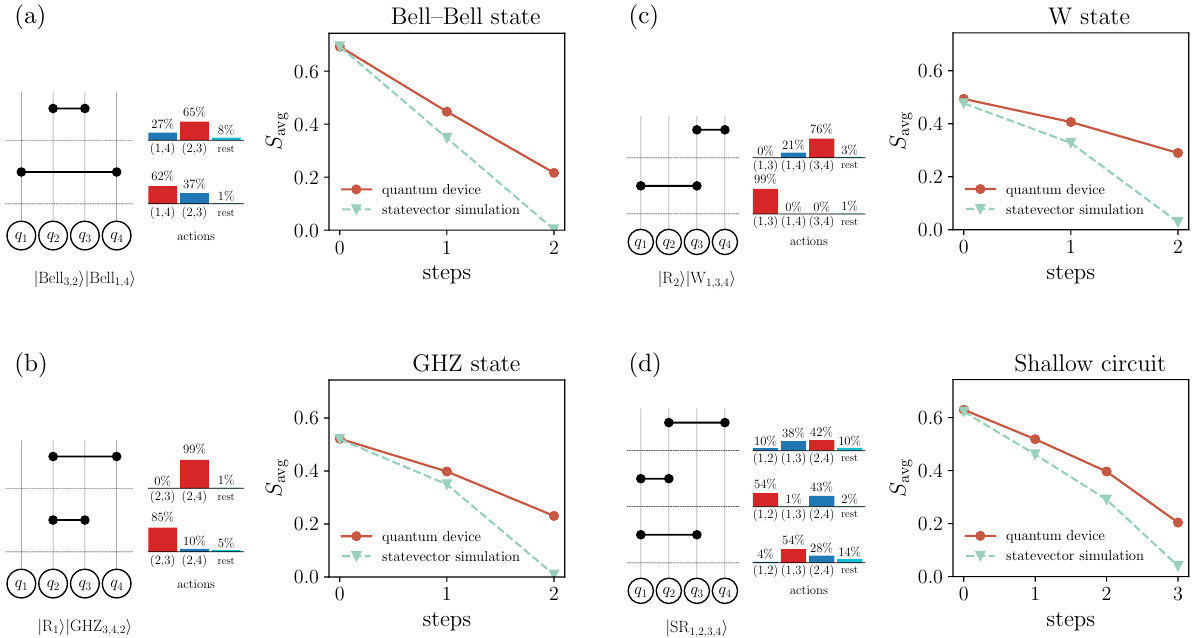}
    \caption{
    Quantum hardware experiment performed on a trapped ion system (described in Appendix~\ref{app:UMDTI_Setup}) for four initial states: (a) Bell-Bell state, (b) GHZ state, (c) W state, and (d) shallow circuit state.
    (Left) The circuit inferred by the RL agent to disentangle the given initial state. The histogram on the right shows the policy with emphasis on the two actions of the highest weight. 
    (Right) The measured averaged von Neumann entanglement entropies during the protocol (red solid line) as well as the corresponding entropies when the protocol is evaluated with a noise-free statevector simulator (blue dashed line). 
    The residual entanglement observed in the experiment can mostly be attributed to decoherence. See Fig.~\ref{fig:app:nisq_1} and Fig.~\ref{fig:app:nisq_compare} for additional hardware results.
    }
    \label{fig:nisq_4}
\end{figure*}

We tested our agent on real quantum hardware and report the results in Fig.~\ref{fig:nisq_4} of an experiment performed on a trapped ion system (further experiments and additional details are discussed in Appendix~\ref{app:noise}). 

We prepare the quantum device in four initial states: a Bell-Bell state, a GHZ state, a W state, and a shallow circuit state, shown in the four panels of Fig.~\ref{fig:nisq_4}. We show the resulting gates inferred by the RL algorithm, as well as the corresponding action probabilities at every step of the circuit on the left. The agent identifies the correct qubit indices for disentangling the state using a minimal number of steps.
On the right, we plot an average of the measured single-qubit entanglement entropies (red solid line) which is reduced at the end of the protocol. However, we find that a residual von Neumann entropy remains in the state; hence, we also show the entropies obtained from a statevector simulation using the same gates as before (blue dashed line); the latter entropy is brought down close to zero as expected. 

In Fig.~\ref{fig:app:nisq_1} we also discuss the entanglement of formation during the protocol, which quantifies the amount of entanglement for mixed states. 
We provide additional results obtained on another trapped ion system as well as one of IBM Quantum's superconducting qubit devices. In Fig.~\ref{fig:app:nisq_compare} we show comparisons of four different quantum hardware devices for two initial states. For both cases we present both the experimentally measured von-Neumann entanglement entropies as well as a noise-free statevector simulator von-Neumann entanglement entropies. Overall, we find that the RL agent can successfully disentangle low-depth states provided in Fig.~\ref{fig:nisq_4}, that can be prepared (and disentangled) with a few gates. For more complex states the agent quickly fails since noise becomes dominant resulting in maximally mixed states for which any selected two-qubit unitary fails. However, these results are expected to improve in the future as devices with lower error rates and larger coherence times become available.

\section{\label{sec:outro}Discussion \& Outlook}

We trained an RL agent in a simulator to disentangle arbitrary 4-, 5-, and 6-qubit states and weakly entangled 12- and 16-qubit states using an actor critic (AC) algorithm. 
The agent is given access only to two-qubit reduced density matrices of the quantum state as partial observations, which makes it applicable on NISQ devices. For a fixed pair of qubits, we propose an analytical way to compute locally optimal two-qubit gates that minimize the entanglement of formation between the qubits; however, the general structure of our RL agent allows this routine to be replaced by alternatives, e.g., Ref.~\cite{hauschild2018finding}.  

In particular, our RL agent is model-free, and, as discussed in Sec.~\ref{subsec:scaling_laws}, scales quadratically with the number of qubits, which makes training and running on NISQ devices possible.
We use a permutation equivariant transformer architecture which adjusts the policy of the agent to permutations of the qubits in the input state: hence, the agent is able to `recognize' when qubits have been swapped and produces a `swapped' sequence accordingly. We directly exploit its capability to identify entanglement structure from state observations -- a hallmark feature of deep learning algorithms -- whose potential application in quantum technologies remains to be fully utilized. Compared to conventional quantum optimal control algorithms, an additional prominent feature of deep RL is that, once trained, the agent readily produces solutions for arbitrary initial states without any additional re-training or further optimization; this makes it appealing to deploy in experiments.

We benchmark the agent on arbitrary 4-qubit states with and without obvious local entanglement structure. For 5- (6-) qubit Haar-random states, our RL agent takes on average 20 (56) two-qubit unitaries to bring the entanglement per qubit below a fixed threshold value. Intriguingly, it uncovers correlated patterns in the learned gate sequence (both among the qubits and in between consecutive gates).
For large systems of $L=12$ and $L=16$ qubits we benchmark the agent on weakly entangled subsystems of Haar-random states (App.~\ref{app:large_systems}) and find that it takes on average $20.1$ ($L{=}12$) and $30.7$ ($L{=}16$) unitaries to bring the total entanglement $S_{\text{tot}}$ below the threshold, cf.~\eqref{eq:max_sqe}.

We analyzed in detail the statistics of disentangling protocols for various combinations of product states of Haar-random states on various subsystems, and demonstrated that the information from partial observations of the state can be utilized to reduce the number of both two-qubit and CNOT gates required, as compared to state-of-the-art deterministic disentangling algorithms.

We also quantified the resilience of our trained agents to both shot and environmental noise which can corrupt the observations; remarkably, our agent can tolerate moderate levels of distinct noise sources, even though it was trained in a noise-free environment -- another salient feature of RL algorithms.

Last but not least, we deployed the trained RL agent on a real trapped-ion quantum processor, where it successfully disentangled several nontrivial states, demonstrating robust transfer from simulation to realistic noisy hardware.

Analyzing optimal 4-qubit sequences, we identified a circuit with at most five 2-qubit gates that disentangles any 4-qubit state. Importantly, the unitaries in this circuit depend on the state itself, cf.~App.~\ref{subsec:small_sys_exact}. The proof reveals that this circuit can be implemented using at most ten CNOT gates. Turning this result around, it follows that one can prepare any 4-qubit state starting from the product state $|0\rangle^{\otimes 4}$ with no more than five 2-qubit unitary (ten CNOT) gates. 

RL agents capable of disentangling states find natural practical applications on NISQ devices. In App.~\ref{app:nisq}, we present results from simulations on noisy superconducting qubits and trapped ion platforms. However, the concept of learning information about the distribution of entanglement in a state and using it to improve the corresponding disentangling circuit, is generic: it can also be applied to circuits in quantum optics~\cite{wang2018multidimensional} (where entangling gates are implemented by beam splitters) and generalized to higher-dimensional local Hilbert spaces (i.e., qudit systems).

For the system sizes within reach, a natural generalization of the problem considered in this work is circuit synthesis~\cite{fosel2021quantum,rietsch2024unitary}. Indeed, decomposing a unitary gate into its constituents using as few two-qubit gates as feasible, finds a wide application in quantum computing. As a special case, the optimal implementation of Haar-random unitaries is the subject of intense research~\cite{alagic2020efficient,koh2023measurement}.  

Although optimal disentangling sequences composed of two-qubit gates generally require resources that scale exponentially with system size, our analysis reveals that this exponential cost originates primarily from the quantum-state simulation itself, whereas the forward and backward passes of the RL agent scale only quadratically with the number of qubits. A practical route toward larger systems is to focus on families of states that can be disentangled in relatively few simulation steps, such as weakly entangled multipartite states. These results motivate future investigations of physically relevant restricted classes of states, such as area-law-entangled states, or states that can be reached in finite time via unitary evolution generated by physical Hamiltonians.

Since unitary evolution is invertible, when run backwards, state disentangling sequences can be used to prepare complex many-body states. The disentangling procedure can thus be seen as state compression, wherein a quantum state is compressed into an initial state and a sequence of unitary gates. In this context, particularly intriguing would be ground states of correlated quantum many-body systems in two dimensions where matrix-product-state techniques are known to struggle. 

When it comes to scaling up to larger system sizes, an exciting future direction is to design the learning architecture with the help of tensor networks~\cite{metz2022self,gillman2022reinforcement}: indeed, the MERA ansatz~\cite{vidal2007entanglement} is known to capture critical states whose entanglement entropy grows logarithmically with the system size.
On the RL side, one can imagine defining compound actions, e.g., adding the reported optimal 4-qubit sequence as an available action; this type of coarse-graining may provide a way to eliminate some of the training complexity of the problem. 
Larger system sizes can also be reached if one restricts the dynamics and the states of interest to those generated by the Clifford group~\cite{cemin2025learning}. 

Finally, another practical application of the framework we developed is found in the recently introduced unitary circuit games~\cite{morral2023entanglement,morral2025disentangling}. Casting these zero-sum games within the RL framework shows that an informed agent can alter the critical behavior of entanglement dynamics~\cite{cemin2025learning}. 

\section{\label{sec:video}Training Video}

The paper is accompanied by a supplementary video displaying excerpts from the training process for a 4-qubit agent [see ancillary files on arXiv]. The movie shows how the RL agent learns to construct optimal disentangling circuits starting from Haar-random initial states, starting with no prior knowledge. 

The agent selects the pair of qubits, corresponding to the red-colored action, according to its policy [bottom right]. Then it computes the locally optimal two-qubit gate, and appends it to the disentangling circuit [top left]; the percentage next to the gate marks the probability of selecting the red-colored action, while the color of each circle in between the gates shows the range of the entanglement between that qubit and the rest [see legend].  The average single-qubit von Neumann entanglement entropy is shown at each time step right below the circuit (it shares the episode step axis); it allows us to directly monitor the quality of each applied gate. 
The training (return) curve is shown in the top right corner, with the red sniper dot denoting the current iteration; this gives an overview of which stage of the training process the displayed iteration stems from. 

The video shows select iterations representative of stages that display qualitative changes in the learning process: 
(i) disentangling a pair of qubits from the rest;
(ii) successfully disentangling all qubits (i.e., bringing the average single-qubit entanglement entropy to zero);
(iii) reducing the number of gates required (circuit depth). 
In doing so one can observe how the RL agent learns as its policy evolves to complete the task.
Finally, we show a high-speed preview of the entire training process. 

\begin{acknowledgments}

This work is supported by a collaboration between the US DOE and other Agencies. 
We would like to thank Ra\'ul Morral Yepes and Benedikt Placke for insightful discussions, and Giovanni Cemin for providing comments on the manuscript. We also thank Anton Trong Than and Xingxin Liu for support with the trapped-ion experiments. 
This material is based upon work supported by the U.S. Department of Energy, Office of Science, National Quantum Information Science Research Centers, Quantum Systems Accelerator under Award No. DE-FOA-0002253. Additional support is acknowledged from the National Science Foundation STAQ project No. PHY-2325080. 
Funded by the European Union (ERC, QuSimCtrl, 101113633). Views and opinions expressed are however those of the authors only and do not necessarily reflect those of the European Union or the European Research Council Executive Agency. Neither the European Union nor the granting authority can be held responsible for them.
This work is supported by the Okinawa Institute of Science and Technology Graduate School (OIST). 
FM acknowledges support by the NCCR MARVEL, a National Centre of Competence in Research, funded by the Swiss National Science Foundation (grant number 205602).
The authors gratefully acknowledge UNITe -- Universities for Science, Informatics and Technology in the e-Society for granting access to the GPU computational cluster located at Sofia University (Sofia, Bulgaria) for this project.
Classical simulations were performed on the high-performance computing cluster (Deigo) provided by the Scientific Computing and Data Analysis section at OIST. 
We acknowledge the use of IBM Quantum services for this work. The views expressed are those of the authors, and do not reflect the official policy or position of IBM or the IBM Quantum team.
We acknowledge support from Microsoft's Azure Quantum for providing credits and access to the IonQ Harmony quantum hardware used in this paper.

\end{acknowledgments}

\section*{Code Availability}

The source code is available on GitHub~\cite{github_code}. 
A Jupyter notebook is also available to test our trained RL agents on user-specified states, to apply mid-circuit perturbations to the state and study the reaction of the agent, and to monitor the attention heads of the underlying transformer network.

\clearpage
\appendix

\begin{widetext}

\begin{center}
\section*{Appendix}
\end{center}

\end{widetext}

\let\addcontentsline\oldaddcontentsline
\tableofcontents

\section{ \label{app:phys_details}Details of the disentangling physics}

\subsection{\label{sec:two_qubit_U}Locally optimal two-qubit disentangling gates}

Since a two-qubit unitary acting on qubits $(i,j)$ can only change the entanglement between subsystems each containing one of these two qubits, it is sufficient for the analysis below to consider the two-qubit reduced density matrix $\rho^{(i,j)}$, obtained by tracing out all other qubits from the pure state. We emphasize that the unitary we seek, $U^{(i,j)}$, will be applied on the pure state $|\psi\rangle$ of the entire system, and we only consider $\rho^{(i,j)}$ for simplicity. 
In general, $\rho^{(i,j)}$ is a mixed state; its von Neumann entropy measures the entanglement between qubits $(i,j)$ and the rest of the system. 

By contrast, here we are interested in constructing a unitary gate $U^{(i,j)}$ which minimizes the entanglement between qubits $(i,j)$. For mixed states $\rho^{(i,j)}$, this amounts to reducing the entanglement of formation $E_f$ which, for a two-qubit system, is given by 
\begin{equation}
\label{eq:ent_formation}
    E^f[\rho^{(i,j)}] {=}h(x),\quad h(x){=}{-}x\log x {-} (1-x)\log(1-x),
\end{equation}
with $x{=}\left(1{+}\sqrt{1{-}C^2(\rho^{(i,j)})}\right)/2$. $C(\rho^{(i,j)})$ denotes the concurrence which can be computed as 
$$
C(\rho^{(i,j)}) = \max\{0,\mu_1{-}\mu_2{-}\mu_3{-}\mu_4\}
$$
where $\mu_j$ denote the eigenvalues of the matrix 
$$
R {=} \sqrt{\sqrt{\rho^{(i,j)}}\;Y^{(i)}Y^{(j)} (\rho^{(i,j)})^* Y^{(i)}Y^{(j)} \sqrt{\rho^{(i,j)}}}
$$ 
with $Y^{(i)}$ the Pauli-$y$ gate acting on qubit $i$, and $\ast$ denotes complex conjugation.

It is straightforward to see that if we take $U^{(i,j)}$ to diagonalize $\rho^{(i,j)}$, the resulting diagonal density matrix after applying the gate, 
\begin{equation}
\label{eq:Uloc_def}
    \tilde\rho^{(i,j)}=U^{(i,j)}\rho^{(i,j)}[U^{(i,j)}]^\dagger = \text{diag}(\lambda_1,\lambda_2,\lambda_3,\lambda_4),
\end{equation}
describes a separable state, and thus has a vanishing entanglement of formation. Therefore, applying the diagonalizing gate minimizes the entanglement between qubits $(i,j)$. 

Furthermore, imposing (without loss of generality)
\begin{equation}
\label{eq:lambda_order}
\lambda_1\geq\lambda_2\geq\lambda_3\geq\lambda_4, \qquad \lambda_j\in[0,1]    
\end{equation}
we conjecture that the same diagonalizing gate minimizes the sum of the single-qubit entanglement entropies~\footnote{In general, for two-qubit mixed states the von Neumann entropy of the single-particle density matrix of a given qubit is not a measure of entanglement between the two qubits. However, recalling that the entire $L$-qubit system is in a pure state $|\psi\rangle$, the single-qubit von Neumann entropy gives the entanglement between the single qubit and all $L-1$ remaining qubits.} $S_\text{ent}[\tilde\rho^{(i)}]{+}S_\text{ent}[\tilde\rho^{(j)}]$, where 
\begin{eqnarray}
    \tilde\rho^{(i)}&=&\text{tr}_{\{j \}}\tilde\rho^{(i,j)}= 
    \text{diag}(\lambda_1+\lambda_2, \lambda_3+\lambda_4)
    , \nonumber\\
    \tilde\rho^{(j)}&=&\text{tr}_{\{i \}}\tilde\rho^{(i,j)}= 
    \text{diag}(\lambda_1+\lambda_3, \lambda_2+\lambda_4).
\end{eqnarray} 
The resulting single-qubit entanglement entropies are computed as $S_\text{ent}[\tilde\rho^{(i)}] {=} h(\lambda_1{+}\lambda_2)$ and $S_\text{ent}[\tilde\rho^{(j)}] {=} h(\lambda_1{+}\lambda_3)$. 
An analytical proof that no other two-qubit gate $U^{(i,j)}$ can achieve a smaller value for the sum $S_\text{ent}[\tilde\rho^{(i)}]{+}S_\text{ent}[\tilde\rho^{(j)}]$, can be obtained following the arguments of Ref.~\cite{jetvic2012quantum}; here, we confirmed this conjecture using numerical optimization over the space of two-qubit unitaries, for an exhaustively large number of random states $\rho^{(i,j)}$. Hence, the diagonalizing gate also minimizes the sum of the two single-particle entanglement entropies (i.e., the entanglement between each of the qubits considered separately, and the remaining $L-1$ qubits). 

Once found, the physical implementation of $U^{(i,j)}$ is straightforward.  
A method to construct an optimal quantum circuit for a general two-qubit gate that requires at most 3 controlled-NOT (CNOT) gates and 15 elementary one-qubit gates, is presented in Refs.~\cite{zhang2003geometric,shende2004minimal,vatan2004optimal,vidal2004universal}.

Let us make a few remarks: 
(i), first, the optimal disentangling two-qubit gate depends on the state $\rho^{(i,j)}$; thus, changing the pair of qubits will result in a different gate in general. 
(ii), notice that the optimal disentangling two-qubit gate is not unique: even after fixing the order of the eigenvalues, cf.~Eq.~\eqref{eq:lambda_order}, we can still multiply $U^{(i,j)}$ by any single-qubit gate without changing the entanglement structure of either of $\tilde\rho^{(i)}$, $\tilde\rho^{(j)}$, and $\tilde\rho^{(i,j)}$. However, doing this in general breaks the property that these reduced density matrices are diagonal in the computational basis. 
In fact, (iii), due to the diagonalizing property and the ordering relation in Eq.~\eqref{eq:lambda_order}, we see that the weight of the resulting state after the application of the gate is shifted towards the $|00\rangle\langle00|$-component; a repeated application of the diagonalizing two-qubit gate on different pairs of qubits within the pure $L$-qubit state $|\psi\rangle$ then serves to ultimately shift the weight of the probability amplitudes to the target state $|0\rangle^{\otimes L}$ after sufficiently many iterations. 
Last, (iv), although the diagonalizing gate is locally optimal in terms of reducing both the entanglement of formation and the sum of the single-qubit entanglement entropies, it need not be optimal in terms of the desired global disentangling process [cf.~two-qubit gate introduced in Ref.~\cite{hauschild2018finding} which uses non-local information]; therefore, a shorter circuit to disentangle an $L$-qubit quantum state $|\psi\rangle$ may still exist, which does not use the locally optimal disentangling gates defined above.

\subsection{Optimal disentangling protocols for 3-qubit and 4-qubit systems} \label{subsec:small_sys_exact}

Any two-qubit pure state $|\psi_{1,2}\rangle$ is, by construction, always disentangled by a single optimal gate $U^{(i,j)}$. 

\subsubsection{Disentangling sequences for arbitrary three-qubit states}

\begin{figure}[t!]
    \centering
    \includegraphics[width=0.3\textwidth]{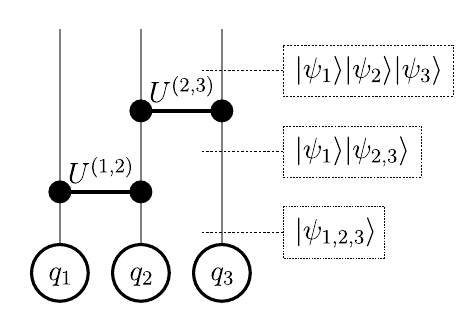}
    \caption{
    Optimal sequence to disentangle an arbitrary 3-qubit state $|\psi_{1,2,3}\rangle$ using two consecutive state-dependent locally-optimal gates. The structure of the state after the application of each gate is displayed on the right, emphasizing its product structure. 
    Qubits are denoted by $q_j$ placed in a circle; 
    qubit lines (thin grey) run vertically; 
    locally optimal two-qubit gates are denoted by black horizontal lines ending in filled black circles that designate the two qubits involved.
    }
    \label{fig:3_qubits}
\end{figure}

Curiously, it turns out that, for generic three-qubit systems, two locally-optimal gates suffice to disentangle an arbitrary 3-qubit state $|\psi_{1,2,3}\rangle$ into the product $|0\rangle^{\otimes 3}$, using the circuit shown in Fig.~\ref{fig:3_qubits}: 
\begin{equation}
  U_\ast = U^{(2,3)}U^{(1,2)}.  
\end{equation}
To see this, we write down an exact representation of the arbitrary initial state $|\psi_{1,2,3}\rangle$ in terms of its Schmidt decomposition $ |\psi_{1,2,3}\rangle = \sum_{\alpha=1}^2 \Lambda_{\alpha}|u_\alpha\rangle_{1,2}|v_\alpha\rangle_{3}$; importantly, the corresponding Schmidt rank is at most two. Thus, the two-qubit reduced density matrix $\rho^{(1,2)}$ has a rank of at most two as well. 
As a result, after applying the locally-optimal gate $U^{(1,2)}$, we necesarily have $\lambda_{3,4}=0$ in Eq.~\eqref{eq:Uloc_def}, and hence $S_\text{ent}[\tilde\rho^{(1)}]=0$, which factors out the state of qubit $1$. The resulting state, $|\psi_1\rangle|\psi_{2,3}\rangle$ is effectively a two-qubit state from the perspective of entanglement, and hence can be disentangled by applying just one more (but different) unitary. 

In general, for an arbitrary three-qubit state, there are three pairs of qubits to apply the gate on. The procedure described above works irrespective of which two pairs are chosen; this qubit-permutation symmetry of the problem leads to a 3-fold degeneracy of the optimal disentangling sequence. To illustrate how this works in practice, consider as a specific example the initial state $|\psi\rangle = |\psi_2\rangle|\psi_{1,3}\rangle$ with $|\psi_{1,3}\rangle$ a Bell state, and the optimal sequence $U_\ast = U^{(2,3)}U^{(1,2)}$. In this case, although there is no entanglement between qubits $(1,2)$, the first gate $U^{(1,2)}$ will still swap the two qubits due to the eigenvalue ordering operation [cf.~Sec.~\ref{sec:two_qubit_U}], thus separating the state of qubit 1 from the remaining qubits; $U^{(2,3)}$ will then disentangle the remaining two qubits. 

For generic initial states, qubits $(1,2)$ are also entangled, and $U^{(1,2)}$ implements a dressed SWAP gate. The eigenvalue ordering breaks the qubit-exchange symmetry, and gives the dressed SWAP gate a preferred direction. Therefore, in some cases, it may be useful to think of the locally optimal two-qubit gate as a directed dressed SWAP operation.  

\subsubsection{Disentangling sequences for arbitrary four-qubit states}

\begin{figure}[t!]
    \centering
    \includegraphics[width=0.5\textwidth]{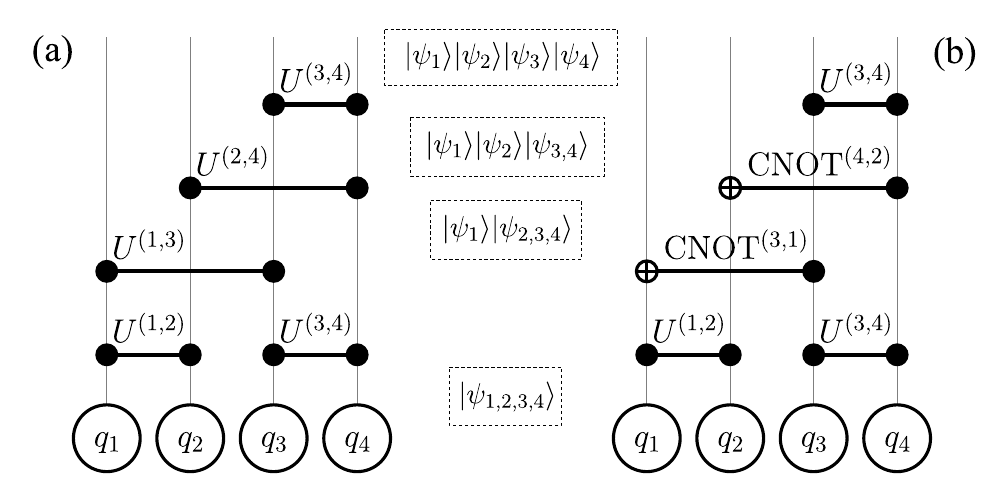}
    \caption{
    Optimal sequence to disentangle an arbitrary 4-qubit state $|\psi_{1,2,3,4}\rangle$ using:
    (a) five state-dependent locally-optimal gates, and
    (b) three state-dependent locally-optimal gates and two CNOT gates.
    The structure of the state after the application of each gate is displayed in the boxes.
    The circuit depth is $4$ in both cases. 
    }
    \label{fig:4_qubits}
\end{figure}

The situation is more intricate for 4-qubit states. Intriguingly, we find that an arbitrary 4-qubit state can be transformed into the state $|0\rangle^{\otimes 4}$ and thus disentangled completely by using a sequence of at most five locally-optimal two-qubit gates, e.g., 
\begin{equation}
\label{eq:4q_optimal}
  U_\ast = U^{(3,4)}U^{(2,4)}U^{(1,3)}U^{(3,4)}U^{(1,2)}.  
\end{equation}
Starting from an arbitrary 4-qubit state $|\psi_{1,2,3,4}\rangle$, the first three unitaries in this sequence, $U^{(1,3)}U^{(3,4)}U^{(1,2)}$, separate out qubit 1 from the rest: $|\psi_{1}\rangle|\psi_{2,3,4}\rangle$, see Fig.~\ref{fig:4_qubits}(a); the remaining state is (at most) 3-qubit-entangled and the protocol follows the optimal 3-qubit sequence discussed above (Fig.~\ref{fig:3_qubits}).

To understand why any four-qubit state can be disentangled using at most five locally optimal gates, we consider a generic 4-qubit state written in its Schmidt decomposition between subsystems $(1,2), (3,4)$:
\begin{equation}
    |\psi\rangle = |\psi_{1,2,3,4}\rangle = \sum_{\alpha=1}^4 \Lambda_{\alpha}|u_\alpha\rangle_{1,2}|v_\alpha\rangle_{3,4},
\end{equation}
where $\{|u_\alpha\rangle_{1,2}\}$ and $\{|v_\alpha\rangle_{3,4}\}$ are orthonormal bases for the corresponding two-qubit subsystems, and the Schmidt values $\Lambda_\alpha$ correspond to the eigenvalues of the reduced density matrices $\rho^{(1,2)}$ and $\rho^{(3,4)}$. 
Since the locally optimal unitary $U^{(1,2)}$ is designed to diagonalize the reduced two-qubit density matrix, it necessarily maps the states to the $z$-eigenstates $U^{(1,2)}|u_\alpha\rangle_{1,2} = |\alpha\rangle_{1,2}$, and similarly: $U^{(3,4)}|v_\alpha\rangle_{3,4} = |\alpha\rangle_{3,4}$. 

Hence, after the application of the first layer of gates, the state takes the form
\begin{eqnarray}
    U^{(3,4)}U^{(1,2)}|\psi\rangle &{=}& 
    \Lambda_1|0000\rangle {+}
    \Lambda_2|0101\rangle {+} 
    \Lambda_3|1010\rangle {+} 
    \Lambda_4|1111\rangle  \nonumber\\
    &=&\phantom{+} |00\rangle_{1,3}\left( \Lambda_1 |00\rangle_{2,4} + \Lambda_2 |11\rangle_{2,4}   \right) \nonumber\\
    &&+
    |11\rangle_{1,3}\left( \Lambda_3 |00\rangle_{2,4} + \Lambda_4 |11\rangle_{2,4}   \right).
\end{eqnarray}
It is now obvious that qubit 1 can be separated from the rest by applying a $U^{(1,3)}=\text{CNOT}^{(3,1)}$ gate; similarly, we can factor out the state of qubit 2 by applying a $U^{(2,4)}=\text{CNOT}^{(4,2)}$ gate. This gives
\begin{eqnarray}
&&\text{CNOT}^{(4,2)}\text{CNOT}^{(3,1)}U^{(3,4)}U^{(1,2)}|\psi\rangle =
    |0\rangle_{1}|0\rangle_{2
    }\otimes\\
    &&\otimes\left( 
    \Lambda_1 |0\rangle_{3}|0\rangle_{4} + 
    \Lambda_2 |0\rangle_{3}|1\rangle_{4} +
    \Lambda_3 |1\rangle_{3}|0\rangle_{4} + 
    \Lambda_4 |1\rangle_{3}|1\rangle_{4} 
    \right). \nonumber
\end{eqnarray}
Note that $\Lambda_i\in\mathbb{R}$ are all real-valued.
Therefore, the remaining two-qubit state on subsystem $(3,4)$ can be disentangled by applying a real-valued gate $U^{(3,4)}$, cf.~Fig.~\ref{fig:4_qubits}(b):
\begin{equation}
\label{eq:4q-seq}
U^{(3,4)}\text{CNOT}^{(4,2)}\text{CNOT}^{(3,1)}U^{(3,4)}U^{(1,2)}|\psi\rangle = |0\rangle^{\otimes 4}.
\end{equation}

We stress that Eq.~\eqref{eq:4q-seq} (or a suitably qubit-permuted version of it, see comment below) holds for any $4$-qubit state $|\psi\rangle$, and thus also gives a decomposition of an arbitrary 4-qubit gate. Since an arbitrary complex-valued 2-qubit gate requires at most 3 CNOT gates~\cite{vatan2004optimal}, and an arbitrary real-valued 2-qubit rotation gate requires at most 2 CNOT gates
\footnote{An arbitrary \textit{real-valued} two-qubit \textit{rotation} gate $U^{(1,2)}{\in}\text{SO}(4)$ contains at most 2 CNOT gates, since it can be parametrized as 
\begin{eqnarray*}
    U^{(1,2)}&=&Y^{(1)}(\alpha)Y^{(2)}(a)\;\text{CNOT}^{(1,2)}\times \\
    &&\times Y^{(1)}(\beta)Y^{(2)}(b)\;\text{CNOT}^{(1,2)}Y^{(1)}(\gamma)Y^{(2)}(c),    
\end{eqnarray*}
at Euler angles $\alpha,\beta,\gamma,a,b,c\in[0,2\pi)$, where $Y^{(j)}(\alpha)=\exp(-i\alpha S^y_j)$ is a single-qubit rotation about the $y$-axis; this identity follows from the Lie algebra decomposition $\mathfrak{so}(4){=}\mathfrak{so}(3){\oplus}\mathfrak{so}(3)$, together with the defining properties of Euler angles. Real-valued gates with a negative determinant, like the SWAP gate, cannot be decomposed in this way.}, 
it follows that any 4-qubit unitary gate can be decomposed using at most $1_\text{re}{\times}2{+}2_\text{cpx}{\times} 3{+}2{=}10$ CNOT gates. 
Remarkably, this reduces the number of CNOT gates by about $50\%$ compared to the state-of-the-art present-day algorithm used in Qiskit \cite{shende2006} (Qiskit v0.20.0-v1.0.0 requires 22 CNOT gates to disentangle an arbitrary 4-qubit state). This reveals the potential usefulness of using locally optimal disentangling two-qubit unitaries for circuit transpilation.

Again, counting the degeneracy due to the qubit-permutation symmetry shows that there are a total of 36 (almost) degenerate disentangling sequences. This time, however, for a \textit{fixed} one of these sequences, it is easy to construct 4-qubit states that cannot be disentangled: for instance, if the two-qubit reduced density matrices $\rho^{(1,2)}, \rho^{(3,4)}$ are maximally mixed, i.e., proportional to the identity, the first two unitaries $U^{(3,4)}U^{(1,2)}$ in Eq.~\eqref{eq:4q_optimal} will be ineffective.
Nevertheless, for any arbitrary but fixed initial state, there will be at least one of the 36 permutation-equivalent sequences to fully disentangle that state in at most 5 steps. This is intimately related to the fact that no absolutely maximally entangled states exist for four-qubit states~\cite{rather2022thirtysix}. 

State manipulation of arbitrary 4-qubit states was recently demonstrated in Ref.~\cite{zhang2022spin}; the reverse of the protocol sequence we report in Fig.~\ref{fig:4_qubits} can also be used to prepare arbitrary $4$-qubit states.

\subsubsection{Disentangling sequences found by the RL agent for 5-qubit states}

\begin{figure*}[t!]
    \centering    
    \includegraphics[width=1.0\textwidth]{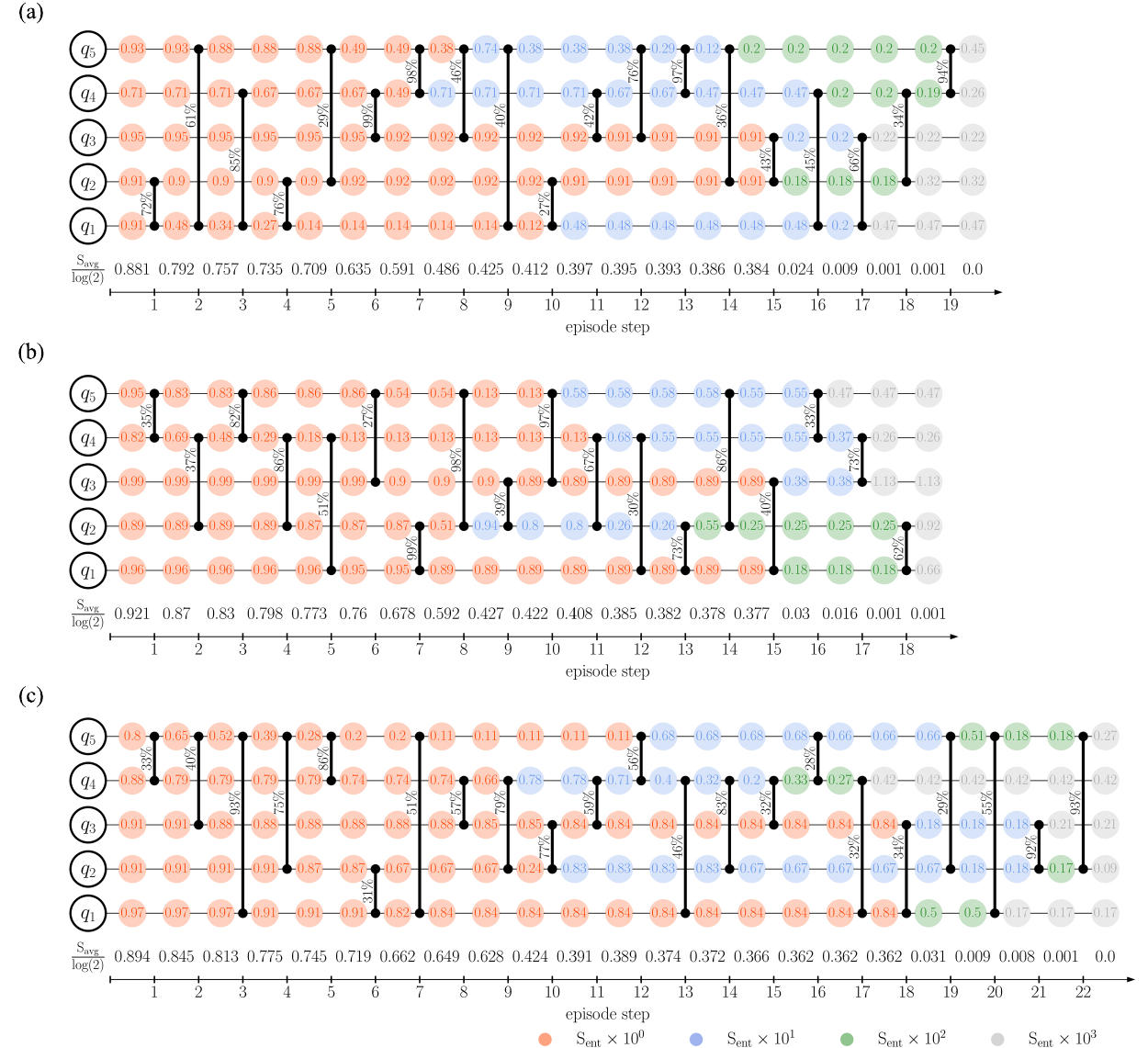}
    \caption{
    Examples of efficient $5$-qubit sequences found by the trained RL starting from Haar-random initial states $|\text{R}_{12345}\rangle$, see Fig.~\ref{fig:validate_agent_5q} for  a detailed description. We distinguish two types of sequences, depending on how they end:
    (a) and (b) show examples of protocols of type (1) (see Sec.~\ref{sec:disent_5q_haar_random}). In panel (a) the agent "factors out" two qubits at once (step 17) and then disentangles the system using only 2 more gates;
    (b) shows an example that factors out qubit $q_5$ in such a way that the subsystem $|\varphi_{1,2,3,4}\rangle$ can be disentangled with only 2 gates -- more efficiently than the generic 5-step four-qubit protocol;
    (c) shows an example of a type (2) divide-and-conquer sequence that factors out qubit $q_4$ first (at step 17), and then applies a generic 5-step four-qubit sequence on subsystem $|\varphi_{1,2,3,5}\rangle$.
    }
    \label{fig:5q-seqs}
\end{figure*}

As explained in the main text, we distinguish two types of 5-qubit sequences found by the RL agent, depending on how they end: 
(1) those that induce interactions between all 5 qubits and use less than 5 gates after bringing the entanglement of one of the qubits below the threshold value (earliest grey circle); 
(2) those that follow a divide-and-conquer strategy that first factors out a qubit and then applies an optimal five-gate sequence to the remaining 4-qubit state.
Examples are shown in Fig.~\ref{fig:5q-seqs}.

\subsection{Details of simulations on NISQ devices} \label{app:nisq}

\subsubsection{\label{app:qst}Quantum state tomography}

At each step of the RL protocol, the agent requires us to feed all pairs of two-qubit reduced density matrices to the policy network. On a quantum computer, the reduced density matrices are not readily accessible and one has to perform quantum state tomography (QST) to reconstruct them. The reconstruction works the same way for all two-qubit pairs and hence, in the following we outline the idea of QST and its steps for a general two-qubit density matrix. We call the true unknown density matrix $\rho_0$ and its reconstruction $\rho$. 

First, note that any two-qubit density matrix can be written in the Pauli basis $\mathcal{P} = \left(\frac{1}{\sqrt{2}}\{I, X, Y, Z\}\right)^{\otimes 2}$ which is informationally complete and has $4^2$ elements. 
The first step in QST is to measure the expectation values of all Pauli basis operators $P_i\in \mathcal{P}$ on the unknown quantum state $\rho_0$, i.e., $m_i = \text{Tr}[ P_i\rho_0]$. In principle, this amounts to measuring all $4^2$ expectation values. However, the expectation values of some Pauli strings such as $Z\otimes \mathbb{I}$ or $\mathbb{I} \otimes Z$ can for example be inferred from measurements of $Z\otimes Z$ which leads to a reduction in the overall number of circuits that need to be run. Thus, effectively only $3^2$ distinct circuits have to be evaluated.

The measured expectation values $m_i$ are naturally subject to noise as they are computed via sampling. Assuming additive Gaussian noise with variance $\nu$, we express the probability of obtaining an outcome $m_i$ when measuring the observable $P_i$ as $p^{(i)}(m_i|\rho) = \frac{1}{\sqrt{2 \pi \nu}} e^{-\left[m_{i}-\operatorname{Tr}\left(P_i \rho\right)\right]^2 /(2 \nu)}$. Therefore, our goal is to find the density matrix $\rho$ that maximizes the likelihood $\mathcal{L}$
\begin{align}
    \mathcal{L}=&\prod_{i} p^{(i)}\left(m_{i} | \rho\right)=\prod_{i} \frac{1}{\sqrt{2 \pi \nu}} e^{-\left[m_{i}-\operatorname{Tr}\left(P_i \rho\right)\right]^2 /(2 \nu)}.
\end{align}
Instead of maximizing the likelihood, it is usually easier to minimize the negative log-likelihood
\begin{align}
    -\log(\mathcal{L})&\propto \sum_i\left[m_i-\operatorname{Tr}\left(P_i \rho\right)\right]^2 + \text{const.}\\
    &=\operatorname{Tr}\left[(\mu-\rho)^2\right] + \text{const.}\\
    &= \|\mu-\rho\|_2^2 + \text{const.}
\end{align}
where we have used the fact that the first line can be expressed as the Hilbert-Schmidt norm (2-norm) of the difference of two matrices: the matrix $\rho$ which is to be found, and the matrix $\mu=\sum_i m_i P_i$ \cite{smolin2012}. Thus, we reduced the problem of QST to that of least squares minimization, i.e., we seek the hermitian matrix $\rho$ that is closest to the matrix $\mu$.

The expression $\|\mu-\rho\|_2^2=\sum_{i j}\left|\mu_{i j}-\rho_{i j}\right|^2$ can be further simplified by working in the eigenbasis of $\mu$ (the 2-norm is independent of the choice of basis). Thus, after diagonalization we now seek the eigenvalues $\lambda_i$ of $\rho$ that minimize $\sum_i\left(\mu_i - \lambda_i\right)^2 $ with the constraints that  $\sum_i \lambda_i=1  $ and $ \lambda_i \geq 0$. These constraints on $\lambda_i$ are necessary to obtain a well-defined density matrix.

The resulting least squares problem can be solved with any standard optimization package. We specifically use a least squares solver from CVXPY \cite{agrawal2018rewriting,diamond2016cvxpy} through the QST routine provided in Qiskit \cite{Qiskit,smolin2012}.

\begin{figure*}[t!]
    \centering
    \includegraphics[width=0.8\textwidth]{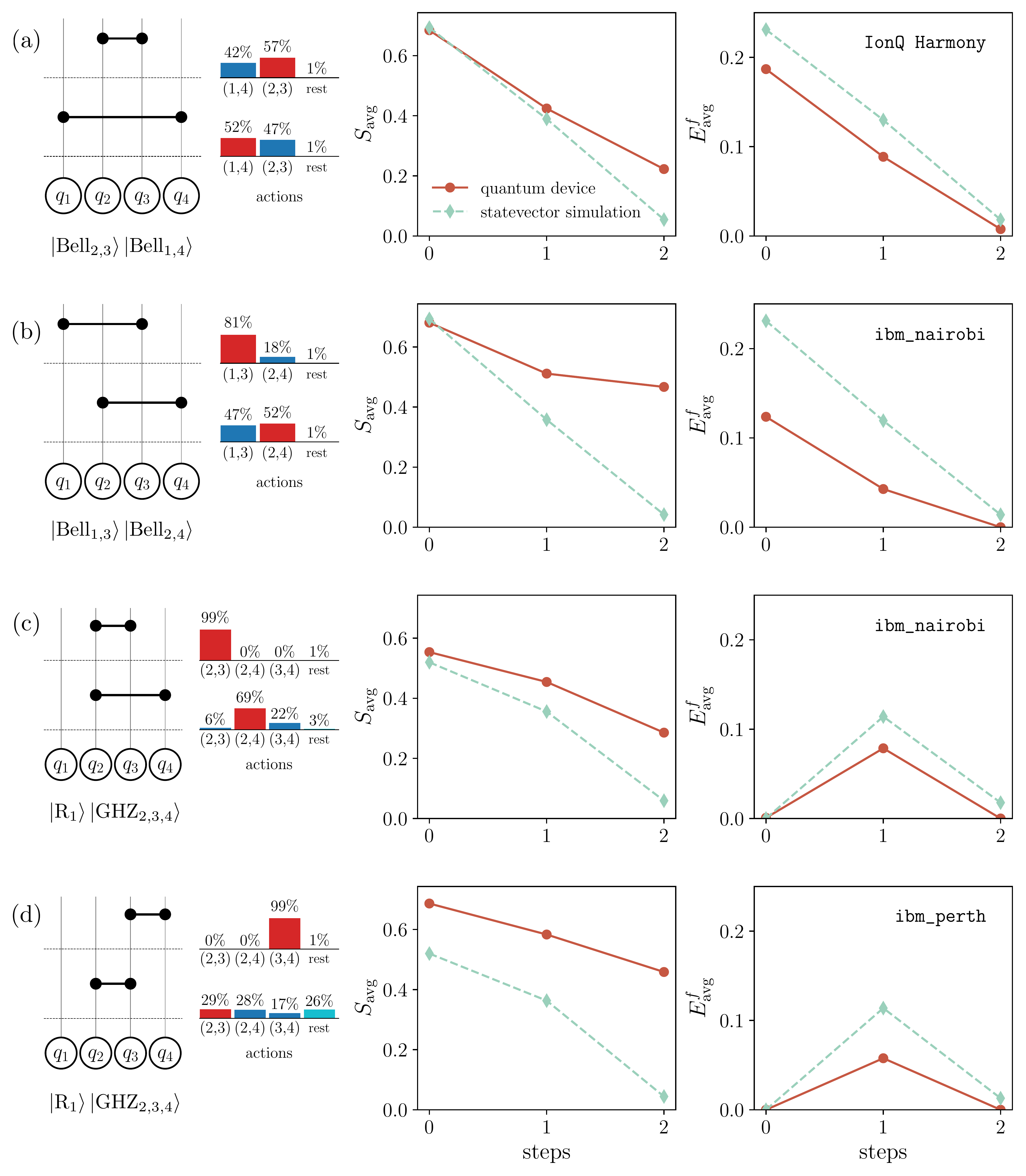}
    \caption{Quantum hardware experiments performed on a trapped-ion device from IonQ (a) and superconducting qubit devices from IBM Quantum (b)-(d). The left panel shows the initial state, the disentangling circuit inferred by our RL framework, and the corresponding probabilities of the 2(3) actions with the highest weight. The middle panel displays the measured averaged von-Neumann entanglement entropies during the protocol (red solid line) as well as the corresponding entropies when the protocol is evaluated with a noise-free statevector simulator instead (blue dashed line). The right panel shows the averaged entanglement of formation between all pairs of qubits, computed again from the noisy density matrices (red solid line) and from statevector simulations (blue dashed line). In all four cases the agent chooses the correct actions and the qubits are close to being disentangled from each other at the end of the protocol. However, a residual entanglement remains with the external environment due to decoherence.}
    \label{fig:app:nisq_1}
\end{figure*}

\subsubsection{5-qubit Ion Trap Hardware}
\label{app:UMDTI_Setup}

The agent is implemented on a trapped ion quantum computer upgraded from the system described in~\cite{debnath2016demonstration}, configured to use ${}^{171}{\rm Yb}^+$ ions with two hyperfine ground states $|F{=}0, m_F{=}0\rangle$ and $|F{=}1, m_F{=}0\rangle$ in the ${}^{2}S_{1/2}$ manifold used as the computational basis states of the qubits. For this work, we load a seven-ion chain with the middle five used as qubits. Coherent operations are implemented using Raman transitions, driven by counter-propagating beams from a mode-locked pulsed 355-nm laser. A single-qubit gate is implemented by adjusting the frequency difference between the two Raman beams to match the frequency splitting between the two computational basis states. Two-qubit gates are implemented by driving motional blue and red sideband transitions to implement a M{\o}lmer-S{\o}rensen (MS) gate that have fidelity's of at least $98\%$~\cite{debnath2016demonstration}. A multi-channel acousto-optic modulator is used for individual addressing and driven by an Arbitrary Waveform Generator (AWG) which is used to modulate the Raman lasers that the individual ions are driven by to implement two qubit gates with pulses shapes derived from techniques in~\cite{blumel2021efficient}. 

The data at each step was SPAM (State preparation and measurement) error corrected before being fed to the agent to determine the next unitary to apply to disentangle the state.

\subsubsection{\label{app:noise}Quantum hardware experiments}

In Sec.~\ref{sec:fake_noise} we studied the performance of the RL agent when the quantum state is subject to simulated noise that mimics the noise model encountered in one of IBM’s real quantum devices. In the following we go one step further and show that the agent can directly be applied on real quantum hardware. To that end, we ran experiments on four different quantum computers: a 5-five qubit trapped-ion device (see Appendix~\ref{app:UMDTI_Setup} for detials), an 11-qubit trapped-ion device from IonQ \cite{ionq}, and two superconducting 7-qubit devices from IBM Quantum \cite{ibmq}. As initial states we chose either a Bell-Bell state pair or a 3-qubit GHZ state which can both be prepared, and thus disentangled, using exactly 2 two-qubit gates. For all initial state realizations we also randomly permuted the qubits and applied a layer of random single-qubit rotations. After preparing the respective states on the quantum devices, we perform QST with $10^3$ shots per circuit, $2*10^3$ for the UMD-TI device, to retrieve all two-qubit reduced density matrices which were subsequently fed to the agent’s policy network. The algorithm then returns a unitary and qubit indices specifying where to apply it. The hardware experiment is rerun with the additional unitary gate applied to the end of the circuit. Note that on the actual quantum computer each two-qubit gate is first transpiled (decomposed) to the native gates of the corresponding device. All the steps outlined above are repeated twice until the state should be fully disentangled.

Fig.~\ref{fig:app:nisq_1} displays the results of 4 exemplary hardware experiments with initial states specified on the left. The circuit diagram shows the corresponding 2-step disentangling circuit found by our algorithm. The 2(3) largest action probabilities that are output by the agent are illustrated on the right next to the unitary. We find that in all of the considered cases the RL agent determines the correct qubit indices and thus, recognizes the entanglement structure of the initial states.

The middle panel displays the average single-qubit von-Neumann entanglement entropy as a function of the protocol step. Note that the entropy can be easily inferred from the experimentally obtained two-qubit reduced density matrices by tracing out the additional qubit. While the average entropy is reduced at the end of the protocol, it is still considerably above zero. This residual entropy can stem from the system qubits not being fully disentangled or from decoherence that effectively entangles the system with its environment. Hence, we also plot the corresponding entanglement entropies computed in a noise-free statevector simulation which uses the same unitaries as in the hardware experiments (dashed blue line). The noise-free entropies nearly vanish at the end of the protocol which suggests that there is indeed no or very little entanglement between the qubits left. The nonzero entropies in the hardware experiments can therefore mostly be attributed to decoherence. This conclusion is also supported when looking at the average entanglement of formation (right panel) which approaches zero at the end of the protocol in both, hardware experiments and statevector simulations.

\begin{figure*}[!]
    \centering
    \includegraphics[width=0.8\textwidth]{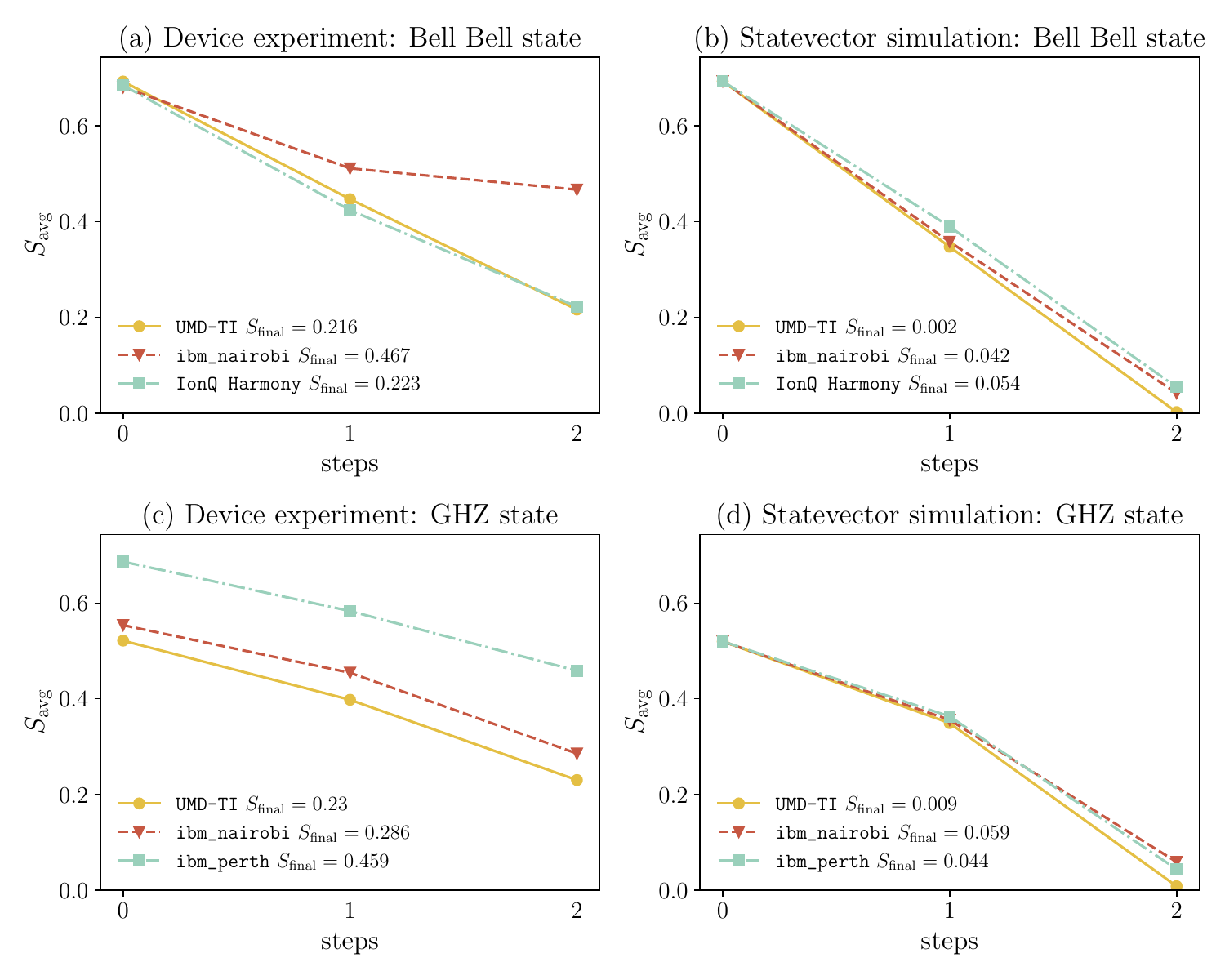}
    \caption{Quantum hardware experiments on a trapped-ion device (described in Appendix~\ref{app:UMDTI_Setup}), a trapped-ion device from IonQ, and superconducting qubit devices from IBM Quantum. Plots (a) and (c) display the measured averaged von-Neumann entanglement entropies during the protocol for an initial Bell Bell state and GHZ state respectively. Plots (b) and (d) display the corresponding entropies when the protocol is evaluated with a noise-free statevector simulator.}
    \label{fig:app:nisq_compare}
\end{figure*}

Fig.~\ref{fig:app:nisq_compare} displays the results of initial states being either Bell Bell states or 3-qubit GHZ states for four different quantum hardware devices. Comparing the four quantum devices employed in this work, we obtained the smallest entropies on UMD-TI's trapped ion device. While having smaller error rates, the device also allows for an all-to-all qubit connectivity which allows direct control over the RL action taken. IonQ's quantum device also possesses this boon. On the other hand, the superconducting qubit devices from IBM feature a T-shaped connectivity which necessitates performing additional SWAP operations when gates are applied to non-adjacent qubits. These two-qubit SWAP gates present another source of noise during a quantum computation and consequently aggravate the results.

We found that the major bottleneck in the quantum device experiments was the pervasive noise, specifically the large two-qubit gate errors. As a result, the quantum states quickly approach a maximally mixed state with increasing circuit depths. Once such a maximally mixed state is reached, the information of the original state is fully lost, and the qubits cannot be disentangled anymore. We also attempted to apply our agent to deeper initial state-circuits on the hardware. However, already for depth-3 states (such as the 4-qubit GHZ state or the shallow random circuit shown in Fig.~\ref{fig:nisq_4}) the agent was unable to infer the correct protocol as the quantum state became too close to such a maximally mixed state. The publicly available devices used in this work belong to an older generation of quantum devices. Thus, we expect our results to improve, i.e., to reach smaller entanglement entropies, as the coherence times of newer devices increase.

\begin{figure*}[t!]
    \centering
    \includegraphics[width=1.0\textwidth]{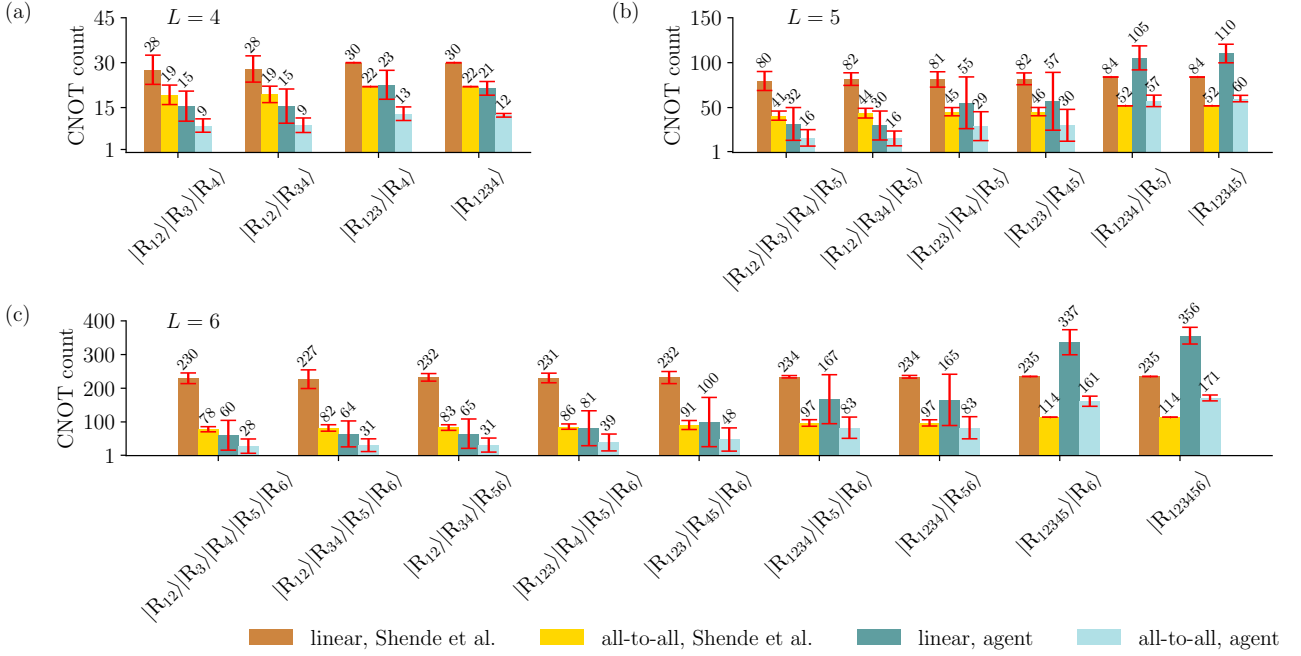}
    \caption{
    Similar to Fig.~\ref{fig:nisq_0} of the main text, we plot the number of CNOT gates in the transpiled circuit using the Qiskit transpiler for different $L=4,5,6$-qubit initial states. However, in this case, we applied an additional layer of CNOT gates to each initial state such that it becomes fully entangled and is no longer separable [see text]. Nevertheless, the RL agent (blue/green bars) can still reduce the overall number of CNOT gates compared to the deterministic algorithm by Shende et al.~(orange/yellow bars) with the exception of 4 of the considered state ensembles:~the two rightmost examples for $L=5,6$ which are (close to) fully Haar-random states.
    }
    \label{fig:app_nisq_0}
\end{figure*}

\subsubsection{\label{app:transpile}Further results on RL-informed circuit transpilation}

In Sec.~\ref{sec:cnot_count} of the main text we showed that using the RL agent as a circuit pre-transpilation step can drastically reduce the number of CNOT gates in the final circuit if the corresponding states have a specific entanglement structure. In particular, we considered states with one or more subsystems that were not entangled with each other and are thus, separable. However, in this case, one can argue that to disentangle (prepare) such a quantum state an even simpler algorithm can be used: By analyzing all two-qubit reduced density matrices of the system we can easily deduce all subsystems of the $L$-qubit state that are not entangled with each other. Hence, each of these subsystems can be prepared independently by either using an RL agent trained on the corresponding subsystem size or by using the universal sequence if the subsystem size is smaller than 5.

To showcase that our agent also outperforms the algorithm of Shende et al.~\cite{shende2006} on states with a less trivial entanglement structure, we repeat the experiment in the main text; however, we apply a final layer of CNOT gates to all states within each ensemble. More specifically, we first generate a random realization of a state with a separable form, e.g., $\ket{\text{R}_{12}}\ket{\text{R}_3}\ket{\text{R}_4}$, and then apply $\prod_{i=1}^{L-1}\text{CNOT}^{(i,i+1)}$. The resultant state, denoted by $\ket{\text{R}_{12} \text{R}_3 \text{R}_4}$, is now fully entangled and cannot be written as a product of subsystem states anymore. At the same time, these quantum states are not Haar-random $L$-qubit states either and thus, still feature a non-trivial entanglement structure.

In Fig.~\ref{fig:app_nisq_0} we show the updated graph of the number of CNOT gates needed to prepare the states specified above. As expected, we find that the CNOT count increases as compared to Fig.~\ref{fig:nisq_0} of the main text, both when employing the deterministic algorithm and when using the RL agent. However, overall the RL agent can still reduce the number of CNOT gates compared to the Shende et al.~algorithm for 15 out of the 19 considered state ensembles. The 4 states for which the agent performs worse are precisely the $L\!=\!5,6$-qubit Haar random states and the states $\text{CNOT}^{(4,5)}\ket{\text{R}_{1234}}\ket{\text{R}_5} = \ket{\text{R}_{1234}\text{R}_5}$ and $\text{CNOT}^{(5,6)}\ket{\text{R}_{12345}}\ket{\text{R}_6} = \ket{\text{R}_{12345}\text{R}_6}$ which are close to being Haar random. This showcases the potential advantage of using the RL agent as a pre-transpilation routine for preparing states that are not Haar-random but still fully entangled.

\section{Details of the RL framework} \label{app:RL}

\subsection{RL environment} \label{app:env_details}

\subsubsection{Vectorized Environment} \label{app:vector_env}

Most on-policy RL algorithms, including Proximal Policy Optimization (PPO, see App.~\ref{app:ppo_details}) need a large number of
environment steps before updating the parameters of the policy;
PPO specifically makes multiple updates after collecting trajectories.
Thus, it is very important that simulating (i.e., ``stepping'')
the environment is computed efficiently.
For systems of $L=16$ qubits, the simulation of the environment is carried out on the GPU.

The state of the environment is represented as a vector containing the $2^L$
wavefunction amplitudes of the simulated quantum state. Applying an action to
that quantum system is implemented as a matrix-vector multiplication between
the vector quantum state and the matrix representation of the acting gate.
Importantly, the 2-qubit unitary is kept as a $4\times 4=(2\times 2)\times(2\times 2)$
matrix for efficiency; instead, we reshape the state into a
$2\times\cdots\times 2$ ($L$ times) tensor
before applying the gate.

Modern compute architectures are
optimized for SIMD-style operations. Hence it is more efficient to batch
multiple computations into a single matrix-matrix multiplication
(so-called vectorization). To do this we instantiate a vectorized
environment consisting of $B$ independent states all of which can
be collected as a single state array $s$ of shape $B \times 2\times\cdots\times 2$.
The agent-environment interaction loop then proceeds as follows:
\begin{enumerate}
  \item draw a vector of $B$ actions $(i,j)$ of pairs of qubits using the current policy.

  \item permute the $2^L$ dimensions for each of the $B$ state arrays individually, so that the indices corresponding to the two qubits the action is applied on, come first. Record the permutation. 
  
  \item compute a three-dimensional tensor of shape $B \times 4 \times 4$ containing $B$ optimal two-qubit gates $U^{(i, j)}$ obtained from diagonalizing the reduced two-qubit density matrices $\rho^{(i,j)}$ (one for each batch element) as described in Sec.~\ref{subsec:small_sys_exact}.

  \item reshape each $4\times 4$ action gate from the batch as a $2\times 2\times 2\times 2$ tensor: $U_{\alpha}^{(i,j)}= U_{\alpha}^{(i_1,j_1),(i_2,j_2)}$, where $\alpha=1,\dots,B$ and $i_1,i_2,j_1,j_2\in\{0,1\}$. 
  
  \item compute the new batch of environment states $s_\text{next}$ using
  tensor-matrix multiplication: $(s_{\alpha}^{i_1\dots i_L})_\text{next} = \sum_{k_1,k_2} U_{\alpha}^{(i_1,k_1),(i_2,k_2)} 
  (s_{\alpha}^{k_1,k_2,i_3\dots i_L})_\text{current}$.

  \item reverse the qubit permutation above to restore the original positions of all qubits. 
  
\end{enumerate}
The vectorized environment allows us to simulate hundreds of agent-environment
interactions in parallel, and can be efficiently implemented in practice~\cite{github_code}. 

\subsubsection{The environment step} \label{app:environment_step}

At each environment step, besides having to compute the next state $s_\text{next}$, we also need the single-qubit entropies
\begin{equation}
    S_\text{ent}[\rho^{(j)}_\alpha] = -\text{tr} \rho^{(j)}_\alpha \log \rho^{(j)}_\alpha,
\end{equation}
for every state in the batch $B$, in order to compute the average and the maximum single-qubit entanglement, cf.~Eq.~\eqref{eq:avg_entanglement}.

Since a two-qubit gate acting on qubits $(i,j)$ can only change the value of the single-qubit entanglement entropies $S_\text{ent}[\rho^{(i)}]$, $S_\text{ent}[\rho^{(j)}]$ and leaves the remaining ones unchanged, we only need to compute these two at each step. Moreover, recall that by definition the gate $U^{(i,j)}$ diagonalizes the two-qubit reduced density matrix $\rho^{(i,j)}$; therefore, we can use the eigenvalues from the diagonalization routine to compute the required single-qubit entropies:
\begin{eqnarray}
    S_\text{ent}[\rho^{(j)}_\alpha] &=& -(\lambda_1+\lambda_2)\log(\lambda_1+\lambda_2) \nonumber\\
    && - (1-\lambda_1-\lambda_2)\log(1-\lambda_1-\lambda_2),    
\end{eqnarray}
where $\tilde\rho^{(i,j)}=U^{(i,j)}\rho^{(i,j)}[U^{(i,j)}]^\dagger = \text{diag}(\lambda_1,\lambda_2,\lambda_3,\lambda_4)$, see Eq.~\eqref{eq:Uloc_def}.

There are a few technical details regarding the computation of $U^{(i,j)}$.
To have reproducible actions, it is important to define a deterministic procedure to compute the gate. 
For instance, one case where this may be challenging is if the reduced density matrix $\rho^{(i,j)}$ is not full rank: in this case the order of the states corresponding to zero-eigenvectors in $U^{(i,j)}$ is undefined; more generally, such ambiguity occurs whenever $\rho^{(i,j)}$ has degenerate eigenvalues.  
We lift this degeneracy by adding small but different numbers close to machine precision to the diagonal entries of $\rho^{(i,j)}$.
A second issue is related to the columns of $U^{(i,j)}$ being defined only up to a global phase (the vectors in the columns correspond to eigenstates of $\rho$). Here, we fix the global phase of the column vectors
of $U^{(i,j)}$ so that the element with the greatest absolute value is real:
$U^{(i,j)}_{:, n} \rightarrow U^{(i,j)}_{:, n}e^{-i\phi_n}$, with the phase $\phi_n = \text{arg}(\text{max}_m(U^{(i,j)}_{m,n}))$ where the $\text{max}$ is computed w.r.t.~the absolute value $|U^{(i,j)}_{m,n}|$. 

Since there is no guarantee that the average entanglement of an arbitrary multiqubit state can be reduced to strictly zero by using only the locally optimal two-qubit gates as defined in Sec.~\ref{subsec:small_sys_exact}, we need to define a stopping criterion for our trajectories. We compute the single-qubit average entanglement for each state in the batch and compare it against a predefined threshold $\epsilon$. A trajectory terminates if the minimum singe-qubit entanglement falls below $\epsilon$ (stopping criterion). The value of $\epsilon$ we use is shown in Table~\ref{table:hyperparams}. In addition, in order to restrict prohibitively long episodes, we introduce a restriction on the maximum number of episode steps $T_\text{trunc}$. If an agent reaches this limit the environment is truncated, i.e., the trajectory is terminated. Once the trajectory of one of the environments from the batch of vectorized environments has been terminated, we re-initialize the state of this environment with a new initial state (see Sec.~\ref{app:init_state_gen}). In doing so we effectively concatenate trajectories into a single continuous segment (cf.~Fig.~\ref{fig:training}).

\subsubsection{Generation of 4-, 5- and 6-qubit Haar-random initial states} \label{app:init_state_gen}

\begin{algorithm}[t!]
    \caption{Initial Quantum State Generation}
    \label{alg:cap}

    \SetKwProg{Procedure}{GenerateState}{}{}
    \SetKwComment{Comment}{/* }{ */}

    \Procedure{($L, q$)}{
        \Comment{
            \textit{L} - number of qubits \\
            \textit{p} - minimum support \\
            returns $\ket{\psi}$ generated state
        }

        $q \gets \textsc{DiscreteUniform}[p, L]$\;
        $\ket{\psi} \gets \textsc{HaarRandomState}(q) $\;
        $r \gets L - q $\;
        \While{ $r > p$ }{
            $q \gets \textsc{DiscreteUniform}[p, r] $\;
            $\ket{\phi} \gets \textsc{HaarRandomState}(q) $\;
            $\ket{\psi} \gets \ket{\psi}\ket{\phi} $\;
            $r \gets r - q $\;
        }
        $\ket{\phi} \gets \textsc{HaarRandomState}(r) $\;
        $\ket{\psi} \gets \ket{\psi}\ket{\phi} $\;
        $ \textbf{return} \ket{\psi} $\;
    }

\end{algorithm}

In this subsection, we walk the reader through the procedure for generating initial states for the environment. Since we are interested in training an agent that can disentangle an arbitrary initial state, a natural choice is to consider initializing the state of the environment as a Haar-random state at the beginning of every agent-environment interaction loop.

The problem that we encountered with this approach is that the performance of the agent on slightly entangled quantum states was very poor (e.g., the agent needed 18 steps to disentangle the state $|W\rangle|\text{Bell}\rangle$ which requires no more than $2{+}1{=}3$ steps, cf.~App.~\ref{subsec:small_sys_exact}).
One reason for this might be the fact that when starting with Haar-random states of full support, the agent can observe states of different support only at the end of a successful trajectory. Thus, the agent was likely provided with very limited experience data for these types of states, and could not learn to generalize during the training procedure. A straightforward brute-force way out of this problem would be to increase the number of training iterations to allow the agent to sample more of these states. 

An alternative solution is to generate the initial states as Haar-random states of different support, e.g., $|\text{R}_{1,\dots,L}\rangle$, $|\text{R}_1\rangle|\text{R}_{2,\dots,L}\rangle$, $|\text{R}_{12}\rangle|\text{R}_{3,\dots,L}\rangle$, etc. Together with the permutation equivariant policy of the agent [cf.~Sec.~\ref{app:perm_equi}], this pool of initial states ensures that the agent visits enough states with sufficiently distinct entanglement distributions among the qubits during training. The exact procedure that we adopt for generating the initial state of the environment is shown in Algorithm~\ref{alg:cap}. The initial state is constructed as a product state between Haar-random states with lower support. The procedure iteratively samples the size of the Hilbert space from which a Haar-random state is drawn. These subsystem states are then tensored together to produce a multi-qubit state. At every step, we choose the size of the support uniformly at random. The minimum support size is denoted as $p$. Using this procedure for generating initial states for the environment improved the agent's performance on recognizing local entanglement structure in slightly entangled quantum systems (see Fig.~\ref{fig:compare_agents}).

Finally, we mention in passing that the choice of distribution for sampling different support sizes is largely based on an empiric guess. Other distributions (e.g.~geometric) may also be useful.

    \subsubsection{Generating large quantum systems of weakly entangled subsystems} \label{app:large_systems}

    To construct quantum states of weakly entangled subsystems, we begin by independently initializing each subsystem as a Haar-random pure state. The global state is then formed by taking the tensor product of these subsystem states.
    
    Weak entanglement between neighboring subsystems is subsequently introduced within the matrix-product-state (MPS) representation of the global wavefunction. After decomposing the product state into left-canonical MPS form via successive singular value decompositions, we replace the Schmidt spectra on the bonds connecting adjacent subsystems by an exponentially decaying distribution parameterized by the dimensionless number $\eta$. Concretely, for Schmidt index $k$, the singular values are chosen as
    \begin{equation*}
        \lambda_k \propto \eta\; \mathrm{e}^{-\eta k},
    \end{equation*}
    followed by normalization. The parameter $\eta$ therefore directly controls the amount of inter-subsystem entanglement: large $\eta$ produces rapidly decaying Schmidt spectra and yields states close to product states, while smaller $\eta$ generates more strongly entangled states.
    
    Finally, after reconstructing the full wavefunction from the modified MPS tensors, a random permutation of qubits is optionally applied in order to remove any residual bias associated with the subsystem ordering.

    The exact procedure used for generating large quantum systems of weakly entangled subsystems is shown in Algorithm~\ref{alg:large_system}.

    \begin{algorithm}[t!]
        \caption{
            Generating large systems of weakly entangled subsystems
        }
        \label{alg:large_system}

        \SetKwComment{Comment}{/* }{ */}
        \SetKwProg{Procedure}{GenWeaklyEntangState}{}{}
        \Procedure{($L, s_{\min}, s_{\max}, \eta_{\min}, \eta_{\max})$}{
            \Comment{
                $L$ - number of qubits\\
                $s_{\min}, s_{\max}$ - min and max subsys sizes\\
                $[\eta_{\min}, \eta_{\max}]$ - subsys bond ent range\\
                returns $\ket{\psi}$ generated state
            }
            
            \BlankLine
            
            $ S \gets \{s_1, s_2, \dots, s_k\}$ \Comment{partition L qubits into subsystems of random sizes; $\sum_i^k s_i = L$}

            $ \mathcal{B} \gets \text{set of bond indices between subsystems}$

            \BlankLine

            \Repeat{\text{SVD succeeds}}{
                \Comment{
                    Construct a product state of independent Haar-random states.
                }

                $ \ket{\psi} \gets \ket{0}$

                \For{$i=1$ \KwTo $k$}{
                    $\ket{\psi} \gets \ket{\psi} \otimes \textsc{HaarRandomState}(s_i)$
                }

                \BlankLine

                \For{$j=1$ \KwTo $L$}{
                    \Comment{
                        Convert $\ket{\psi}$ into an MPS representation and define the Schmidt spectrum.
                    }
                    $(U, \Lambda, V^{\dagger}) \gets \textsc{SVD}(\ket{\psi})$

                    \BlankLine

                    \Comment{
                        Sample $\eta$ and replace the original Schmidt coeff on the bond.
                    }
                    \If{$j+1 \in \mathcal{B}$}{
                        $\eta \sim \mathcal{U}(\eta_{\min}, \eta_{\max})$

                        $\Lambda_k \gets \eta e^{-\eta k}$

                        $\Lambda \gets \textsc{Normalize}(\Lambda)$
                    }
                }
            }

            \BlankLine

            $\ket{\psi} \gets \textsc{ReconstructFromMPS}(\Lambda)$

            \Return $\ket{\psi}$            
        }
    \end{algorithm}

\begin{figure}[t!]
        \centering
        \includegraphics[width=0.5\textwidth]{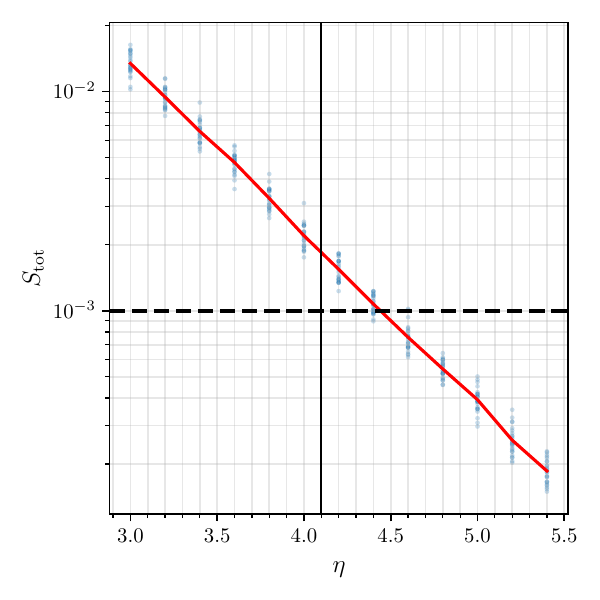}
        \caption{
        Remaining system entanglement $S_\text{tot}$ as a function of the inter-subsystem entanglement parameter $\eta$ for $L=12$ qubits and $L=16$ qubits (lines coincide). For each value of $\eta$, 25 weakly entangled states composed of 4-qubit Haar-random subsystems were generated and independently disentangled using the optimal 4-qubit protocol. Individual points correspond to the residual entanglement measured in each realization, while the solid curve denotes the mean over all samples. The horizontal dashed line denotes the threshold of disentanglement $\eta$, the vertical solid line denotes the chosen value of $\eta$.
        }
        \label{fig:app_eta_vs_entanglement}
    \end{figure}

    To select an appropriate value of the inter-subsystem entanglement parameter $\eta$, we begin from a product state of Haar-random subsystems with support 4 and introduce weak entanglement across the subsystem boundaries using a fixed value of $\eta$. After that, the disentangling sequence optimized for isolated 4-qubit subsystems is independently applied to each subsystem. In the limit $\eta \to \infty$, corresponding to an exact product state, this procedure fully disentangles the global system. For finite $\eta$, however, residual entanglement remains due to the weak coupling between subsystems. A plot showing the remaining entanglement $S_\text{tot}$ as a function of $\eta$ for $L=12,16$ qubits is shown on Fig.~\ref{fig:app_eta_vs_entanglement}.

    By measuring the remaining entanglement after the subsystem-wise disentangling procedure as a function of $\eta$, we obtain a monotonic relation between the inter-subsystem coupling strength and the residual global entanglement. We then select the largest value of $\eta$ such that the remaining entanglement still exceeds the disentangling threshold $\epsilon$ as defined in Sec.~\ref{app:environment_step}. This yields states that remain non-trivially entangled, while still being amendable to efficient disentangling strategies. We set $\eta=4.1$ for both $L=12,16$.

    \subsubsection{Observation space} \label{app:obs_space}

    As discussed in Sec.~\ref{sec:RL_framework}, instead of working with the full state space $\mathbb{C}^{2^L}$, we are working with an observation space consisting of the two-qubit reduced density matrices $\rho^{(i,j)}, 1 \leq i,j \leq L, i \neq j$. However, note that the matrices $\rho^{(i,j)}$ and $\rho^{(j,i)}$ are not equivalent -- although both matrices contain the same elements, $\rho^{(j,i)}$ has the second (corresponding to $|01\rangle$) and third (corresponding to $|10\rangle$) rows as well as the second and third columns of $\rho^{(i,j)}$ swapped.
    Thus, the agent does not receive any new information from observing both reduced density matrices, and it should be enough to work only with one of them. 
    
    A naive approach would be to simply reduce the observation space in half and work with the reduced density matrices $\rho^{(i,j)}$, such that $1 \leq i < j \leq L$. However, this approach would lead to an issue: permuting the qubits of the quantum state will not result in permuting the input for the agent's policy network.
    Let us illustrate the issue using a simple example: consider the three-qubit state $\ket{\psi_{1,2,3}}$ and its permutation $\ket{\psi_{2,1,3}}$. The inputs for the policy network for $\ket{\psi_{1,2,3}}$ consist of the reduced two-qubit density matrices $s=\{ \rho^{(1,2)}, \rho^{(1,3)}, \rho^{(2,3)} \}$, while the inputs for $\ket{\psi_{2,1,3}}$ are $s'=\{ \rho^{(2,1)}, \rho^{(2,3)}, \rho^{(1,3)} \}$. Clearly, the elements of $s'$ are not a permutation of the elements in $s$ since $\rho^{(1,2)}\neq \rho^{(2,1)}$. 
    
    To fix this issue and at the same time reduce the observation space, we need to find a transformation
    \begin{eqnarray*}
       g: \left(\mathbb{C}^{4 \times 4}, \mathbb{C}^{4 \times 4}\right) &\rightarrow&  \mathbb{C}^{4 \times 4},
    \end{eqnarray*}
    which, given two reduced density matrices $\rho_1$ and $\rho_2$, produces an output that is invariant under exchanging the inputs, i.e., $g(\rho_1,\rho_2){=} g(\rho_2,\rho_1)$. The simplest such function is the average $g(\rho_1,\rho_2) = (\rho_1+\rho_2)/2$. Hence, we define an observation to comprise all symmetrized two-qubit reduced density matrices of the current quantum state $\rho$:
    \begin{equation*}
    \label{eq:mean_rhos}
        o(\rho) = \left\{ 1/2(\rho^{(i,j)} + \rho^{(j,i)}) \quad | \quad 1 \leq i < j \leq L \right\}.
    \end{equation*}

    It should be noted here that when performing quantum state tomography (see Sec.~\ref{sec:applns_noisy}) only $\rho^{(i,j)}, 1 \leq i < j \leq L$, is computed, while $\rho^{(j,i)}$ is reconstructed by transforming $\rho^{(i,j)}$ as explained above.

    \subsubsection{Action space} \label{app:act_space}
    
    In Sec.~\ref{sec:two_qubit_U} we propose an analytical solution to find locally-optimal two-qubit disentangling gates. Thus, for any pair of qubits $(i,j)$ we have a prescription to directly compute the optimal gate $U^{(i,j)}$ to be applied to those qubits. This implies that the action set of our agent consists of all unordered pairs of indices $(i,j)$:
    \begin{equation*}
        \mathcal{A}_{\text{full}} = \{ (i,j) \quad | \quad 1 \leq i < j \leq L \}.
    \end{equation*}
    Similar to Sec.~\ref{app:obs_space}, we would like to reduce the action space, and consider only the unordered pairs:
    \begin{equation*}
        \mathcal{A} = \{ (i,j) \quad | \quad 1 \leq i,j \leq L, i \neq j \}.
    \end{equation*}
    However, recall that $U^{(i,j)}$ is computed as a function of $\rho^{(i,j)}$, and, similarly, $U^{(j,i)}$ is computed as a function of $\rho^{(j,i)}$. Thus, in general, $U^{(i,j)} \neq U^{(j,i)}$, and we cannot simply reduce the action space.

    To see why this would be an issue, let us again consider the three-qubit state $\ket{\psi_{1,2,3}}$ and its permutation $\ket{\psi_{2,1,3}}$. Selecting the action $(1,2)$ for the state $\ket{\psi_{1,2,3}}$ will result in applying the gate $U^{(1,2)}$ to qubits $1$ and $2$ (in this order), while selecting the same action $(1,2)$ for the state $\ket{\psi_{2,1,3}}$ will result in applying the gate $U^{(2,1)}$ to qubits $2$ and $1$. Thus, transposing the qubits of the state would actually result in a different action being applied, which would make the agent sensitive to qubit permutations.

    To resolve this issue we choose the action $(i,j)$ in the following way:
    \begin{equation} \label{eq:app_action_definition}
        a(i,j) {=}
        \begin{cases}
            U^{(i,j)} \; \text{applied on} \; (i,j), \;\;
                                \text{if} \;\; S_\text{ent}[\rho^{(i)}] {>} S_\text{ent}[\rho^{(j)}]; \\
            U^{(j,i)} \; \text{applied on} \; (j,i), \quad \text{otherwise}.
        \end{cases}
    \end{equation}

    To implement this procedure, when the agent selects action $(i,j)$, before computing $U^{(i,j)}$, we compare the entanglement entropies of qubits $i$ and $j$ and swap those qubits if $S_\text{ent}[\rho^{(i)}] < S_\text{ent}[\rho^{(j)}]$. Once the quantum gate has been applied we undo the swap, if it was applied in the first place. 
    
    Working with this reduced action space we did not observe any degradation in the agent's performance. Moreover, training and inference speed was increased by about 40\% compared to using the full action space.

    The application of the locally optimal gate $U$ can sometimes lead to large reductions in the entanglement entropy of one of the qubits. In rare cases, it can somewhat increase the entanglement entropy of the other qubit. To make the disentangling protocols produced by the agent (cf.~Fig.~\ref{fig:5q-seqs}) more easily readable we decided to impose the following convention:
    \textit{The ordering relation between $S_\text{ent}[\rho^{(i)}]$ and $S_\text{ent}[\rho^{(j)}]$ before applying the gate $U^{(i,j)}$ should remain the same after applying the gate}:
    \begin{equation*}
        S_\text{ent}[\rho^{(i)}]>S_\text{ent}[\rho^{(j)}]\quad \Longrightarrow \quad S_\text{ent}[\tilde \rho^{(i)}]>S_\text{ent}[\tilde \rho^{(j)}],
    \end{equation*}
    where the tilde denotes the state after applying the gate.
    To comply with this convention we introduce an additional swap operation at the end, that is applied to qubits $i$ and $j$ only if the convention is not met.

    \subsubsection{Reward function} \label{app:reward_fn}

    The reward function that we use during training, cf.~Eq.~\eqref{eq:reward_fn}, is given by:
    \begin{equation} \label{eq:reward_fn2}
        \displaystyle \mathcal{R}(s_t, a_t, s_{t+1}) =
        \sum_{j=1}^L \frac
            {S_\text{ent}[\rho_t^{(j)}] - S_\text{ent}[\rho_{t+1}^{(j)}]}
            {\max (S_\text{ent}[\rho_t^{(j)}], S_\text{ent}[\rho_{t+1}^{(j)}])}
        - n(s_{t+1}),
    \end{equation}
    where $n(s_{t+1})$ is the number of entangled qubits in the state $s_{t+1}$.
    
    To motivate this choice, below we walk the interested reader through the steps that led us to this definition of the reward. 
    During the study, we experimented with several (worse-performing) reward functions; by identifying their shortcomings and addressing them one by one, we finally settled on the version given in Eq.~\eqref{eq:reward_fn2}. 
    
    Our first choice was a simple sparse reward function:
    \begin{equation*}
        \mathcal{R}(s_t, a_t, s_{t+1}) = 
        \begin{cases}
            0 \quad \text{if} \; s_{t+1} \; \text{is terminal}, \\
            -1 \quad \text{otherwise}.
        \end{cases}
    \end{equation*}
    We penalize the agent for every step that it takes until it disentangles the quantum state. This reward function is an exact translation of the task that we are trying to solve: \textit{disentangle the given state using as few steps as possible}. However, in order for the agent to learn, our rollout data has to contain terminal states. If we have a segment of an episode that has not finished yet (see Fig.~\ref{fig:training}), then there is no reward information in this segment.
    
    To fix this issue we augmented the reward function with an additional term that depends on the entanglement entropy of the next state:
    \begin{align*}
        \mathcal{R}(s_t, a_t, s_{t+1}) 
            &= -S_\text{avg}(s_{t+1}) - 1 \\
            &= -\sum_{j=1}^L S_\text{ent}[\rho_{t+1}^{(j)}] -1.
    \end{align*}
    The rationale here is that the agent is penalized more heavily if it brings the environment into states with a high entanglement entropy. With this reward function, we actually bias the agent to act more greedily and to prefer actions that would cause a larger reduction in the entanglement entropy of the state. Note that, with this choice, the reward function is not sparse; hence, unfinished trajectory segments provide a better signal for learning.
    
    However, one problem now is that the value of the entanglement entropy quickly decreases and the reward is dominated by the constant term $-1$. As the episode progresses the entanglement entropy of the state changes from high to low, and the agent receives rewards at different scales. However, we would like to achieve a high reward if the action is good, and a low reward if the action is bad, independent of the time step.
    
    We note in passing that the entanglement reduction  per qubit, given by:
    \begin{equation*}
        \Delta = \sum_{j=1}^L
        \Big(
            S_\text{ent}[\rho_t^{(j)}] - S_\text{ent}[\rho_{t+1}^{(j)}]
        \Big),
    \end{equation*}
    also suffers from this problem, as quantum states with low entanglement entropy naturally undergo lower reductions, regardless of how good the action is.
    
    Thus, instead, we consider the relative reduction of entanglement per qubit that was induced by the action:
    \begin{equation} \label{eq:rel_delta}
        \Delta_\text{rel} = \sum_{j=1}^L \frac
            {S_\text{ent}[\rho_t^{(j)}] - S_\text{ent}[\rho_{t+1}^{(j)}]}
            {\max (S_\text{ent}[\rho_t^{(j)}], S_\text{ent}[\rho_{t+1}^{(j)}])}.
    \end{equation}
    
    Notice that we need the reward function to be negative so that the agent has an incentive to use as few steps as possible. In an idealistic scenario, the largest value of the quantity given in Eq.~\eqref{eq:rel_delta} is attained when an action disentangles all of the qubits:
    
    \begin{align*}
        \displaystyle \sum_{j=1}^L \frac
            {S_\text{ent}[\rho^{(j)}] - 0}
            {S_\text{ent}[\rho^{(j)}]} = n(s_{t+1}),
    \end{align*}
    where $n(s_{t+1})$ is the number of qubits in the current state for which $S_\text{ent}[\rho^{(j)}] > \epsilon$. The value of $\epsilon$ used is given in App.~\ref{app:hyperparams}. 
    
    Thus, to ensure that the reward is always negative we subtract the number of entangled qubits from the relative entanglement reduction. In doing so, we arrive at the expression for the reward function given in Eq.~\eqref{eq:reward_fn2}.

\subsection{Reinforcement Learning Optimization} \label{app:ppo_details}

    \subsubsection{Proximal policy optimization} \label{app:ppo_clip_details}

    As explained in Sec.~\ref{subsec:RL_algo} we are using a variant of the policy gradient algorithm -- Proximal Policy Optimization (PPO) \cite{ppo}. Here we explain the implementation details so that the results can be reproduced. For the specific choice of hyper-parameters, we refer the reader to Sec.~\ref{app:hyperparams}.
    
    One downside of standard policy gradient algorithms is that they are on-policy. This means that we can only update the policy parameters using data collected with the latest policy. If we update the parameters even once, the data becomes off-policy and, strictly speaking, we have to throw it away.
    
    Proximal policy optimization algorithms were introduced to allow for multiple update steps of the policy parameters to be performed before throwing out the collected rollout data during the agent-environment interaction. In order to compute the correct gradient update estimates we have to make sure that the current policy $\pi_\theta$ does not deviate too much from the policy $\pi_{\theta_\text{old}}$ that was used to collect the data.
    
    In our implementation, we use the algorithm PPO-CLIP (see Sec. 3 of \cite{ppo}) that clips the objective function if $\pi_\theta$ deviates too much from $\pi_{\theta_\text{old}}$. The clipped objective has the form:
    \begin{align} \label{eq:loss_clip}
        J_{\text{clip}}(\theta)& = \\ \nonumber
        =& \underset{s, a \sim \pi_\theta}{\mathbb{E}}
        \bigg[
            \min \big(
                \rho(\theta) A_t, \space \text{clip}(\rho(\theta), 1-\epsilon_\pi, 1+\epsilon_\pi) A_t
            \big)
        \bigg],
    \end{align}
    where $\displaystyle \rho(\theta) = \frac{\pi_\theta(a_t|s_t)}{\pi_{\theta_\text{old}}(a_t|s_t)} $, and $\epsilon_\pi$ is a hyper-parameter for clipping (see Table~\ref{table:hyperparams}). Here $A_t=A(s_t,a_t)$ is the advantage function (see below).
    
    In addition, we augment the algorithm with a check for early stopping. If the mean \textit{KL divergence} between $\pi_\theta$ and $\pi_{\theta_\text{old}}$ grows beyond a given threshold, then we prematurely stop taking gradient steps and collect new rollout data.

    \subsubsection{Entropy regularization} \label{app:entropy_reg}

    The clipped objective is further augmented with an entropy regularization term:
    \begin{equation} \label{eq:entropy_reg_obj}
        J(\theta) = J_{\text{clip}}(\theta) + \beta^{-1}\mathcal{H}(\pi_\theta),
    \end{equation}
    where $\beta^{-1}$ is a temperature parameter that controls the level of regularization, and $\mathcal{H}(\pi_\theta)$ is the statistical entropy of the policy distribution, which is given by:
    \begin{align}
        \mathcal{H}(\pi_\theta) \notag
        &= \underset{s \sim \pi_\theta}{\mathbb{E}} \bigg[
            \sum_{a \in \mathcal{A}} \bigg(\pi_\theta(a|s) \log \pi_\theta(a|s) \bigg)
        \bigg] \\
        &= \underset{s, a \sim \pi_\theta}{\mathbb{E}} \bigg[ -\log \pi_\theta(a|s) \bigg]. \label{eq:policy_entropy}
    \end{align}
    
    Using Eq.~\eqref{eq:policy_entropy} we can easily estimate the current entropy of the policy from the data sampled during rollout. Trying to maximize the entropy has the effect of pushing the policy distribution to be more random, preventing it from becoming a delta function; thus, it increases exploration during training.

    \subsubsection{Training algorithm for the policy} \label{app:train_algo_ppo}

    The training algorithm consists of two stages: \textit{Rollout stage} and \textit{Learning stage}.
    \begin{figure}[t!]
        \centering
        \includegraphics[width=0.45\textwidth]{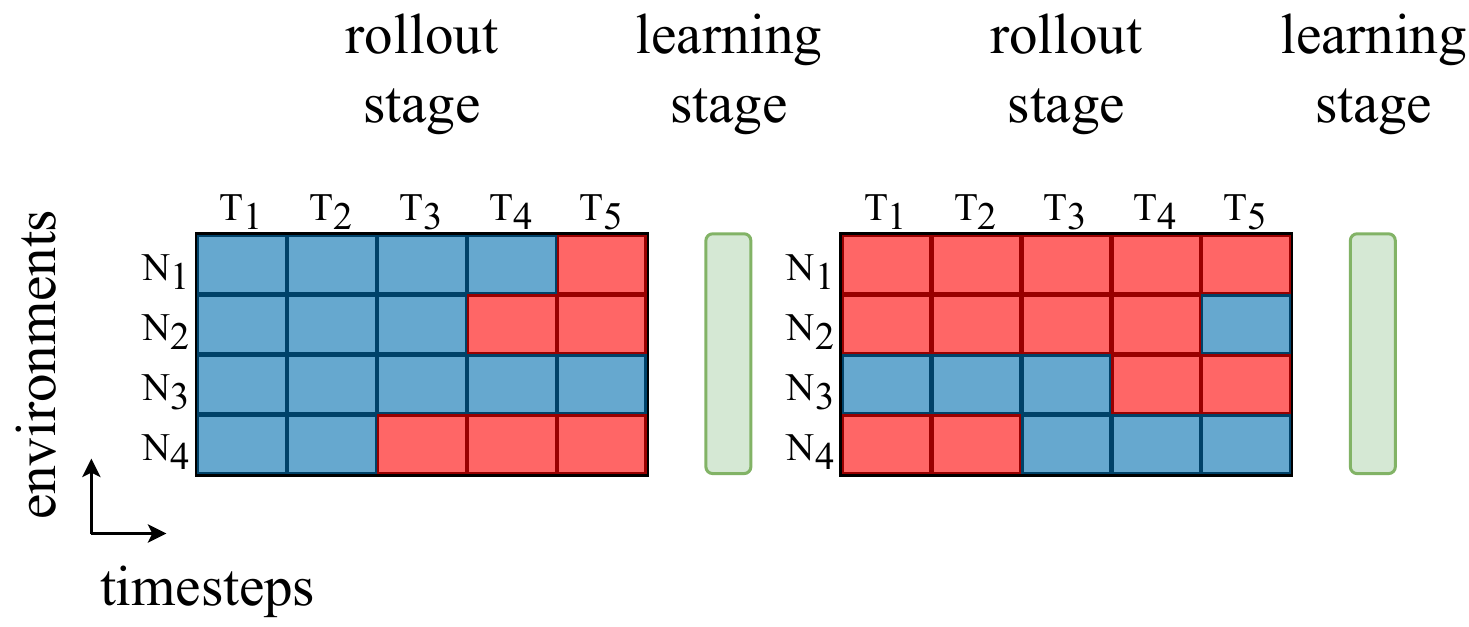}
        \caption[Training pipeline]{Training pipeline. Training consists of a \textit{rollout stage} (showing collected rollout trajectories horizontally) and a \textit{learning stage} (represented by vertical green bars) when model parameters are updated. During rollout we perform a fixed number $T_{\text{seg}}$ of steps on multiple states in a vectorized environment in parallel. We indicate the end/beginning of episodes by changing the color of the box (blue and red). Once the learning stage is over the next rollout stage starts and the agent continues to step the environments from where it left off. For the number of states in the vectorized environment and the length of the trajectory segments please refer to Table~\ref{table:hyperparams}.}
        \label{fig:training}
    \end{figure}
    
    For our rollout stage, we use \textit{fixed-length trajectory segments}, defined as follows. The agent performs a fixed-length $T$-step rollout (segment length $T_{\text{seg}}$) collecting (state, action, reward) triples. Using the collected data the agent updates its policy during the learning stage. Once the learning stage is over the agent starts a new rollout stage but continues to step the environment from where it left off, i.e., the environment is not reset at the beginning of the rollout stage. Note that the fixed-length segment could contain environment transitions from multiple episodes that are concatenated one after another. However, it could also contain only a fragment of an episode that was started in a previous rollout stage and will end in a future rollout stage. To allow for a more efficient data collection we perform rollouts on several environments in parallel, cf.~Fig.~\ref{fig:training} for a schematic representation, and Sec.~\ref{app:env_details} for a discussion.
    
    During the training stage the agent performs multiple updates to the policy weights using the collected data. The data is split into mini-batches of size $B_{\text{PPO}}$ and we optimize the objective for $K$ epochs using the Adam optimizer.
    
    In order to compute the advantage $A(s_t, a_t)$ in Eq.~\eqref{eq:loss_clip} we need an advantage estimator that does not look beyond the final time-step $T_{\text{seg}}$ of the fixed-length segment. To address this issue we make use of a second neural network (critic) which is used to approximate the value function -- a function that estimates how good it is for the agent to be in a given state. Thus, we will bootstrap the estimation at time-step $T_{\text{seg}}$ by using an approximation for the value function~\footnote{A simple $n$-step bootstrapped estimation of the advantage takes the form 
    $A(s_t, a_t) = r_t + r_{t+1} + \dots + r_{T-1} + V_\phi(s_T)$.
    }.
    The estimator that we use is a truncated version of the generalized advantage estimator \cite{gae}:
    
    \begin{equation} \label{eq:gae}
        A(s_t, a_t) = \delta_t + (\gamma \lambda) \delta_{t+1} +
            \cdots + (\gamma \lambda)^{T-t+1} \delta_{T-1},
    \end{equation}
    where $\delta_t = r_t + \gamma V_\phi(s_{t+1}) - V_\phi(s_t)$; $\gamma$ is the discount factor; $\lambda$ is the weight averaging factor, and $V_\phi$ is the critic neural network approximating the value function (see Fig.~\ref{fig:policy_network}, Table~\ref{table:hyperparams}).

    At every iteration, before computing the objective, the advantages are standardized on the mini-batch level to have zero mean and unit variance:
    \begin{equation} \label{eq:adv_norm}
        \hat{A}(s_t, a_t) = \frac{A(s_t, a_t) - \mu_{A}}{\sigma_{A}}.
    \end{equation}
    This modification makes use of a constant baseline for all (state, action) pairs in the batch and effectively re-scales the learning rate by a factor of $1 / \sigma_{A}$.
    
    Note that the entropy regularization term is also computed over the mini-batch. This is done because after every update we change the policy weights and hence the entropy.
    
    Finally, at the end of every epoch we check the \textit{KL divergence} between the newest and the original policy and stop the learning phase if a given threshold ($KL_{\text{lim}}$) is reached. The $KL$ divergence can be calculated as follows:
    \begin{align} \label{eq:kl_div_calc}
        D_\text{KL}(\pi_\text{old} \space || \pi)
        &= \underset{a \sim \pi_\text{old}}{\mathbb{E}} \bigg[ \log \frac{\pi_\text{old}(a|s)}{\pi(a|s)} \bigg] \\ \notag
        &= \underset{a \sim \pi_\text{old}}{\mathbb{E}} \big[ \log \pi_\text{old}(a|s) - \log \pi(a|s) \big].
    \end{align}
    This is needed, because simply clipping the objective might not be enough to capture the divergence between the current policy $\pi_\theta$ and the old policy $\pi_{\theta_\text{old}}$. Having this additional check also allows us to terminate the learning phase if the new policy starts to deviate too much. For more details see Algorithm~\ref{alg:training_ppo}.

    \begin{algorithm}[t!]
        \caption{Training with PPO}
        \label{alg:training_ppo}
    
        \SetKwInOut{Input}{input}
        \SetKwInOut{Output}{output}
        \SetKwComment{Comment}{/* }{ */}
        \SetKwComment{InlineComment}{$\triangleright$ }{}
    
        \Input{\textit{obs} - environment observations}
        \Input{\textit{acts} - agent actions}
        \Input{\textit{returns} - obtained returns}
        \Input{\textit{K} - number of training epochs}
        \Input{\textit{KL}$_{\text{lim}}$ - $D_{\text{KL}}$ threshold for stopping}
    
        $ adv \gets \textsc{GAE}(returns) $ \InlineComment*{Eq.~\eqref{eq:gae}}
        $ p\_old \gets \textsc{Policy}(obs) $ \InlineComment*{calc.~$\pi$ before updates}
        
        $ i \gets 1 $\;
        \For{$i \leq K$}{
            \Comment{
                $\footnotesize \text{sample mini-batches from the experiences}$ \\
                $\footnotesize o   \text{ :~a mini-batch of observations}$ \\
                $\footnotesize a   \text{ :~corresp.~selected actions}$ \\
                $\footnotesize ad  \text{ :~corresp.~calculated advantages}$\\
                $\footnotesize p_o \text{:~corresp.~probs using } \pi \text{ before update}$
            }
            \For{$o, a, ad, p\_o \in \textsc{Batch}(obs, acts, adv, p\_old)$}{
                $ p \gets \textsc{Policy}(o) $\;
    
                $ ad \gets \displaystyle \frac{ad - ad.\textsc{mean}()}{ad.\textsc{std}()} $ \InlineComment*{Eq.~\eqref{eq:adv_norm}}
    
                $ \rho \gets \displaystyle \frac{p\_{old}}{p} $\;
                $ loss \gets \textsc{min} \displaystyle \big( \rho \times ad, \space \textsc{clip}( \rho, 1 -\epsilon, 1 + \epsilon) \times ad \big) $\;
                $ loss \gets loss.\textsc{mean}() $ \InlineComment*{Eq.~\eqref{eq:loss_clip}}
    
                $ loss \gets loss -\log \textsc{Policy}(a|o) $ \InlineComment*{Eq.~\eqref{eq:policy_entropy}}
    
                \Comment{$\footnotesize \text{backward pass to update policy network}$}
            }
    
            $ i \gets i + 1 $\;
            $ p \gets \textsc{Policy}(obs) $\;
            $ KL \gets \log p\_old - \log p $\ \InlineComment*{Eq.~\eqref{eq:kl_div_calc}}
            \If{ $ KL > KL_{\mathrm{lim}} $}{
                $ \textsc{Break} $\;
           }
       }
    
    \end{algorithm}

    \subsubsection{Value function optimization} \label{app:ppo_value_clip_details}

    In addition to optimizing the policy we also need to optimize the weights of the value network (value net). The value network is trained by minimizing the mean squared error between its predictions and the computed returns from the trajectories, which are given by
    \begin{equation*}
        V_\text{target} = A(s_t, a_t) + V_\phi (s_t).
    \end{equation*}
    The value network (critic) is updated simultaneously with the policy network (actor) using the same mini-batching strategy, i.e., we train in $K$ epochs with mini-batches of size $B$. If the early stopping criterion is met then we stop updating the value network as well.
    
    Just like the clipped objective for the policy network, we also clip the value loss before updating the parameters:
    \begin{align*}
    V_\text{clip} &= \text{clip}(V_\phi, V_{\phi_\text{old}}-\epsilon_V, V_{\phi_\text{old}}+\epsilon_V) \\
    L^V &= \max[(V_\phi - V_\text{target})^2, (V_\text{clip} - V_\text{target})^2],
    \end{align*}
    where $\epsilon_V$ is a hyper-parameter for clipping the value objective.
    The policy network and the value network use different learning rate parameters. The corresponding objectives are clipped using different clipping values. All hyper-parameters used for training the models are given in Table~\ref{table:hyperparams}.

    \subsubsection{Staged Training} 
    \label{app:staged_training}

    As the size of the system grows, training agents to disentangle the system becomes increasingly challenging. For large systems composed of weakly entangled subsystems (cf.~App.~\ref{app:large_systems}), we employ a two-stage training procedure inspired by transfer learning.

    Stage 1. The agent is first trained on product states composed of smaller subsystems, where the subsystem size varies between 2 and 4 qubits, with checkpoints saved every $500$ iterations. As training progresses, the policy gradually becomes more deterministic and exploitative, whereas earlier checkpoints retain stronger exploratory behavior, reflecting the standard exploration–exploitation tradeoff in RL.

    The saved checkpoints are subsequently evaluated on the target weakly entangled states according to both disentangling accuracy and average episode length. The checkpoint with the best overall performance is then selected as the initialization for the second training stage.

    Stage 2. Starting from the selected checkpoint, the agent is further trained directly on weakly entangled subsystems of maximally entangled Haar-random states supported on smaller subsystems. Both stages of training proceed according to Algorithm 3.

\subsection{Policy and value network architectures} \label{app:net_arch}

    In what follows, we will describe the architectures of the neural networks used to approximate the policy and value functions of our agents.

    \begin{figure}[t!]
        \centering
        \includegraphics[width=0.45\textwidth]{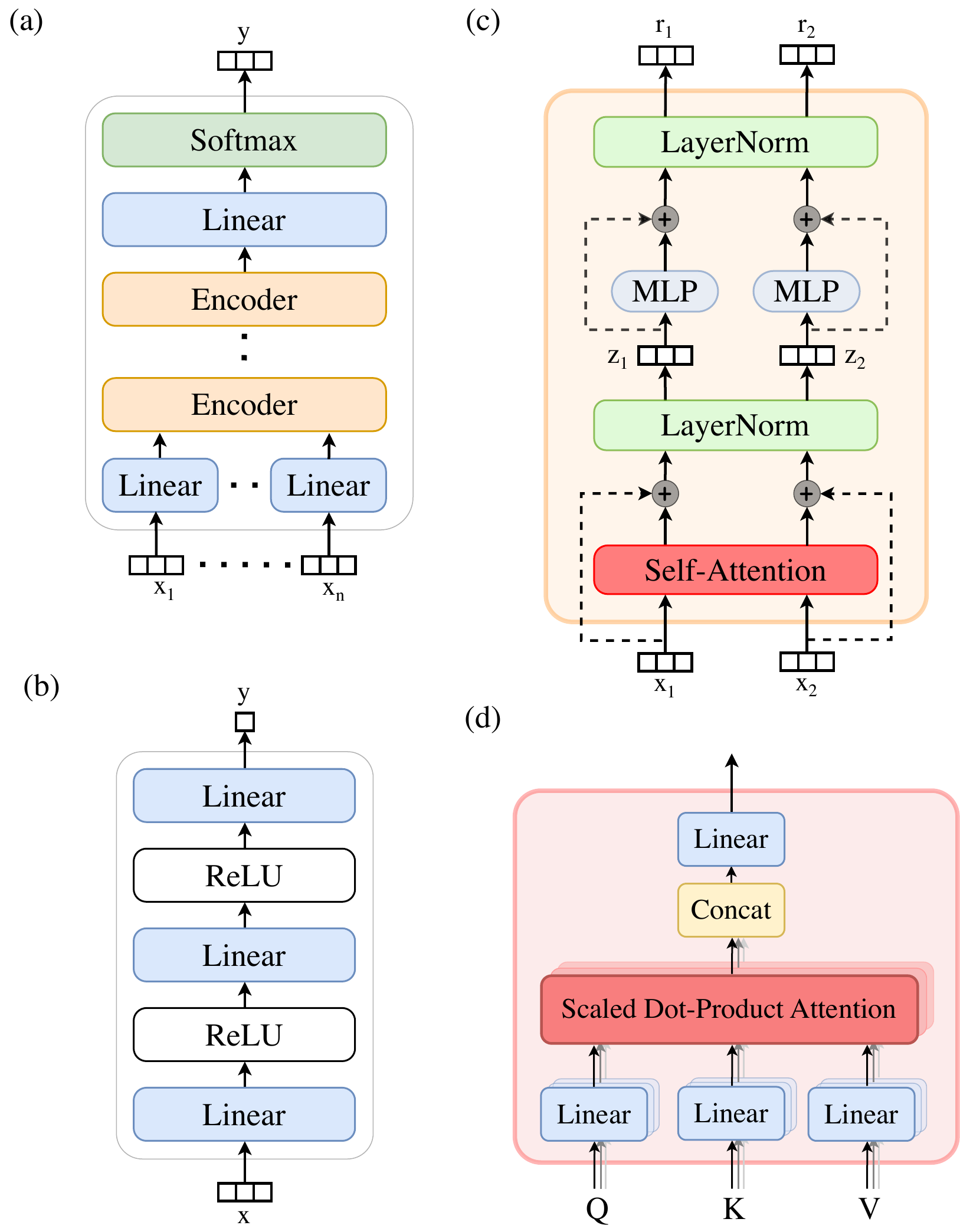}
        \caption[Policy and Value networks architecture]{Policy and Value networks architecture. The policy network (a) accepts as inputs the symmetrized two-qubit reduced density matrices of a quantum state and outputs a probability distribution over the actions. The policy is a permutation-equivariant network consisting of a linear layer, a stack of transformer encoder blocks, and a final linear layer with softmax non-linearity. The value network (b) is a simple three-layer fully-connected network that accepts the same input flattend and returns a single number -- the value for that state. Both networks use `ReLU' non-linearities.
        The Encoder block (c) has a self-attention layer followed by a position-wise fully-connected network. The block accepts a sequence of vectors $\mathrm{x_1 \dots x_n}$ as input and outputs the encoding vectors $\mathrm{r_1 \dots r_n}$. We make use of a multi-headed self-attention layer (d) allowing each element to "pay attention" to multiple elements. The attention layer accepts three different inputs, namely keys ($\mathbf{K}$), queries ($\mathbf{Q}$), and values ($\mathbf{V}$). For self-attention we have $\mathbf{K} = \mathbf{Q} = \mathbf{V} = \mathbf{X}$ (see Eq.~\eqref{eq:transformer_layer}).
        }
        \label{fig:policy_network}
    \end{figure}

    The policy network is designed to be permutation-equivariant (see App.~\ref{app:perm_equi}). It is a stack of $N_{\text{layer}}$ identical transformer encoder blocks (\cite{vaswani2023attention}) that are applied one after another (see Fig.~\ref{fig:policy_network}a). Each encoder block has a multi-headed self-attention layer followed by a position-wise fully-connected network (multilayer perception). After each layer, a residual connection is applied followed by a normalization layer (Fig.~\ref{fig:policy_network}c). We do not use dropout in the policy network. In each of the encoder blocks the self-attention layer has queries, keys, and values of dimensionality $D_{\text{qkv}}$ split into $N_{\text{heads}}$ attention heads (Fig.~\ref{fig:policy_network}d). The fully connected network has one hidden layer with dimensionality $D_{\text{mlp}}$ and `ReLU' non-linearity. Both the input and output dimensions of the fully connected network are $D_{\text{qkv}}$.
   
    Note that the inputs and the outputs of each of the encoder blocks have the same dimensionality, i.e., the transformer encoders operate in an embedding space with a fixed dimension (in this case $D_{\text{qkv}}$). 

    The policy network receives as input the symmetrized two-qubit density matrices (see App.~\ref{app:obs_space}). Before being forwarded through the encoder blocks, the reduced density matrices are forwarded through a position-wise linear layer with no non-linearities; they are embedded into $D_{\text{qkv}}$-dimensional space.

    The output of the policy network should be a vector containing the probability of taking each of the actions from the action set (see App.~\ref{app:act_space}), representing the probability of choosing that action. Thus, the outputs of the encoder blocks are forwarded through another position-wise linear layer and are embedded in one-dimensional space. After that, we apply a softmax non-linearity to get the probability scores.

    The value network is a simple three-layer fully connected network that accepts the same input as the policy network but flattened (Fig.~\ref{fig:policy_network}b). The network uses `ReLU' non-linearities and each of the hidden layers has size $h_{\text{hid}}$. The output is a single number -- the value for the given quantum state.

    It is worth mentioning here that we do not require that the value network exhibits the permutation-equivariance/invariance property. Note that the value network is only used for training the agent when computing the advantages [Eq.~\eqref{eq:gae}]. During inference, solely the design of the policy is important so that the agent remains insensitive to qubit permutations. In our experiments, we did not find any improvement in the policy training when using a permutation-insensitive architecture for the value network. On the downside, however, training took longer due to the increased computational complexity.

\subsection{Permutation-equivariant deep learning architecture} \label{app:perm_equi}
    
    In this section, we briefly discuss the inherent permutation equivariance property of the self-attention layer. The equations governing the self-attention layer (see~Fig.~\ref{fig:policy_network}c) are:
    \begin{align}
    \label{eq:transformer_layer}
        \mathbf{Z}
        &= \textsc{SelfAttention}(\mathbf{X}) \nonumber \\
        &= \text{Softmax} \bigg( \frac{\mathbf{XW_QW_K^TX^T}}{\sqrt{d_k}} \bigg) \mathbf{XW_V},
    \end{align}
    where $\mathbf{W_Q} \in \mathbb{R}^{d_\text{model} \times d_k}$, $\mathbf{W_K} \in \mathbb{R}^{d_\text{model} \times d_k}$, $\mathbf{W_V} \in \mathbb{R}^{d_\text{model} \times d_\text{v}}$ are the query, key, and value weight matrices respectively, $\mathbf{X} \in \mathbb{R}^{N \times d_\text{model}}$ is the sequence of inputs to the layer, and $\mathbf{Z} \in \mathbb{R}^{N \times d_\text{v}}$ is the sequence of output embeddings
    \footnote{In our case, we have $d_v = d_k = d_q = d_\text{model} = D_\text{qkv}$. However, the expressions here hold in general, i.e., for any $d_k$, $d_v$.}.

    Let us consider how an arbitrary embedding $z_i$ is computed. First, the attention probability matrix ($\mathbf{A} \in \mathbb{R}^{N \times N}$) is computed, by matrix-vector multiplying the query embeddings with the key embeddings and applying the \textsc{Softmax} function on the result:
    \begin{align} \label{eq:attn_dist}
        \mathbf{A} &= \text{Softmax} \bigg( \frac{\mathbf{XW_QW_K^TX^T}}{\sqrt{d_k}} \bigg) \\
        &= \text{Softmax} \bigg( \mathbf{XWX^T} \bigg).
    \end{align}
    The row-vector $a_i$ of $\mathbf{A}$ contains the attention probabilities with which element $x_i$ attends to all the elements in the sequence:
    \begin{equation}
    \label{eq:attn_softmax}
        a_i = \text{Softmax} \bigg( x_i\mathbf{WX^T} \bigg).
    \end{equation}
    In particular, the entry $a_{ij}$ gives us the attention probability with which element $x_i$ attends to element $x_j$:
    \begin{equation} 
    \label{eq:attn_prob}
        a_{ij} = \frac{\text{exp}(x_iWx_j^T)}{\sum_k \text{exp}(x_iWx_k^T)}.
    \end{equation}   
    The embedding $z_i$ of $x_i$ is then computed as:
    \begin{align} \label{eq:app_perm_equi}
        z_i &= a_i \mathbf{XW_V} \\
        &= \sum_j a_{ij} x_j \mathbf{W_V} \notag\\
        &= \sum_j \bigg( \frac{\text{exp}(x_iWx_j^T)}{\sum_k \text{exp}(x_iWx_k^T)} x_j \mathbf{W_V} \bigg). \notag
    \end{align}
    
    Let us now assume that the order of the vectors $\{x_0, x_2, \dots, x_{N-1}\}$ is permuted according to $i \rightarrow i'$; i.e., $i'$ is the new position corresponding to the vector $x_i$ from the original sequence. Consulting Eq.~\eqref{eq:app_perm_equi} we can see that the embedding $z_{i'}=z_i$, implying that permuting the input sequence results in an equivariant permutation of the output sequence.

    Alternatively, one can denote the permutation by a matrix $\mathbf{P}$, and describe its action on the input data by $\mathbf{X}\to\mathbf{PX}$. Substituting this into Eq.~\eqref{eq:transformer_layer}, we find:
    \begin{align*}
        & \textsc{SelfAttention}(\mathbf{PX}) =\nonumber \\
        &= \text{Softmax} \bigg( \frac{\mathbf{PXW_QW_K^TX^TP^T}}{\sqrt{d_k}} \bigg) \mathbf{PXW_V} \nonumber \\
        &= \mathbf{P}\; \text{Softmax} \bigg( \frac{\mathbf{XW_QW_K^TX^T}}{\sqrt{d_k}} \bigg) \mathbf{P^T}\mathbf{PXW_V} \nonumber \\
        &= \mathbf{P}\; \text{Softmax} \bigg( \frac{\mathbf{XW_QW_K^TX^T}}{\sqrt{d_k}} \bigg) \mathbf{XW_V} \nonumber \\
        &= \mathbf{PZ},
    \end{align*}
    where we used that $\text{Softmax}(\mathbf{PXP^T}){=}\mathbf{P}\text{Softmax}(\mathbf{X})\mathbf{P^T}$, and  $\mathbf{PP^T}=\mathbf{1}$ since permutations are unitary maps. Hence, the output of the self-attention layer is correctly transformed when permuting its input.

\subsubsection{Analyzing Trained Attention Heads} 
\label{app:attention_heads}

Section \ref{app:perm_equi} defined the single-head self-attention. The transformer architecture uses multi-head self-attention and each head has its own set of learnable parameters $\mathbf{W_Q}, \mathbf{W_K}, \mathbf{W_V}$. Multi-head attention is a way to linearly project the keys, queries, and values $N_\text{heads}$ time with different $\mathbf{W_Q}, \mathbf{W_K}, \mathbf{W_V}$ matrices. The projections become ($D_\text{qkv}$/$N_\text{heads}$)-dimensional vectors (instead of $D_\text{qkv}$-dimensional) which are then concatenated into a single $D_\text{qkv}$-dimensional output embedding: hence, each head computes only part of the output embedding for each input token (Fig.~\ref{fig:policy_network}(d)). Since heads do not share $\mathbf{W_Q}, \mathbf{W_K}, \mathbf{W_V}$ it follows that each head has an attention distribution $\mathbf{A}$ specific to it.

\begin{figure}[t!]
    \centering
    \includegraphics[width=0.45\textwidth]{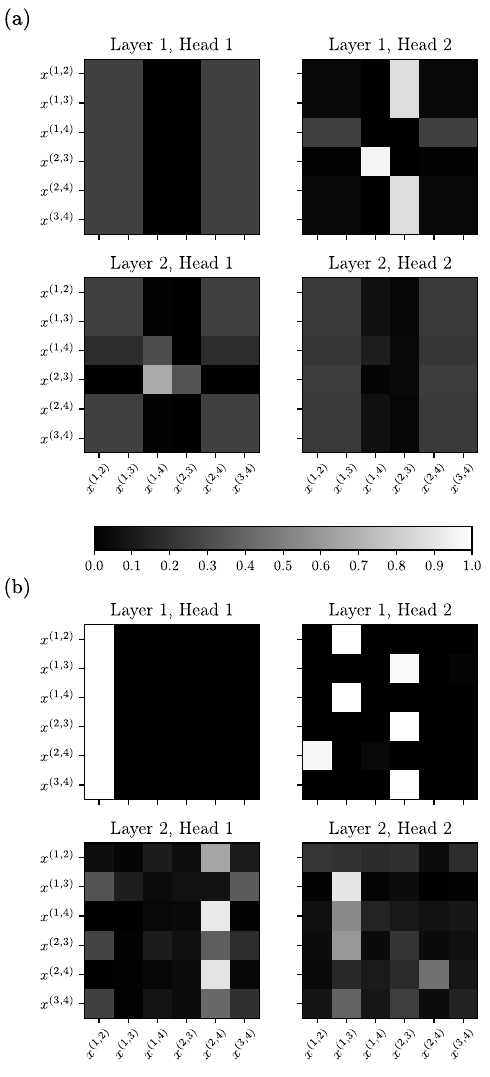}
    \caption[Attention Heads]{ 
    Attention Heads.
    Attention scores for the state $\mathrm{|0_1\rangle|Bell_{23}\rangle|0_4\rangle}$ (a), and for $\mathrm{|R_{12}\rangle|R_{34}\rangle}$ (b), see colorbar.
    Row vectors $a_i$ give the attention scores with which element $x^{(i,j)}$ attends to all elements in the sequence. All elements in the row vector $a_i$ add up to unity, because $a_i$ is normalized to be a probability distribution, cf.~Eq.~\eqref{eq:attn_softmax}. 
    }
    \label{fig:attention_scores}
\end{figure}

The attention distribution $\mathbf{A}$ is an $\mathcal{|A| \times |A|}$ dimensional matrix, and can be plotted as a heatmap. In fact, this is one of the aspiring aspects of the self-attention mechanism: as an architecture that has its roots in machine translation, the attention scores can reveal which words in the source language are being attended to when generating each word in the target language. By visualizing attention distributions, one can identify which source words contribute the most to generating specific target words. Larger attention scores indicate stronger connections between tokens (embeddings), implying that those input tokens are more influential in generating the output tokens. Although the inputs to our multi-head attention layers are linear embeddings of vectors derived from reduced density matrices (see Eq.~\eqref{eq:mean_rhos} and Fig.~\ref{fig:policy_network}(a)) instead of language words, they can nevertheless shed some light on the decision-making process of the agent. Since our results show that the agents are capable of recognizing local entanglement structure (see Fig.~\ref{fig:compare_agents}), one can suppose that the attention distributions of the policy networks may show some information that correlates with the entanglement structure.

We show the attention distributions from our 4-qubit agent for two exemplary initial states in Fig.~\ref{fig:attention_scores}. The policy network of this agent has 2 encoder layers, each with 2 heads [Table \ref{table:hyperparams}].
Each $x^{(i,j)}$ is an embedding of an observation $o^{(i,j)}$ which is a symmetrized reduced density matrix (see Sec.~\ref{app:obs_space}). We use ``token" and ``embedding" interchangeably in the text. Note that each token is specified with two super-indices $^{(i,j)}$, which indicate that the RDMs are computed over qubits $i$ and $j$; each element $a_{kl}$ in the matrix $\mathcal{A}$ is the attention score between two tokens.

Fig.~\ref{fig:attention_scores}(a) shows the attention distribution for state $\mathrm{|0_1\rangle|Bell_{23}\rangle|0_4\rangle}$ with only 2 entangled qubits. 
For this state, connections can be drawn between the attention distribution and the entanglement structure. Each row $a_i$ in Layer 1, Head 1 has nonzero attention only on embeddings $x^{(i,j)}$ such that one of $i$ or $j$ is from the Bell subsystem (qubits $(2,3)$), and the other -- from the $(1,4)$ qubit subsystem. Head 2 in Layer 1 attends primarily to the Bell subsystem, except $x^{(1,4)}$ which attends to itself.
In Layer 2, Head 1 we see that $x^{(2,3)}$ attends only to $x^{(1,4)}$ and $x^{(2,3)}$. As $x^{(1,4)}$ and $x^{(2,3)}$ are embeddings of $o^{(1,4)}$ and $o^{(2,3)}$, they represent the $\mathrm{|0_{1}\rangle}\mathrm{|0_{4}\rangle}$ and $|\mathrm{Bell}_{23}\rangle$ subsystems, respectively. Recall also that the policy network maps each observation $o^{(i,j)}$ to the probability for taking action $(i,j)$ (see Secs.~\ref{app:act_space} and \ref{app:obs_space}). A connection can then be drawn between the selected action and the entanglement structure of the state: the action with the highest probability, in this case, is (2,3) and its corresponding embedding $x^{(2,3)}$ attends only to the embeddings that represent subsystems $\mathrm{|0_{1}\rangle}\mathrm{|0_{4}\rangle}$ and $|\mathrm{Bell}_{23}\rangle$.

Figure~\ref{fig:attention_scores}(b) shows the attention distribution for a state with two pairs of entangled subsystems -- this time the Haar-random state $\mathrm{|R_{12}\rangle|R_{34}\rangle}$. In Layer 1, Head 1 every $x^{(i,j)}$ attends to $x^{(1,2)}$ which represents $\mathrm{|R_{12}\rangle}$. In Head 2 every (row) token attends to only one (column) token; except for $x^{(2,3)}$ which attends to itself, all other tokens $x^{(i,j)}$ attend to a token that includes either qubit $i$ or $j$ in its RDM. In Layer 2, Head 1 most of the tokens attend to $x^{(2,4)}$. Contrary to the $\mathrm{|0_1\rangle|Bell_{23}\rangle|0_4\rangle}$ example, here it is harder to draw a connection between the action the agent selects and the attention distribution.
Interpreting transformer architectures is a difficult yet important open problem at the forefront of modern research~\cite{suresh2024interpretable}.

\subsection{Hyper-parameters \& training times} \label{app:hyperparams}

In the following Table~\ref{table:hyperparams} we provide all of the hyper-parameters needed to reproduce the results from the experiments. The hyper-parameters are split into three categories:
i) hyper-parameters concerning the environment;
ii) hyper-parameters concerning the agent policy architecture;
iii) hyper-parameters concerning the PPO training procedure.
\par Table~\ref{table:train_times} shows the approximate training times of the RL agents for each  system size $L$. 

\begin{table*}[t!]
\setlength{\tabcolsep}{6pt}
\centering
\caption{Hyper-parameters used for training the RL agents on quantum systems of different sizes. The hyper-parameters are split into three categories: environment, policy, and training hyper-parameters.}
\begin{tabular}{ p{7.6cm} p{2cm} p{1cm} p{1cm} p{1cm} p{1.4cm} p{1.4cm}}
    \hline
    Description & Notation & $L=4$ & $L=5$ & $L=6$ & $L=12$ & $L=16$\\
    \hline
    \textsc{Environment} \\
    Number of states in the vectorized environment (Sec.~\ref{app:vector_env})     & $B$                & 64  & 128  & 512 & 64 & 64 \\
    Maximum number of episode steps (Sec.~\ref{app:environment_step})              & $T_{\text{trunc}}$ & 8    & 40   & 90 & 100 & 200 \\
    Disentangling threshold for $S_\text{tot}$ (Sec.~\ref{app:environment_step})   & $\epsilon$         & 1e-3 & 1e-3 & 1e-3 & 1e-3 & 1e-3 \\
    Minimum support for generating initial states  (Sec.~\ref{app:init_state_gen}) & $p$                & 2    & 2    & 3 & 2 & 2 \\
    Inter-subsystem entanglement (Sec.~\ref{app:large_systems}                     & $\eta$             & -    & -    & - & 4.1 & 4.1 \\
    \hline
    \textsc{Policy} (Sec.~\ref{app:net_arch})\\
    Number of transformer encoders                   & $N_{\text{layers}}$ & 2   & 4   & 4 & 8 & 8 \\
    Number of attention heads per layer              & $N_{\text{heads}}$  & 2   & 4   & 4 & 4 & 4 \\
    Embedding dimension for the self-attention layer & $D_{\text{qkv}}$    & 128 & 128 & 256 & 256 & 256 \\
    Fully-connected encoder layer dimension          & $D_{\text{mlp}}$    & 256 & 512 & 1024 & 1024 & 1024 \\
    Value network number of layers                   & $N_\text{value\_layers}$ & 3 & 3 & 3 & 3 & 3 \\
    Value network hidden layers size                 & $h_{\text{hid}}$    & 128 & 256 & 256 & 256 & 256 \\
    \hline
    \textsc{Training} \\
    Number of steps per trajectory segment (Sec.~\ref{app:train_algo_ppo})              & $T_{\text{seg}}$       & 16   & 64    & 90 & 60 & 100 \\
    Number of training iterations                                                       & $N_{\text{iters}}$     & 4000 & 10000 & 12000 & 1000+20000 & 1000+20000 \\
    Policy network learning rate                                                        & $lr_{\pi}$             & 2e-4 & 2e-4  & 2e-4 & 3e-6 & 3e-6 \\
    Clip value for PPO updates of the policy (Sec.~\ref{app:ppo_clip_details})          & $\epsilon_{\pi}$       & 0.2  & 0.2 & 0.2 & 0.2 & 0.2 \\
    Value network learning rate                                                         & $lr_{V}$               & 3e-4 & 3e-4  & 3e-4 & 1e-5 & 1e-5 \\
    Clip value for updates of the value network (Sec.~\ref{app:ppo_value_clip_details}) & $\epsilon_{V}$         & 10.  & 10.   & 10. & 10. & 10. \\
    $D_{\text{KL}}$ threshold for early stopping (Eq.~\eqref{eq:kl_div_calc})           & $KL_{\text{lim}}$      & 0.01 & 0.01  & 0.01 & 0.01 & 0.01 \\
    Discount factor for advantage estimation (Eq.~\eqref{eq:gae})                        & $\lambda_{\text{GAE}}$ & 0.95 & 0.95  & 0.95 & 0.95 & 0.95 \\
    Gradient clipping by norm for both networks \cite{gradclip2012Mikolov}              & $g_{\text{clip}}$      & 1.0  & 1.0   & 1.0 & 1.0 & 1.0 \\
    Temperature for entropy regularization (Eq.~\eqref{eq:entropy_reg_obj})              & $\beta^{-1}$           & 0.1  & 0.1   & 0.1 & 0.001 & 0.001 \\
    Number of PPO updates per iteration (Sec.~\ref{app:train_algo_ppo}) & $n_{\text{ppo\_updates}}$ & 96 & 96 & 96 & 96 & 96 \\
    Batch size for PPO updates (Sec.~\ref{app:train_algo_ppo})                          & $B_{\text{PPO}}$       & 128  & 128   & 512 & 512 & 512 \\
    Discount factor for future rewards (Eq.~\eqref{eq:exp_return})                      & $\gamma$               & 1.0  & 1.0   & 1.0 & 0.99 & 0.99 \\
    \hline
\end{tabular}
\label{table:hyperparams}
\end{table*}


\begin{table}[h]
    \centering
    \begin{tabular}{p{5cm} p{3cm}}
        \hline
        \textbf{Number of qubits $L$} & \textbf{Training time} \\
        \hline
        4  & 0.5 hours \\
        5  & 10 hours \\
        6  & 60 hours \\
        12 & 320 hours \\
        16 & 390 hours \\
        \hline
    \end{tabular}
    \caption{Approximate training times for different system sizes $L$. We used NVIDIA Tesla T4 GPU for $L\leq 6$ and
    NVIDIA 4070 Ti Super GPU for $L\geq 12$. The corresponding number of iterations is listed in Table~\ref{table:hyperparams}. 
    }
    \label{table:train_times}
\end{table}

\section{Beam search for circuit optimization} \label{app:beam_search}

    \subsection{Tree search algorithms}

    As discussed in Sec.~\ref{sec:setup}, the disentangling problem can be reduced to two sub-problems: (1) identifying the sequence of pairs of qubits to apply the two-qubit gates on, and (2) finding the corresponding optimal two-qubit unitary gates. A partial analytical solution to (2) is provided in Sec.~\ref{sec:two_qubit_U}. In this section, we propose an alternative solution to (1) using a tree search algorithm which is very suitable for noise-free environments.

    Given the combinatorial nature of the problem, and more precisely, the fact that different sequences of actions need to be explored before a solution is found, one may argue that other classes of algorithms, such as tree-search algorithms, could also be suitable for this task.

    \begin{figure}[t!]
        \centering
        \includegraphics[width=0.45\textwidth]{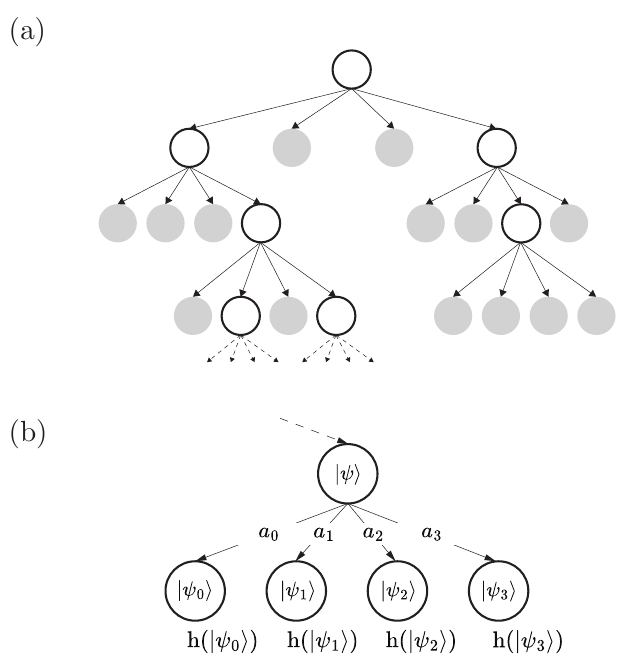}
        \caption[Beam search]{An example execution of the beam search algorithm. Subfigure (a) shows three iterations of the algorithm. At each iteration the current leaf nodes are compared and sorted using a heuristic function $h$. Only the best $k=2$ nodes are expanded and the rest (shown in gray) are pruned. Expanding a node, shown in (b), consists of applying every possible action from the action space $|\mathcal{A}|=4$ to this node - the new states $\ket{\psi_0}, \dots, \ket{\psi_3}$ become the child nodes of the expanded node.
        }
        \label{fig:beam_search}
    \end{figure}

    Indeed, in the case of a noise-free environment, a tree-search procedure may be used to produce the disentangling circuit offline. In this case, a tree-search agent takes as input the initial quantum state $|\psi\rangle$ in its entirety (i.e., no partial observations), and produces the actions that need to be applied for the state to be disentangled. We call the resulting sequence of actions a \textit{plan}. Note that in case the environment is non-deterministic (e.g., due to noise), using the plan is not guaranteed to lead to a solution. (Obviously, deviating from the planned trajectory due to some non-determinism would set the agent off-course and, thus, render the plan useless.)

    The number of possible $T$-step sequences of actions is $|\mathcal{A}|^T$, where $|\mathcal{A}| = L(L-1)/2$, and $L$ is the number of qubits in the quantum state. As discussed in Sec.~\ref{subsec:large_sys_rnd}, the number of actions needed to disentangle a quantum system grows exponentially with the system size, i.e. $T\sim \exp(L)$. Thus, for the size $|\mathcal{S}|$ of the search space of our tree-search algorithm, we have $|\mathcal{S}| \sim \exp(\exp(L))$ -- a super-exponential growth.

    Therefore, in large state spaces global search algorithms are very inefficient to run because they need to explore the entire space to find the optimal solution. Local search algorithms, on the other hand, are usually very efficient and manage to find near-optimal solutions in large state spaces. In what follows we describe how the beam search algorithm works and how it is used to produce disentangling circuits.

    A search algorithm \cite{russell_norvig_aima} works by examining the various paths/branches that are formed from the initial state, trying to find a path that reaches the goal state. Starting with the initial state at the root of a search tree, the algorithm sequentially expands state nodes from the tree until it reaches the goal state. The process of expanding a state node consists of applying all of the actions from the action space separately to the current state and creating child state nodes for each of the produced states. This implies that state nodes that are ready for expansion are leaf nodes, while inner nodes are already expanded.

    A global search algorithm proceeds by expanding every node state in the search tree. Local search algorithms, and beam search in particular, expand only a subset of the nodes, and the rest are pruned. At every iteration of the beam search algorithm, a set of $k$ leaf nodes are selected and expanded, while the rest of the leaf nodes are marked non-expandable, and are irreversibly discarded from the tree-search. The value $k$ is called the \textit{beam size} of the algorithm. An example execution of the beam search algorithm can be seen in Fig.~\ref{fig:beam_search}.

    \subsection{Heuristic functions for evaluating tree nodes}

    The performance of the beam search algorithm depends on the ranking algorithm that is used to pick the $k$-best leaf nodes for expansion. In what follows we describe the ranking heuristic $h$ that we use for producing disentangling circuits.

    The heuristic $h$ is a function that takes as input a pure $L$-qubit quantum state and returns a real number:
    \begin{equation*}
        h: \mathbb{C}^{2^L} \rightarrow \mathbb{R}.
    \end{equation*}
    Note that, besides assuming a noise-free environment, we also assume that at every state node we have access to the full quantum state at this step (the second assumption needs to be relaxed for the algorithm to be applicable in an experiment).

    Our first choice of a heuristic function is the average entanglement of the entire state calculated as given in Eq.~\eqref{eq:avg_entanglement}:
    \begin{equation} \label{eq:app_standard_heuristic}
        h(|\psi\rangle) = S_\text{avg} = \frac{1}{L}\sum_{j=1}^L S_\text{ent}[\rho^{(j)}],
    \end{equation}
    where $\rho^{(j)}$ is the reduced density matrix of qubit $j$ for the pure state $|\psi\rangle$.

    \begin{figure}[t!]
        \centering
        \includegraphics[width=0.45\textwidth]{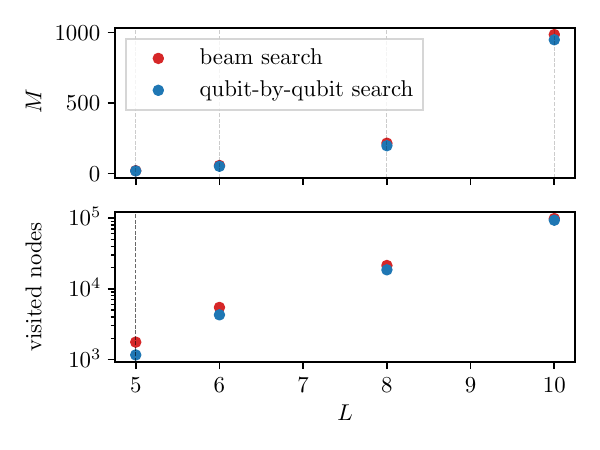}
        \caption[Beam search comparison]{Comparison between the standard beam search algorithm (red) and the modified `qubit-by-qubit' version (blue) [see text].
        (a) Average number of gates $M$ needed to disentangle a quantum state.
        (b) Average number of nodes that the algorithms visits before disentangling a quantum state.
        The modified `qubit-by-qubit' version requires fewer number of visited nodes while also producing shorter disentangling protocols. The modification provides an improvement of around $10\%$ for both metrics. The results are averaged over 100 test simulations with different random initial states. Both algorithms are evaluated on the same set of states.
        We render a state disentangled when $S_\text{tot}<\epsilon$. The value of $\epsilon$ for $L=5,6$ is as per Table~\ref{table:hyperparams}; for $L=8,10$ we set $\epsilon=0.01$.
        }
        \label{fig:beam_search_stats}
    \end{figure}

    However, a very simple improvement can be made by noting that once a qubit is disentangled from the rest, then actions that do not involve this qubit can no longer influence the single-qubit entanglement entropy $S_\text{ent}[\rho^{(j)}]$ of that qubit. Thus, we focus on disentangling the rest of the system, effectively reducing the system size. Instead of aiming to disentangle the entire state as a whole, we run the beam search algorithm with the goal to disentangle a single fixed qubit, say qubit $j$. The heuristic function that we use for this run of the algorithm is then:
    \begin{equation*}
        h(|\psi\rangle) = S_\text{ent}[\rho^{(j)}].
    \end{equation*}

    Once qubit $j$ is disentangled, we arrive at the new state $|\psi_1\rangle = |0_j\rangle|\varphi_{1,\dots,j-1,j+1,\dots,L}\rangle$. We then repeat the procedure but for the state $|\psi_1\rangle$ with the goal to disentangle another qubit, say qubit $k$. The heuristic function that we use for this second run is:
    \begin{equation*}
        h(|\varphi_{1,\dots,j-1,j+1,\dots,L}\rangle) = S_\text{ent}[\rho^{(k)}],
    \end{equation*}
    where $\rho^{(k)}$ is the reduced density matrix of qubit $k$ for the state $|\varphi_{1,\dots,j-1,j+1,\dots,L}\rangle$. 

    We keep repeating this procedure iteratively until the entire system is disentangled qubit by qubit.

    This strategy provides an improvement to the algorithm of around $10\%$ both in terms of number of actions and in terms of speed of execution. A comparison between the standard beam search algorithm using the heuristic function from Eq.~\eqref{eq:app_standard_heuristic} and this modified `qubit-by-qubit' version can be seen in Fig.~\ref{fig:beam_search_stats}. Not only does using the modified heuristic function produce shorter sequences, but the algorithm also runs faster.

    Even though the beam search algorithm manages to find near-optimal solutions to the problem, it is obvious that the time needed to run the algorithm continues to grow exponentially with the number of qubits. Moreover, the search algorithm has no learning component, i.e., it has to be re-run for every different initial state separately. Finally, we mention in passing that the solution produced by the beam search algorithm need not be physically optimal: shorter sequences may exist that bring the system into a product state using fewer gates. For all of these reasons, we focus on models trained on data that can learn the correct sequence of actions without having to perform this expensive look-ahead search. In the future, it will be interesting to combine search algorithms with deep learning~\cite{dalgaard2020global} to solve the state disentangling problem.

\bibliography{bibliography.bib}

\end{document}